\documentclass[a4paper,fleqn,usenatbib]{mnras}

\usepackage{newtxtext,newtxmath}
\usepackage[T1]{fontenc}
\usepackage{ae,aecompl}

\usepackage{epstopdf}
\usepackage{graphicx}	% Including figure files
\usepackage{amsmath}	% Advanced maths commands
\usepackage{amssymb}	% Extra maths symbols
\usepackage{url,bm}
\usepackage{nicefrac}
\usepackage{caption}
\usepackage{subfigure}
\usepackage{CJKutf8}

\usepackage{etoolbox}
\makeatletter
\patchcmd\@combinedblfloats{\box\@outputbox}{\unvbox\@outputbox}{}{%
   \errmessage{\noexpand\@combinedblfloats could not be patched}%
}%
 \makeatother

\newcommand{\eg}{\textit{e.g., }}
\newcommand{\ie}{\textit{i.e. }}

\title[Hierarchical quadruple systems]{Dynamics of Quadruple Systems Composed of Two Binaries: Stars, White Dwarfs, and Implications for Ia Supernovae}

\author[X. Fang et al.]{
Xiao Fang (\begin{CJK*}{UTF8}{gbsn}方啸\end{CJK*}),$^{1,3}$\thanks{E-mail: fang.307@osu.edu}
Todd A. Thompson,$^{2,3}$
and Christopher M. Hirata$^{1,3}$
\\
$^{1}$Department of Physics, The Ohio State University, Columbus, Ohio 43210, USA\\
$^{2}$Department of Astronomy, The Ohio State University, Columbus, Ohio 43210, USA\\
$^{3}$Center for Cosmology and AstroParticle Physics, The Ohio State University, Columbus, Ohio 43210, USA\\
}

\date{Accepted XXX. Received YYY; in original form ZZZ}

\pubyear{2017}

\begin{document}
\label{firstpage}
\pagerange{\pageref{firstpage}--\pageref{lastpage}}
   \maketitle

\begin{abstract}
 We investigate the long-term secular dynamics and Lidov-Kozai (LK) eccentricity oscillations of quadruple systems composed of two binaries at quadrupole and octupole order in the perturbing Hamiltonian. We show that the fraction of systems reaching high eccentricities is enhanced relative to triple systems, over a broader range of parameter space. We show that this fraction grows with time, unlike triple systems evolved at quadrupole order. This is fundamentally because with their additional degrees of freedom, quadruple systems do not have a maximal set of commuting constants of the motion, even in secular theory at quadrupole order. We discuss these results in the context of star-star and white dwarf-white dwarf (WD) binaries, with emphasis on WD-WD mergers and collisions relevant to the Type Ia supernova problem. For star-star systems, we find that more than 30\% of systems reach high eccentricity within a Hubble time, potentially forming triple systems via stellar mergers or close binaries. For WD-WD systems, taking into account general relativistic and tidal precession and dissipation, we show that the merger rate is enhanced in quadruple systems relative to triple systems by a factor of $3.5-10$, and that the long-term evolution of quadruple systems leads to a delay-time distribution $\sim 1/t$ for mergers and collisions. In gravitational wave (GW)-driven mergers of compact objects, we classify the mergers by their evolutionary patterns in phase space and identify a regime in about 8\%\ of orbital shrinking mergers, where eccentricity oscillations occur on the general relativistic precession timescale, rather than the much longer LK timescale. Finally, we generalize previous treatments of oscillations in the inner binary eccentricity (evection) to eccentric mutual orbits. We assess the merger rate in quadruple and triple systems and the implications for their viability as progenitors of stellar mergers and Type Ia supernovae.
\end{abstract}


%%%%%%%%%%%%%%%%%%%%%%%%%%%%%%%%%%%%%%%%%%%%%%%%%%

%%%%%%%%%%%%%%%%% BODY OF PAPER %%%%%%%%%%%%%%%%%%

\section{Introduction}

The dynamics of hierarchical triple systems has long been investigated. This is a special case of triple systems whose tertiary is at a large distance, serving as a perturber of the inner binary. At quadrupole order in the perturbing Hamiltonian, the eccentricity of the inner binary and the mutual inclination between the inner and outer orbit exhibit periodic oscillations, the Lidov-Kozai (LK) oscillations, on a timescale much longer than both of the orbital periods \citep[\eg][]{1962P&SS....9..719L,1962AJ.....67..591K}. Starting with a high tertiary inclination orbit, the initially low eccentricity of the inner binary can reach a very high value. Due to the high sensitivity of tidal interactions to the orbital eccentricity, this phenomenon has many potentially interesting astrophysical implications, such as inducing migration of planets and producing hot Jupiters \citep[\eg][]{2003ApJ...589..605W,2007ApJ...669.1298F,2017MNRAS.464..688H}, tight binaries, stellar mergers, and even blue stragglers \citep[\eg][]{1979A&A....77..145M,2001ApJ...562.1012E,2006A&A...450..681T,2009ApJ...697.1048P,2013ApJ...766...64S,2014MNRAS.439.1079A,2014ApJ...793..137N,2016ApJ...816...65A,2017ApJ...846..146P}. The LK mechanism and its higher-order effects have also found applications in systems with compact objects due to the strong eccentricity dependence of the general relativistic precession and gravitational wave (GW) dissipation. It has been proposed that this mechanism could be relevant to the evolution of intermediate-mass black holes in globular clusters \citep[\eg][]{2002MNRAS.330..232C,2003ApJ...598..419W} and super-massive black holes in the centres of galaxies \citep[\eg][]{2002ApJ...578..775B,2012ApJ...757...27A,2013ApJ...773..187N,2016MNRAS.460.3494S,2017arXiv170609896H,2016ApJ...816...65A}.

There have also been many discussions of white dwarf (WD) mergers as candidate progenitors of Type Ia supernovae (SNe Ia): in this ``double-degenerate scenario'' (DDS), two WDs gradually lose their orbital energy and angular momentum via GWs before merging with each other \citep[\eg][]{1984ApJ...277..355W,1984ApJS...54..335I}. However, the GW energy dissipation rate suggests that the WD binaries have to start with a compact orbit (semi-major axis $a<0.01$ AU) to merge within the Hubble time, calling for a mechanism to produce compact WD binaries. One of the proposals is that the orbit rapidly shrinks during the common envelope phase, as the removal of the common envelope takes away lots of orbital energy. Although some results from binary population synthesis models have shown the possibility of explaining the SN Ia rate with the DDS, they depend on the modelling of the common envelope physics, and recent calculations under-predict the short-time-delay SN Ia rate \citep[\eg][]{2009ApJ...699.2026R}.

Given the uncertainties in the derived SN Ia rate from WD-WD binaries, it is interesting to consider the role of triple systems, where gravitational dynamics could lead to rapid gravitational wave driven mergers. Following work by \cite{2002ApJ...578..775B} and \cite{2002MNRAS.330..232C} in other contexts, \cite{2011ApJ...741...82T} showed that the merger time for WD-WD binaries can be decreased by orders of magnitude by the Lidov-Kozai (LK) eccentricity oscillations caused by the tertiary, and argued that most SNe Ia may occur in hierarchical triple systems. 

However, there are two potential problems with this hypothesis. The first is the ``inclination problem": the LK oscillations that lead to a rapid WD-WD merger would have already led to close encounters of the inner binary stars before they evolved into WDs. Thus, although triple systems may play a role in forming the tight binaries that eventually lead to WD-WD binaries through traditional common envelope evolution, the tertiary would not participate in driving the merger of the two WDs per se. The ``eccentric LK mechanism" --- octupole-order oscillations that can produce much higher eccentricities than quadrupole-order LK oscillations when the components of the inner binary have unequal masses \citep[\eg][]{2000ApJ...535..385F,2011ApJ...742...94L, 2011PhRvL.107r1101K,2013MNRAS.431.2155N} --- could potentially exacerbate this issue by driving more binaries to contact during stellar evolution \citep[\eg][]{2009ApJ...697.1048P,2014ApJ...793..137N}. In addition, \cite{2013ApJ...766...64S} found that mass loss can instigate the eccentric LK mechanism after the first WD forms, potentially increasing the formation rate of tight WD-star systems. In an effort to mitigate this issue, \citet{2016MNRAS.456.4219A} investigated dynamical scattering and flyby encounters as a way to generate high-inclination triple systems after WD binary formation, but significant uncertainty remains about the evolution of triple systems as their components evolve.

The second ``rate problem" with the triple scenario is the same as that for the normal stellar binary channel for tight WD-WD binaries: it is unclear if the observed SN Ia rate can be accommodated. In the triple scenario, in order to reach very high eccentricities, the initial inclinations are limited to a very narrow range in secular theory, potentially making it hard to explain the observed SN Ia rate. This issue was partially addressed by \cite{2012arXiv1211.4584K} who showed with $N$-body simulations that non-secular dynamics can produce ``clean'', head-on collisions of WD-WD binaries in about 5\% of moderately hierarchical triple systems. The possibility of such head-on collisions producing SNe was supported by \cite{2013ApJ...778L..37K}, who computed explosion models for colliding WDs. \cite{2014MNRAS.438.3456P} compared the expected Ia luminosity function in the collision scenario, finding that low-luminosity supernovae are preferred because of the observed strong peak in the WD mass function. In contrast with \cite{2012arXiv1211.4584K},  \cite{2017arXiv170900422T} recently estimated the clean WD-WD collision rate in triples and found it to be $\sim2-3$ orders of magnitude lower than the observed SN Ia rate, with an almost uniform delay-time distribution that is inconsistent with observations. The role of {\it mergers} (rather than collisions) in producing the observed rate and delay-time distribution has not yet been explored.

In this paper, we calculate the secular dynamics of hierarchical triple systems and quadruple systems composed of two binaries with an eye towards addressing these ``inclination" and ``rate" problems.  In particular, \cite{2013MNRAS.435..943P} showed that the fraction of quadruple systems reaching high-eccentricity (high-$e$) is greatly enhanced compared to otherwise identical triple systems, potentially suggesting an increased rate for quadruples. However, because \cite{2013MNRAS.435..943P} used full few-body dynamics, their investigation of the parameter space was necessarily limited.  Here, we derive the equations for the secular evolution of quadruple systems including a treatment of general relativity and tides. We show that quadruple systems exhibit irregular behaviour even at quadrupole order, in contrast to the regular LK oscillations in triples at the same order. We further show that the high-$e$ fraction produced by quadruple systems is large and that it grows steadily in time, producing mergers or collisions of WD-WD binaries. We find that the WD-WD merger rate is $\sim3.5-10$ times larger than for triples, and that the majority of the mergers are highly eccentric inspirals or potentially collisions. The delay-time distribution for both quadruples and triples follows $\sim t^{-1}$. Given the relative fraction of observed triples and quadruples, these findings lead us to propose that quadruples may dominate WD-WD mergers.

We explore how the rate of mergers in quadruple systems depends on the WD masses, separation, and relative inclinations. We classify the mergers by their evolution patterns in phase space right before their orbits rapidly shrink, and identify $\sim 8\%$ of mergers that experience a previously unidentified ``precession oscillation'' phase at the beginning of their orbital shrinking.

An important component of the problem for both triples and quadruples is the role of non-secular dynamics. In particular, rapid eccentricity oscillations occurring on the timescale of the mutual or outer orbit --- ``evection'' --- can cause large perturbations to the angular momentum of the inner binary while it is at high eccentricity \citep[\eg][]{2005MNRAS.358.1361I,2012arXiv1211.4584K,2012ApJ...757...27A,2014MNRAS.438..573B,2014MNRAS.439.1079A}.  Previous treatments have either relied on fully dynamical calculations or analytic expressions derived in limiting cases. We generalize these previous analytic investigations to arbitrary mutual eccentricity, and assess evection for the merger and collision rate of both triple and quadruple systems over the range of semi-major axis ratios we explore. We find that evection slightly enhances the merger rates, but may have a more substantial effect on the nature of the merger (\eg whether head-on collisions or gravitational wave-driven mergers).

The remainder of this paper is organized as follows. In \S\ref{sec:theory}, we discuss the secular effects we have considered in this work. In \S\ref{sec:features}, we compare the secular results from quadruple systems to the triple systems and discuss the new features we find in quadruple systems that could lead to important astrophysical implications. Then, in \S\ref{sec:SN}, we apply our calculations to the ``quadruple scenario'' of WD mergers and show how it can shed light on the SN Ia rate puzzle. We discuss the role of evection in \S\ref{sec:evection}. Finally we summarize our results and discuss the caveats and limitations of our work in \S\ref{sec:conclusion}. We tabulate the coefficients in the octupole perturbation formula in Appendix \ref{App:Octu}. Descriptions and tests of our secular code are presented in Appendix \ref{App:code}. The physics of a ``precession oscillation'' phenomenon in some WD mergers is explained in Appendix \ref{app:rapidwiggle}. Detailed calculations pertaining to evection are described in Appendix \ref{App:Evec}.

\section{Secular Theory}
\label{sec:theory}

\begin{figure}
\centering
\includegraphics[width=\columnwidth]{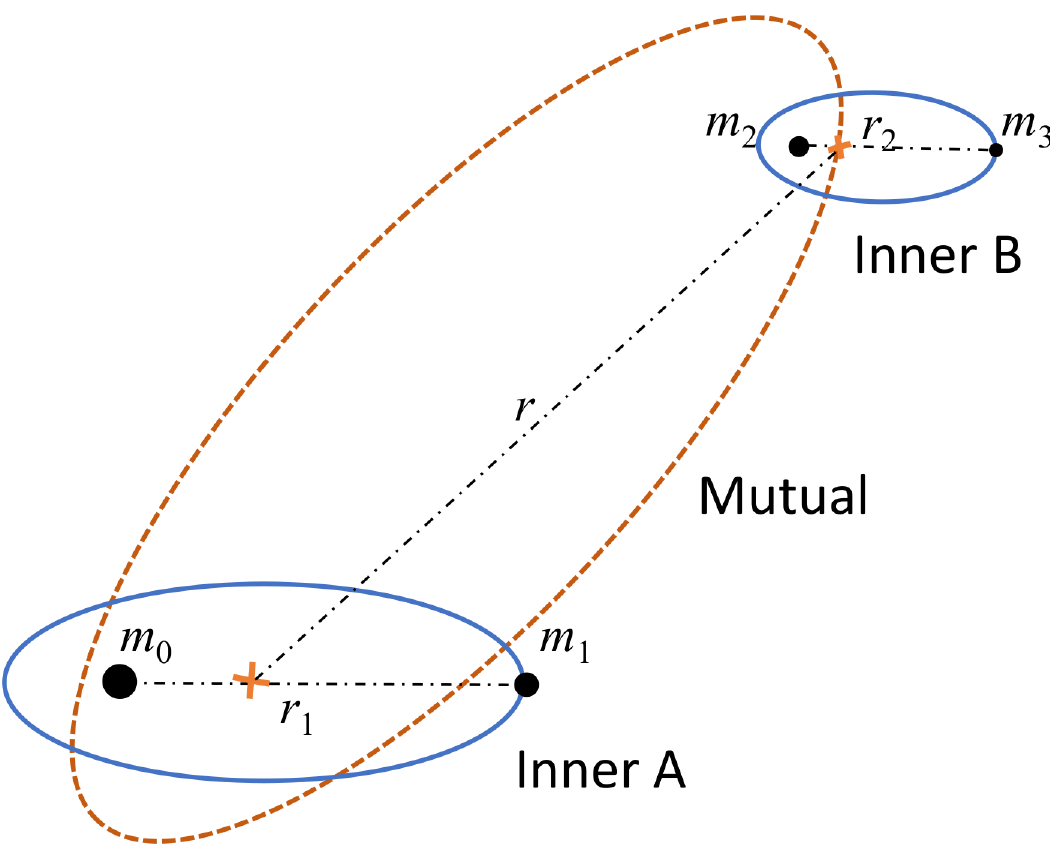}
\caption{Illustration of a ``2+2'' hierarchical quadruple star system. Masses $m_0$ and $m_1$ form ``inner binary A'' with separation $r_1$, $m_2$ and $m_3$ form ``inner binary B'' with separation $r_2$, and their centres of masses orbit each other in the ``mutual'' orbit with separation $r$. We focus on systems where $r_1,r_2\ll r$.}
\label{fig:quad}
\end{figure}

\begin{figure}
\centering
\includegraphics[width=\columnwidth]{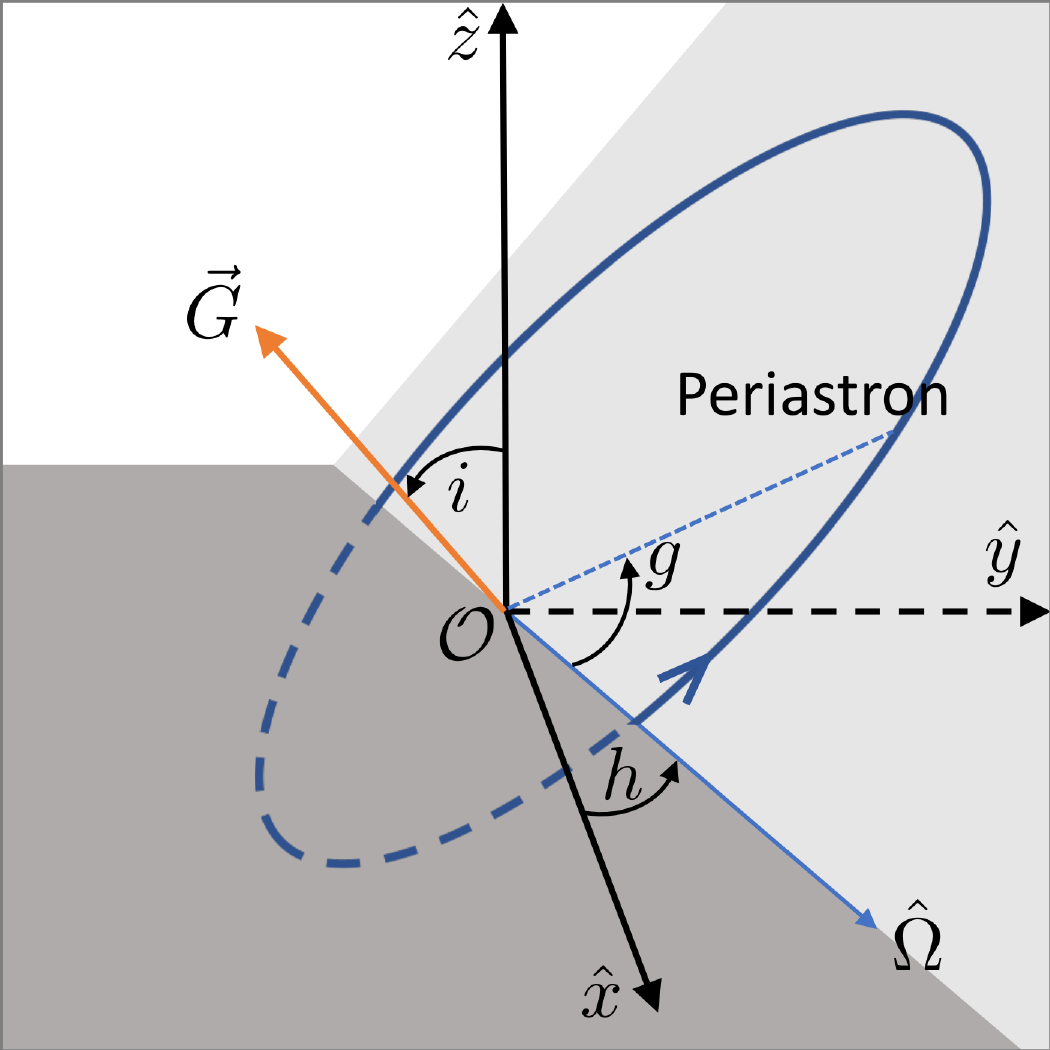}
\caption{Illustration of the orbital elements. The orbital plane intersects the reference plane $\hat{x}-\hat{y}$ along the line of nodes with the direction of ascending node denoted by $\hat{\Omega}$. $h$ defines the argument of ascending node with respect to the reference plane, and $g$ defines the argument of periastron in the orbital plane. The angle between the orbital angular momentum $\bm{G}$ and the $z-$axis defines the inclination $i$, which is also the angle between the reference plane and the orbital plane.}
\label{fig:coordinate}
\end{figure}

The ``$2+2$'' hierarchical quadruple system is composed of two binary systems, with their centres of mass $C_1$ and $C_2$ separated by $r$, as shown in Fig.~\ref{fig:quad}.
The mutual orbit has parameters: $a$, $i$, $e$, $g$, and $h$, representing the semi-major axis, the inclination between the orbit and the reference plane in the rest frame, the eccentricity, the argument of the periastron, and the argument of the ascending node, as illustrated in Figure \ref{fig:coordinate}. Each of the (inner) binary systems has a much smaller orbit. The first one (we call it inner orbit A) is composed of masses $m_0$ and $m_1$ and separation $r_1$, and has orbital parameters: $a_1$, $i_1$, $e_1$, $g_1$, and $h_1$; while the second one (\ie inner orbit B) is composed of masses $m_2$ and $m_3$ and separation $r_2$, and has parameters: $a_2$, $i_2$, $e_2$, $g_2$, and $h_2$. Note that $a_1,a_2\ll a$, as defined by the ``hierarchical'' assumption. We also define the inclinations between the inner orbits and the mutual orbit as $i_A$ and $i_B$, respectively.

In this section, we introduce the secular effects we have considered, including up to octupole order in the expansion of the Hamiltonian (\S\S\ref{subsec:NG},\ref{subsec:Octu}), general relativistic precession (\S\ref{subsec:1PN}) and dissipation (\S\ref{subsec:GW}), and tidal precession (\S\ref{subsec:TidalPrec}) and dissipation (\S\ref{subsec:TidalDissip}). At the end (\S\ref{subsec:OtherEff}), we will discuss other effects that we have ignored and justify why they will not jeopardize our results. The role of the non-secular effect, evection, will be considered in \S\ref{sec:evection}.

\subsection{Newtonian gravity and quadrupole order interactions}
\label{subsec:NG}

In the point-mass limit, the Hamiltonian of the quadruple system is given by
\begin{equation}
\mathcal{H}=-\frac{\mathcal{G}m_0m_1}{2a_1}-\frac{\mathcal{G}m_2m_3}{2a_2}+T_{\rm out}+V_{02}+V_{03}+V_{12}+V_{13}~,
\end{equation}
where $T_{\rm out}$ is the kinetic energy of the mutual orbital motion and $V_{ij}=-\frac{\mathcal{G}m_im_j}{r_{ij}}$ are the gravitational potential energy between two objects in different inner orbits. The first two terms on the right-hand side are equal to the total energy of the inner Keplerian orbits, where $\mathcal{G}=4\pi^2$ AU$^3$ yr$^{-2}$ $M_\odot^{-1}$ is Newton's constant.

We can write the Hamiltonian of the system as an expansion in terms of $\alpha_1\equiv a_1/a$ and $\alpha_2\equiv a_2/a$.
Keeping terms up to quadrupole order $\mathcal{O}(\alpha_1^2),\mathcal{O}(\alpha_2^2)$, the Hamiltonian reduces to
\begin{align}
\mathcal{H}=&-\frac{\mathcal{G}m_0m_1}{2a_1}-\frac{\mathcal{G}m_2m_3}{2a_2}-\frac{\mathcal{G}m_A m_B}{2a}
\nonumber\\
& -\frac{\mathcal{G}}{a}\left(\frac{a}{r}\right)^3
 \Bigl[\alpha_1^2 S_{1}\left(\frac{r_1}{a_1}\right)^2(3\cos^2\Phi_1-1)
\nonumber \\ &~~
+\alpha_2^2 S_{2}\left(\frac{r_2}{a_2}\right)^2(3\cos^2\Phi_2-1)\Bigr],
\end{align}
where $m_A = m_0+m_1$ and $m_B = m_2+m_3$ are the masses of the inner binaries; $S_1 = m_0m_1m_B/m_A$ and $S_2 = m_2m_3m_A/m_B$ are coefficients; and $\cos\Phi_1\equiv\hat{\bm{r}}_1\cdot\hat{\bm{r}}$ and $\cos\Phi_2\equiv\hat{\bm{r}}_2\cdot\hat{\bm{r}}$ represent angles between the separation vectors,
with $\hat{\bm{r}}_1$ being the unit vector pointing from $m_0$ to $m_1$, $\hat{\bm{r}}_2$ pointing from $m_2$ to $m_3$, and $\hat{\bm{r}}$ pointing from $C_1$ to $C_2$.
(See \eg \citealt{1968AJ.....73..190H} and \citealt{2000ApJ...535..385F} for similar derivations.)

We adopt Delaunay's canonical angle variables: $l_1,l_2,l$ (mean anomalies); $g_1,g_2,g$ (arguments of periastron); and $h_1,h_2,h$ (longitudes of ascending node) for both of the inner orbits (with subscripts ``1'' and ``2'' respectively) as well as the mutual orbit (without a subscript). Their conjugate actions are related to the orbital elements via
\begin{align}
\begin{array}{lll}
L_1=m_0m_1\sqrt{\mathcal{G}a_1/m_A}, & G_1=L_1\sqrt{1-e_1^2}, & H_1=G_1\cos i_1,\\
L_2=m_2m_3\sqrt{\mathcal{G}a_2/m_B}, & G_2=L_2\sqrt{1-e_2^2}, & H_2=G_2\cos i_2,\\
L=m_Am_B\sqrt{\mathcal{G}a/M}, & G=L\sqrt{1-e^2},{\rm~and} & H=G\cos i,
\label{eq:canon-variables}
\end{array}
\end{align}
where $M\equiv m_A+m_B$ is the total mass of the system \citep{2000ssd..book.....M}. The Hamiltonian can then be written as
\begin{align}
\mathcal{H}=&-\frac{\beta}{2L^2}-\frac{\beta_1}{2L_1^2}-\frac{\beta_2}{2L_2^2}
\nonumber\\ &
-8B_1\left(\frac{L_1^4}{L^6}\right)\left(\frac{r_1}{a_1}\right)^2\left(\frac{a}{r}\right)^3(3\cos^2\Phi_1-1)\nonumber\\
&-8B_2\left(\frac{L_2^4}{L^6}\right)\left(\frac{r_2}{a_2}\right)^2\left(\frac{a}{r}\right)^3(3\cos^2\Phi_2-1)~,
\label{eq:fullHamil-quad}
\end{align}
where the coefficients are
\begin{align}
\beta_1&=\mathcal{G}^2\frac{(m_0m_1)^3}{m_A}~,~~\beta_2=\mathcal{G}^2\frac{(m_2m_3)^3}{m_B}~,~~\beta=\mathcal{G}^2\frac{(m_Am_B)^3}{M}~,\nonumber \\
B_1&=\frac{\mathcal{G}^2}{16}\frac{(m_Am_B)^7}{(m_0m_1M)^3}~,~{\rm and}~~B_2=\frac{\mathcal{G}^2}{16}\frac{(m_Am_B)^7}{(m_2m_3M)^3}~.
\end{align}

Only the last two terms in Eq.~(\ref{eq:fullHamil-quad}) represent quadrupole order corrections due to the ``monopole-quadrupole'' interaction between the two inner orbits\footnote{The Hamiltonian of each inner orbit part is expanded into a series of terms, where the monopole moment is the Kepler term, and the dipole moment vanishes since it is taken around the centre of mass. Thus, the lowest order perturbation terms are the two ``monopole-quadrupole'' terms and the next order are ``monopole-octupole'' terms, which will be discussed in \S\ref{subsec:Octu}. At higher order there would be two ``monopole-hexadecapole'' terms and one ``quadrupole-quadrupole'' term, the latter of which could produce orbital resonances, but we will leave these terms to a future work.}, \ie
\begin{equation}
\mathcal{H} = \mathcal{H}^{\rm (Kepler)}_{\rm mutual}+\mathcal{H}^{\rm (Kepler)}_1+\mathcal{H}^{\rm (Kepler)}_2+\mathcal{H}^{\rm (quad)}_1+\mathcal{H}^{\rm (quad)}_2~,
\end{equation}
where the Kepler terms do not contribute to the secular equations of motion since the short-period motions will be averaged out and the angles $l_1,l_2,l$ will become cyclic, leaving their conjugate momenta invariant.

After averaging over the inner binary orbits and the mutual orbit, the quadrupole part $\mathcal{H}^{\rm (quad)}_1$ becomes
\begin{align}
&\overline{\mathcal{H}}^{\rm (quad)}_1 = -\frac{B_1L_1^4}{8G^3L^3}\times
\nonumber\\
&\left\lbrace 3\sin^2i \left[10e_1^2(3+\cos 2i_1)\cos 2g_1 + 4(2+3e_1^2)\sin^2 i_1\right]\cos 2\Delta h_1 \right. \nonumber\\
&+(1+3\cos 2i)\left[(2+3e_1^2)(1+3\cos 2i_1)+30e_1^2\cos 2g_1\sin^2 i_1 \right]\nonumber\\
&+12(2+3e_1^2-5e_1^2\cos 2g_1)\sin 2i \sin 2i_1\cos \Delta h_1\nonumber\\
&+120e_1^2\sin 2g_1\sin 2i \sin i_1 \sin\Delta h_1\nonumber\\
&\left.-120 e_1^2\sin 2g_1 \sin^2 i\cos i_1\sin 2\Delta h_1\right\rbrace~,
\end{align}
where $\Delta h_1 \equiv h_1-h~$, and $\overline{\mathcal{H}}_2^{\rm (quad)}$ has the same form with the subscript ``1'' substituted with ``2''.

The equations of motion can be acquired by expressing the averaged Hamiltonian in terms of the canonical variables listed in Eq.~(\ref{eq:canon-variables}), and applying Hamilton's equations. Note that the averaged quadrupole order Hamiltonian\footnote{This is true for any order.} only depends on $\Delta h_1$ and $\Delta h_2$ due to the conservation of the projected total angular momentum $H_{\rm tot}\equiv H_1+H_2+H$, which leads to a simplification to the equations of motion, \ie
\begin{equation}
\dot{H}^{(\rm quad)} = - \dot{H}_1^{(\rm quad)} - \dot{H}_2^{(\rm quad)}~.
\end{equation}

\subsection{Octupole order interactions}
\label{subsec:Octu}

The octupole-monopole interaction between inner orbit A and B vanishes when stars in the inner binary A have the same mass, due to the parity symmetry of the gravitational potential of the inner binary A. (This holds for {\em any} odd-$\ell$ moment of binary A.)

Similar to the treatment of the quadrupole in \S\ref{subsec:NG}, the octupole order gives two additional terms in the Hamiltonian, $\mathcal{H}_1^{\rm (oct)}$ and $\mathcal{H}_2^{\rm (oct)}$. We focus on the first octupole term, which corresponds to the inner orbit A interacting with the mutual orbit ($\mathcal{H}_2^{\rm (oct)}$ is similar). We have that
\begin{equation}
\mathcal{H}_1^{\rm (oct)} = -2 C_1\left(\frac{L_1^6}{L^8}\right)\left(\frac{r_1}{a_1}\right)^3\left(\frac{a}{r}\right)^4\left(5\cos^3\Phi_1-3\cos\Phi_1\right)~,
\end{equation}
where\footnote{Note that $C_1$ here corresponds to $\beta_3$ in, \eg \cite{2000ApJ...535..385F} or \cite{2013MNRAS.431.2155N}, for triple cases.}
\begin{equation}
C_1\equiv\frac{\mathcal{G}^2(m_Am_B)^9(m_0-m_1)}{4(m_0m_1)^5M^4}~.
\end{equation}
After double-averaging, we can write the Hamiltonian in the form of
\begin{align}
\overline{\mathcal{H}}_1^{\rm (oct)} = \lambda_1 f(G)\sum_{m=-3}^{3}&\left(\mathcal{A}^{(m)}\cos mh + \mathcal{B}^{(m)}\sin mh\right)\nonumber\\
&\times\left(\mathcal{A}_1^{(m)}\cos mh_1 + \mathcal{B}_1^{(m)}\sin mh_1\right)~,\label{eq:oct_avg}
\end{align}
where the prefactors are defined as $\lambda_1\equiv{15C_1L_1^6}/{2048L^4}$ and $f(G)\equiv {\sqrt{L^2-G^2}}/{G^5}$.
In the Fourier series, the coefficients $\mathcal{A}^{(m)}$ and $\mathcal{B}^{(m)}$ are functions of $g$ and $i$ only, and $\mathcal{A}_1^{(m)}$, $\mathcal{B}_1^{(m)}$ are functions of $e_1$, $g_1$ and $i_1$ only. Their explicit expressions are listed in Appendix \ref{App:Octu}. Note that for any $m$, we have the relation
\begin{align}
\mathcal{A}^{(m)} = \mathcal{B}^{(-m)}~,
\label{eq:coeff_relation}
\end{align}
which is guaranteed by the fact that the potential is real and rotationally invariant.\footnote{One can rewrite $\overline{\mathcal{H}}_1^{\rm (oct)}$ as $\sum_{m=-\ell}^{\ell}\langle\mathcal{O}^{*}_{\ell m}\rangle \langle\mathcal{O}_{1,\ell m}\rangle e^{im\Delta h_1}$, where $\mathcal{O}_{\ell m}$, $\mathcal{O}_{1,\ell m}$ are moments from mutual and inner orbit A, respectively, and $\ell=3$.}

The corresponding contribution to the equations of motion is easier to evaluate in this ``separated'' form. It is not necessary to rewrite the Hamiltonian in Eq.~(\ref{eq:oct_avg}) solely in the canonical variables. Instead, we can use the Jacobian to show
\begin{align}
&\dot{g}_1^{\rm (oct)} = -\frac{G_1}{e_1L_1^2}\frac{\partial \overline{\mathcal{H}}_1^{\rm (oct)}}{\partial e_1} + \frac{1}{G_1\tan i_1}\frac{\partial \overline{\mathcal{H}}_1^{\rm (oct)}}{\partial i_1}~,\nonumber\\
&\dot{h}_1^{\rm (oct)} = -\frac{1}{G_1\sin i_1}\frac{\partial \overline{\mathcal{H}}_1^{\rm (oct)}}{\partial i_1}~,~{\rm and}\nonumber\\
&\dot{h}^{\rm (oct)} = -\frac{1}{G\sin i}\frac{\partial \overline{\mathcal{H}}_1^{\rm (oct)}}{\partial i}-\frac{1}{G\sin i}\frac{\partial \overline{\mathcal{H}}_2^{\rm (oct)}}{\partial i}~,
\end{align}
while the other equations keep the canonical form. The additional equation is
\begin{equation}
	\dot{G}^{\rm (oct)} = -\frac{\partial \overline{\mathcal{H}}_1^{\rm (oct)}}{\partial g}-\frac{\partial \overline{\mathcal{H}}_2^{\rm (oct)}}{\partial g}.
\end{equation}
Note that this is non-zero, so whereas the magnitude of the angular momentum of the mutual orbit is conserved at quadrupole order, it is not conserved at octupole order.

\subsection{First-order Post-Newtonian (1PN) corrections}
\label{subsec:1PN}

The general relativistic (GR) corrections to a binary star orbit can be expanded in inverse powers of $c$. Expanding the corresponding Hamiltonian of the binary system in such metric up to order $1/c^2$ gives the so-called 1PN correction, which sources the leading part of the GR precession.

In the centre-of-mass frame, the 1PN Hamiltonian correction of the orbit A is given by \citep[\eg][]{2014grca.book..111D}
\begin{align}
c^2\mathcal{H}_1^{\rm (1PN)}= &\frac{\mu_1}{8}(3\nu_1-1)\frac{\bm{p}_0^4}{\mu_1^4}
\nonumber\\
& -\left[(3+\nu_1)\frac{\bm{p}_0^2}{\mu_1^2}+\frac{\nu_1}{\mu_1^2}p_{r0}^2\right]\frac{\mu_1\mathcal{G}m_A}{2r_1} + \frac{\mu_1\mathcal{G}^2m_A^2}{2r_1^2}~,
\end{align}
where $\bm{p}_0$ is the momentum of star ``0'' relative to the centre of mass of the binary, and the radial component is defined as $p_{r0}\equiv -\bm{p}_0\cdot\hat{\bm{r}}_1$. The reduced mass is $\mu_1\equiv m_0m_1/m_A$ and we use the mass parameter \begin{equation}
\nu_1\equiv\frac{\mu_1}{m_A}=\frac{m_0m_1}{m_A^2}\le\frac14~.
\end{equation}
After averaging over the orbit and dropping the constant terms (since they do not affect the equations of motion), we obtain the effective averaged 1PN Hamiltonian
\begin{equation}
\overline{\mathcal{H}}_1^{\rm (1PN, eff)} = -\frac{3\mathcal{G}^2\mu_1m_A^2L_1}{c^2a_1^2G_1}~,
\end{equation}
which leads to an additional orbital precession
\begin{equation}
\dot{g}_1^{\rm (1PN)}=\frac{\partial\overline{\mathcal{H}}_1^{\rm (1PN,eff)}}{\partial G_1}=\frac{3(\mathcal{G}m_A)^{3/2}}{c^2(1-e_1^2)a_1^{5/2}}~.
\end{equation}
The expression for the inner orbit B is similar.

\subsection{Tidal precession}
\label{subsec:TidalPrec}
In the case of stars approaching each other during a close periastron passage, the point-mass assumption is no longer a good approximation, and the initially spherical stars are deformed due to the tidal forces exerted by their companions. This leads to a correction in their gravitational potential, hence in their Hamiltonian.

Let $R_i$ be the radius of star $m_i$. Star $m_0$ develops a quasi-static quadrupole moment $\sim k_0m_1R_0^5/r_1^3$ \citep{2014LRR....17....2B}, where $r_1$ is the distance between the two stars and $k_0$ is the dimensionless Love numbers of the two stars\footnote{The tidal Love number, associated with quadrupole moment, is usually denoted as $k_2$. Here we drop the subscript ``2'' for simplicity. We take $k=0.01$ for WDs \citep{2012ApJ...747....4P} and $k=0.0138$ for main-sequence stars \citep{1995A&AS..109..441C,2011A&A...529A..50L}.}, which depends on their internal structure. The resulting Hamiltonian correction for the orbit A is given by
\begin{equation}
\mathcal{H}_1^{\rm (tide)}=-\frac{\mathcal{G}}{r_1^6}(m_0^2k_1R_1^5+m_1^2k_0R_0^5)~.
\end{equation}
The orbit average is
\begin{equation}
\overline{\mathcal{H}}_1^{\rm (tide)}=-\frac{\mathcal{G}}{a_1^6(1-e_1^2)^{9/2}}\left(m_0^2k_1R_1^5+m_1^2k_0R_0^5\right)\left(1+3e_1^2+\frac{3}{8}e_1^4\right)~,
\end{equation}
which leads to an additional precession rate \citep{2003ApJ...589..605W,2007ApJ...669.1298F}:
\begin{equation}
\dot{g}_1^{\rm (tide)}=\frac{15(\mathcal{G}m_A)^{1/2}}{a_1^{13/2}(1-e_1^2)^5}\left(\frac{m_0}{m_1}k_1R_1^5+\frac{m_1}{m_0}k_0R_0^5\right)\left(1+\frac{3}{2}e_1^2+\frac{1}{8}e_1^4\right)~.
\label{eq:g_tideprec}
\end{equation}

\subsection{Gravitational wave dissipation}
\label{subsec:GW}
Due to the gravitational wave emission, the orbits gradually dissipate their energy and angular momenta. The orbital averaged dissipation rates are given by \cite{1964PhRv..136.1224P}. Converted into our notation, the relevant equations of motion for the inner orbit A are
\begin{align}
\frac{\dot{L}_1^{\rm (GW)}}{L_1} &= - \frac{32\mathcal{G}^3m_0 m_1 m_A}{5 c^5 a_1^4 \left(1-e_1^2\right)^{7/2}}\left(1+ \frac{73}{24}e_1^2 + \frac{37}{96}e_1^4 \right)~,
\nonumber\\
\dot{G}_1^{\rm (GW)} &=  -\frac{32\mathcal{G}^{7/2} (m_0m_1)^2m_A^{1/2}}{5c^5 a_1^{7/2} \left(1-e_1^2\right)^2}\left(1+\frac{7}{8}e_1^2\right)~,
~{\rm and}\nonumber\\
\dot{H}_1^{\rm (GW)} &=\dot{G}_1^{\rm (GW)}\cos i_1~.
\end{align}

\subsection{Tidal dissipation}
\label{subsec:TidalDissip}

Tidal dissipation is much more complicated due to the existence of various types of tidal interaction mechanisms. In principle, the tides are categorized into the ``equilibrium tides'' and the ``dynamic tides.''

In the ``equilibrium tide'' models, the star is deformed to be roughly in equilibrium with the time-dependent potential of the system, and the viscosity of the stars dissipates the energy in the motion of the tides \citep[\eg][]{1880RSPT..171..713D,1973Ap&SS..23..459A,1980A&A....92..167H,1981A&A....99..126H,1982A&A...110...37H,1998ApJ...499..853E}. At the end, the orbit is circularized and the spins of the stars are aligned with the orbital axis. However, the tidal dissipation rate via this channel is very small in well-separated binaries. The circularization timescale can be estimated by \citep[\eg][]{1981A&A....99..126H} 
\begin{align}
\tau_e &\equiv -\frac{e_1}{\dot{e}_1}\sim\frac{R_0^3}{\mathcal{G}m_A\tau}\left(\frac{a_1}{R_0}\right)^8\sim P_1\left(\frac{a_1}{R_0}\right)^5\frac{P_1}{\tau}\sim\frac{Q}{n_1}\left(\frac{a_1}{R_0}\right)^5 \nonumber\\
&>\frac{10^{17}}{n_1},
\end{align}
where $\tau$ is the time lag introduced by tidal dissipation and $P_1$ and $n_1$ are the period and mean motion of inner orbit A. The tidal $Q\sim P_1/\tau$ is of order $10^7$ for C/O WDs \citep[\eg][]{2011ApJ...740L..53P,2013MNRAS.433..332B}. The ratio of orbital sizes $a_1/R_1$ is assumed to be larger than 100 due to the assumption that binaries are well-separate initially. The result shows that the circularization timescale due to equilibrium tides is longer than $10^{17}$ inner orbital periods, \ie much longer than the Hubble time with $a_1\sim$\,AU. Thus, we neglect dissipation via equilibrium tides.

When one of the inner orbits is at high eccentricity, the tidal dissipation via dynamical tides \citep[\eg][]{1975A&A....41..329Z,1975MNRAS.172P..15F,1987ApJ...318..261M,1989ApJ...342.1079G,1992ApJ...385..604K} may become dominant, especially when the tidal capture mechanism proposed by \cite{1975MNRAS.172P..15F} occurs. During the close encounter near periastron, the time-dependent tidal forces will excite non-radial oscillation modes in the stars and transfer energy from the orbit into stellar oscillations. A consequence is that the semi major-axis gradually decays while the orbit is circularized. \cite{1977ApJ...213..183P} derived a general formula for this energy transfer rate during a close periastron passage in the parabolic limit, and numerically computed the results for a polytropic stellar model with index $n=3$, which is appropriate for massive stars (approximately constant entropy, radiation pressure dominated) or WDs near Chandrasekhar limit (\ie relativistic electron gas). For low mass main sequence stars (with approximately fully convective monatomic gas) or normal WDs (\ie non-relativistic degenerate electron gas), $n=3/2$ is a better approximation \citep[\eg][]{1980MNRAS.191..897G,1986AcA....36..181G}. In the work presented here, we implemented the fitting formula provided in Appendix B of \cite{1986AcA....36..181G}.

Our approach to tidal excitation at periastron assumes that the excited modes of the stars decay via either linear or non-linear damping before the next periastron passage so that it does not then feed energy back into the orbit. If this turns out not to be the case for a given system, the next step would be an analysis of coupling the orbit to the dominant modes of the star (see \eg \citealt{2017arXiv170809392V} for a recent exploration of the possible dynamics\footnote{Our assumption is equivalent to Eq.~(26) of \citet{2017arXiv170809392V}.}).

\subsection{Spin}
\label{subsec:OtherEff}
We neglect the spins of the stars due to the dominance of the tidal effects at high eccentricities. We compare the precession rate due to the rotational (oblate spheroid) deformation of the stars to the precession rate caused by tidal deformation, which contains many of the same factors. \cite{2003ApJ...589..605W} provides the rotational precession rate for equatorial orbits and the $m_1\ll m_0$ case:
\begin{equation}
\dot{g}_1^{(\rm rot)} = \frac{1}{2}n_1\frac{k_1}{(1-e_1^2)^2}\left(\frac{\Omega_1}{n_1}\right)^2\frac{m_0}{m_1}\left(\frac{R_1}{a_1}\right)^5~,
\label{eq:g_rot}
\end{equation}
where $n_1$ is the mean motion of inner orbit A and $\Omega_1$ is the rotation angular frequency of the star $m_1$. In general there is a similar term corresponding to the distortion of the star $m_0$ due to its companion. At high-$e$, the ratio between the rotational precession rate (Eq.~\ref{eq:g_rot}) and the tidal precession rate (Eq.~\ref{eq:g_tideprec}) is estimated by
\begin{equation}
\frac{\dot{g}_1^{(\rm rot)}}{\dot{g}_1^{(\rm tide)}} \sim \frac{\Omega_1^2}{n_1^2}(1-e_1)^3 \sim \frac{\Omega_1^2}{\dot{f}_p^2}~,
\end{equation}
where $\dot{f}_p$ is the orbital frequency at the periastron. Since the tidal effects are only important at very high-$e$, $\Omega_1$ can be orders of magnitude smaller than $\dot{f}_p$. For tidal effects to be important in stellar binaries, we take $e_1=0.997$, orbital period $P_1=10$ yrs, then $\dot{f}_p^{-1}\sim 0.6$ day, which is much smaller than the rotation period of most of solar-like stars ($\sim 24$ days). For WDs, which are about $10^{-2}$ times smaller than the Sun in radius, we take $1-e_1=0.003\times10^{-2}$ and the same orbital period, then $\dot{f}_p^{-1}\sim 50$ s, still smaller than the typical WD rotation period, \ie $\sim 10^2 - 10^3$ s \citep[\eg][]{2004IAUS..215..561K}. For these reasons, we can neglect the stellar spins.

During the close passage, WDs can be spun up by dynamical tides due to the angular momentum transfer associated with energy injection. Since we used a non-spinning calculation of the tidal excitation during the encounter, we have to check that the WD rotation velocity remains small compared to the pattern speed of the excited modes (mainly the $f$-modes). During each passage, the energy injected to some oscillation mode (with frequency $\omega$, moment $m$, pattern frequency $\Omega_p=\omega/m$) is of order $\Delta E_{\rm mode}\sim \mathcal{G}m_0m_1\Delta(1/a_1)$, corresponding to an angular momentum change $\Delta G_{\rm mode}=\Delta E_{\rm mode}/\Omega_p$, hence a spin angular velocity change of Star ``1'' by $\Delta\Omega_1\sim\Delta G_{\rm mode}/(m_1R_1^2)$. Thus, we have
\begin{equation}
\frac{\Delta\Omega_1}{\Omega_p}\sim\frac{\mathcal{G}m_0\Delta(1/a_1)}{(R_1\Omega_p)^2}\sim\frac{R_1}{a_1}\frac{\Delta a_1}{a_1}\ll\frac{\Delta a_1}{a_1}~,
\end{equation}
where we have used in the second step that the pattern speed for the $f$-mode is of order the Keplerian speed $[\mathcal{G}m_0/R_1^3]^{1/2}$.
We conclude that in the time it takes to dissipate the orbital energy, the WD is spun up to a speed $\ll\Omega_p$, and it is indeed safe to neglect its spin.

\subsection{Non-secular effects}

Recent work has shown the ``double-averaging'' of the Hamiltonian can fail to describe the long-term evolution of LK cycles in moderately hierarchical systems \citep[\eg][]{2012arXiv1211.4584K,2014MNRAS.438..573B,2014MNRAS.439.1079A,2016MNRAS.458.3060L}.
The failure of double averaging has been historically important in the problem of lunar motion, where the ratio of outer to inner periods is $P_{\rm Sun-Earth}/P_{\rm Earth-Moon}\sim 12$. For example, the rate of precession of the Moon's perigee due to solar perturbations is roughly twice that predicted by double-averaging \citep[\eg][]{2010PhT....63a..27B}. We will discuss non-secular effects in \S\ref{sec:evection}, with a particular emphasis on evection (a short-term variation of the inner binary's eccentricity) due to its potential impact on close encounters.

\section{Secular Evolution of Quadruple versus Triple Systems}
\label{sec:features}
The evolution of quadruple systems is generally much more irregular than triple systems \citep[\eg][]{2015MNRAS.449.4221H,2016MNRAS.461.3964V,2017MNRAS.tmp...42H,2017MNRAS.470.1657H}. To explore the secular evolution of these systems, we wrote our secular code, which is described and tested in Appendix \ref{App:code}. In this section, we first use three special systems with increasing complexity to show the qualitatively different evolution patterns of quadruple systems from that of triple systems (\S\ref{subsec:evo}). Then, we run systems with random orientations and highlight some important features of the evolution in quadruple systems (\S\ref{subsec:enhanced_frac}-\ref{subsec:safe}), which illuminate our explorations of the astrophysical implications presented in the next section. Finally, we run systems with orbital sizes and shapes sampled from given distributions, confirming our results over a large range of parameter space (\S\ref{subsec:4star_full}).

Only taking quadrupole order terms in the Hamiltonian and ignoring any other effect such as GR and tides, the additional degrees of freedom introduced by the second inner binary system makes the evolution of the whole system irregular. It is well-known that in the secular + quadrupole approximation, the triple problem is integrable (see \eg \citealt{1968AJ.....73..190H}, but also the discussion in \S\ref{sec:features} of \citealt{2013MNRAS.431.2155N}). This is because in the orbit-averaged problem, where each orbit has 2 non-trivial degrees of freedom\footnote{We count an angle and its conjugate action as a single degree of freedom, as usual in Hamiltonian mechanics.}, the triple system has 4 degrees of freedom and 4 commuting constants of the motion: the perturbation Hamiltonian $\overline{\mathcal{H}}^{\rm (quad)}_1$; the $z$-angular momentum $H+H_1$; the squared total angular momentum\footnote{The explicit expression in terms of actions and angles can be built from the law of cosines:\\$G_{\rm tot}^2 = G^2 + G_1^2 + 2[HH_1 + \sqrt{(G_1^2-H_1^2)(G^2-H^2)}\cos (h_1-h)]$.} $G_{\rm tot}^2$; and the outer angular momentum $G$. The first three of these are conserved due to time and rotational symmetry, and the last is due to the accidental axisymmetry of the quadrupolar tidal field of a Keplerian orbit. The fourth star adds two degrees of freedom but no new commuting constants of the motion. Since the additional precession and dissipation effects are only important at high-$e$, we can safely ignore them first and get a general understanding of how the quadruple systems could behave differently from the triple systems before the high eccentricities are reached. As a summary, we find a much enhanced high-$e$ fraction in quadruple systems comparing to its triple limit. This result holds for different mass ratios, orbital sizes and initial shapes.

\subsection{Examples}
\label{subsec:evo}
In this subsection, we explore three types of systems: (1) triple systems (\S\ref{subsubsec:triple}); (2) [Star-Planet]-[Star-Star] systems (\S\ref{subsubsec:planet}); (3) ``4-Star'' systems (\S\ref{subsubsec:4star}). In each case, we will only include their secular effects from the Hamiltonian expansion up to octupole order and ignore the GR and tidal effects.

\subsubsection{Triple systems}
\label{subsubsec:triple}
\begin{figure}
\includegraphics[width=\columnwidth]{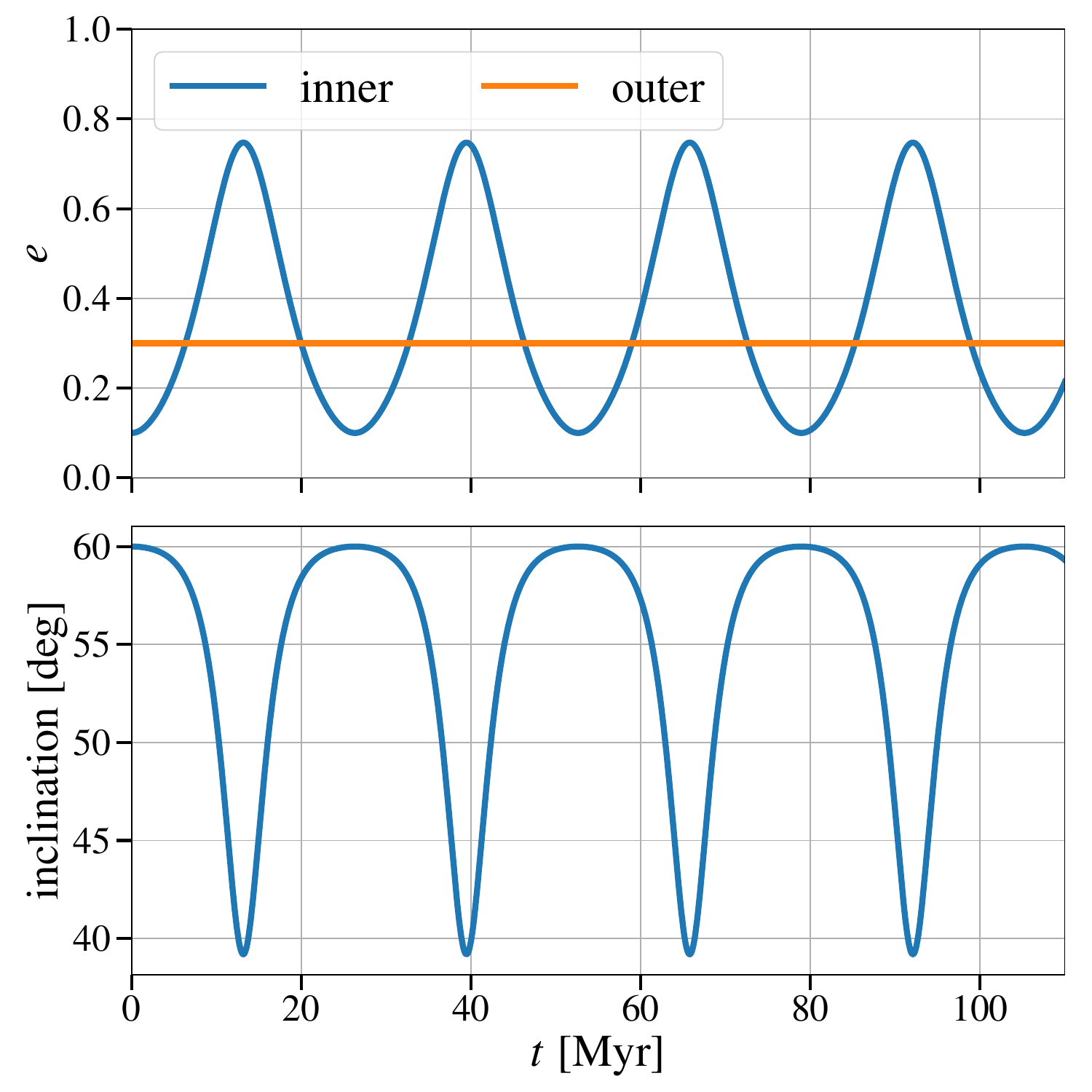}
\caption{The evolution of the triple system discussed in \S\ref{subsubsec:triple}. The upper panel shows the eccentricities of the inner and outer orbits, while the lower panel shows the inclination between the inner and outer orbits. The system exhibits the regular LK oscillation. The initial orbital elements of this example system are listed in Table \ref{Tab:Triple}.}
\label{fig:triple_e_inc}
\end{figure}

Hierarchical triple systems have a rather regular secular evolution. Here we assume the system consists of three 1 M$_\odot$ stars with initial orbital elements listed in Table \ref{Tab:Triple}. The eccentricities of the inner and outer orbit and their mutual inclination are shown in Figure \ref{fig:triple_e_inc}. The periodic oscillations of the inner eccentricity and the inclination are due to quadrupole order Hamiltonian, which leads to LK oscillations. The oscillations are in antiphase due to the conservation of the total angular momentum. The octupole order effect vanishes since the inner binary stars have equal masses. As a result, the outer orbit eccentricity is unchanged.

\begin{table}
\centering
\begin{tabular}{| c | c | c |}
 \hline
 \rule{0pt}{2.5ex} Elements & Inner Orbit & Outer Orbit \\
 \hline
 $e$ & 0.1 & 0.3 \\
 $a$ & 10AU & 1000AU \\
 $i$& 50$^\circ$ & 10$^\circ$ \\
 $g$& 0  & 0  \\
 $h$& 0  & 180$^\circ$  \\
 \hline
\end{tabular}
\caption{The initial orbital elements of a hierarchical triple system consisting of three 1M$_\odot$ stars, discussed in \S\ref{subsubsec:triple}. The initial inclination between the inner and outer orbit is 60$^\circ$.}
\label{Tab:Triple}
\end{table}

\subsubsection{[Star-Planet]-[Star-Star] systems}
\label{subsubsec:planet}
\begin{figure*}
\includegraphics[width=\textwidth]{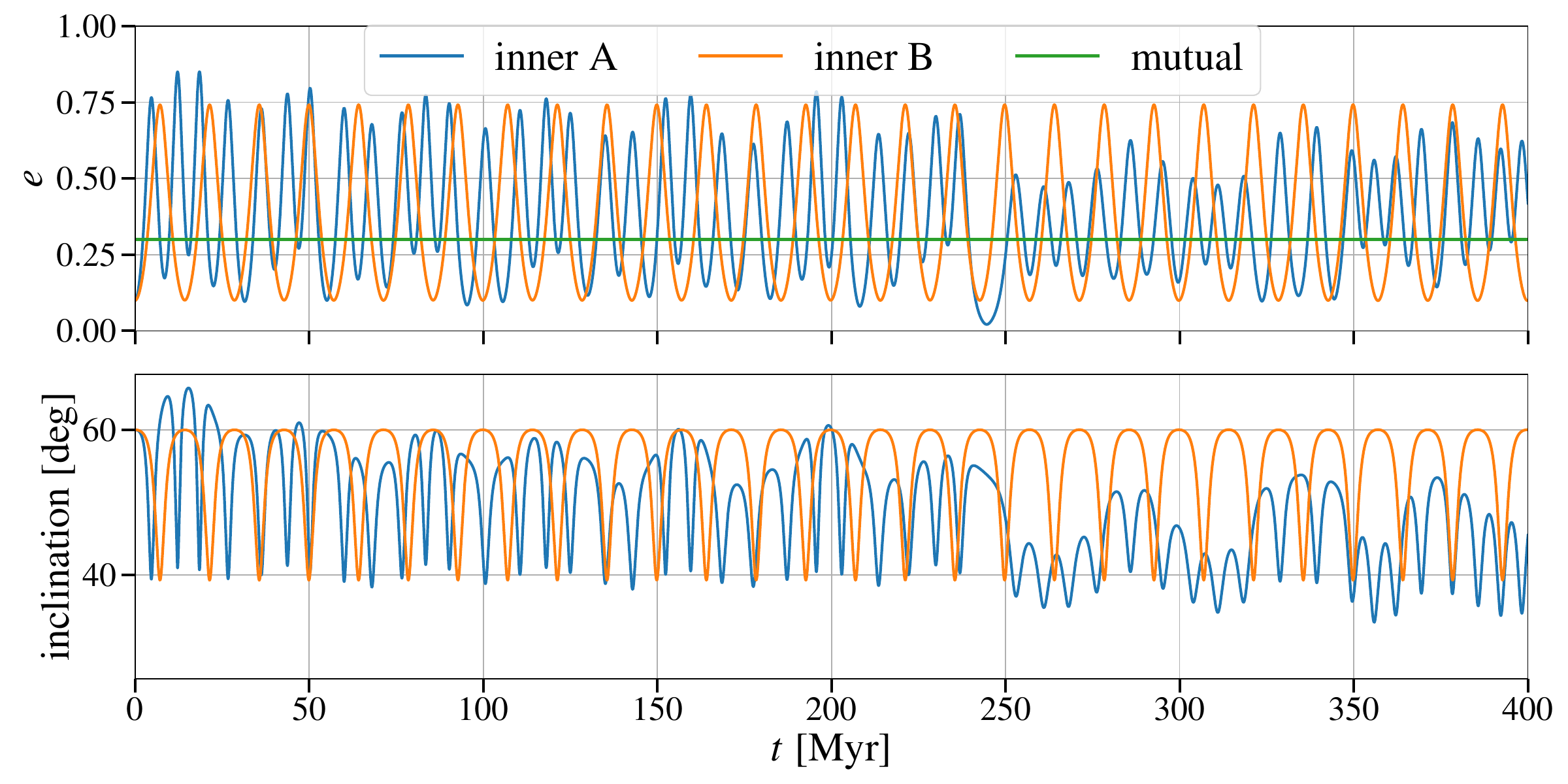}
\caption{The evolution of the quadruple system ([Star-Planet]-[Star-Star]) discussed in \S\ref{subsubsec:planet}. The upper panel shows the eccentricities of the inner and outer orbits, while the lower panel shows the inclinations between the two inner orbits and the outer orbit. The inner orbit B exhibits the regular LK oscillation, while orbit A evolves irregularly. The initial orbital elements of this example system are listed in Table \ref{Tab:Quadruple_planet}.}
\label{fig:HJ_e_inc}
\end{figure*}

The simplest non-trivial quadruple system is the [Star-Planet]-[Star-Star] system, \ie adding a planet (nearly a test particle) to the triple stellar system. We assume the [Star-Planet] pair as the inner orbit A and the stellar pair as B. The stars are solar-mass and the planet has one Jupiter mass, \ie 0.001M$_\odot$. The initial orbital elements are listed in Table \ref{Tab:Quadruple_planet}. The eccentricities of the two inner orbits and outer orbit and their mutual inclinations are shown in Figure \ref{fig:HJ_e_inc}. 
Since the planet mass is negligible comparing to the stars, the orbital evolution of the stellar binary is expected to behave like that in a triple stellar system, \ie exhibiting the regular LK oscillation as in \S\ref{subsubsec:triple}. However, the planet evolves rather irregularly, due to the fact that its ``Kozai action'' is not constant even at the test particle limit (TPL) and quadrupole order \citep[\eg][]{2017MNRAS.tmp...42H,2017MNRAS.470.1657H} (see \S\ref{app:code_kozai} for the conservation of Kozai action in triple systems).

\begin{table}
\centering
\begin{tabular}{| c | c | c | c |}
 \hline
 \rule{0pt}{2.5ex} Elements & Inner Orbit A & Inner Orbit B & Mutual Orbit\\
 \hline
 $e$ & 0.1 & 0.1 & 0.3 \\
 $a$ & 10AU & 15AU & 1000AU \\
 $i$& 50$^\circ$ & 50$^\circ$ &10$^\circ$ \\
 $g$& 0 & 0 & 0  \\
 $h$& 0 & 0 & 180$^\circ$  \\
 \hline
\end{tabular}
\caption{The initial orbital elements of a hierarchical quadruple system, discussed in \S\ref{subsubsec:planet}. The inner orbit A consists of a solar-mass star and a Jupiter-mass planet and the orbit B consists of a pair of solar-mass stars.}
\label{Tab:Quadruple_planet}
\end{table}

\subsubsection{``4-Star'' systems}
\label{subsubsec:4star}
\begin{figure*}
\includegraphics[width=\textwidth]{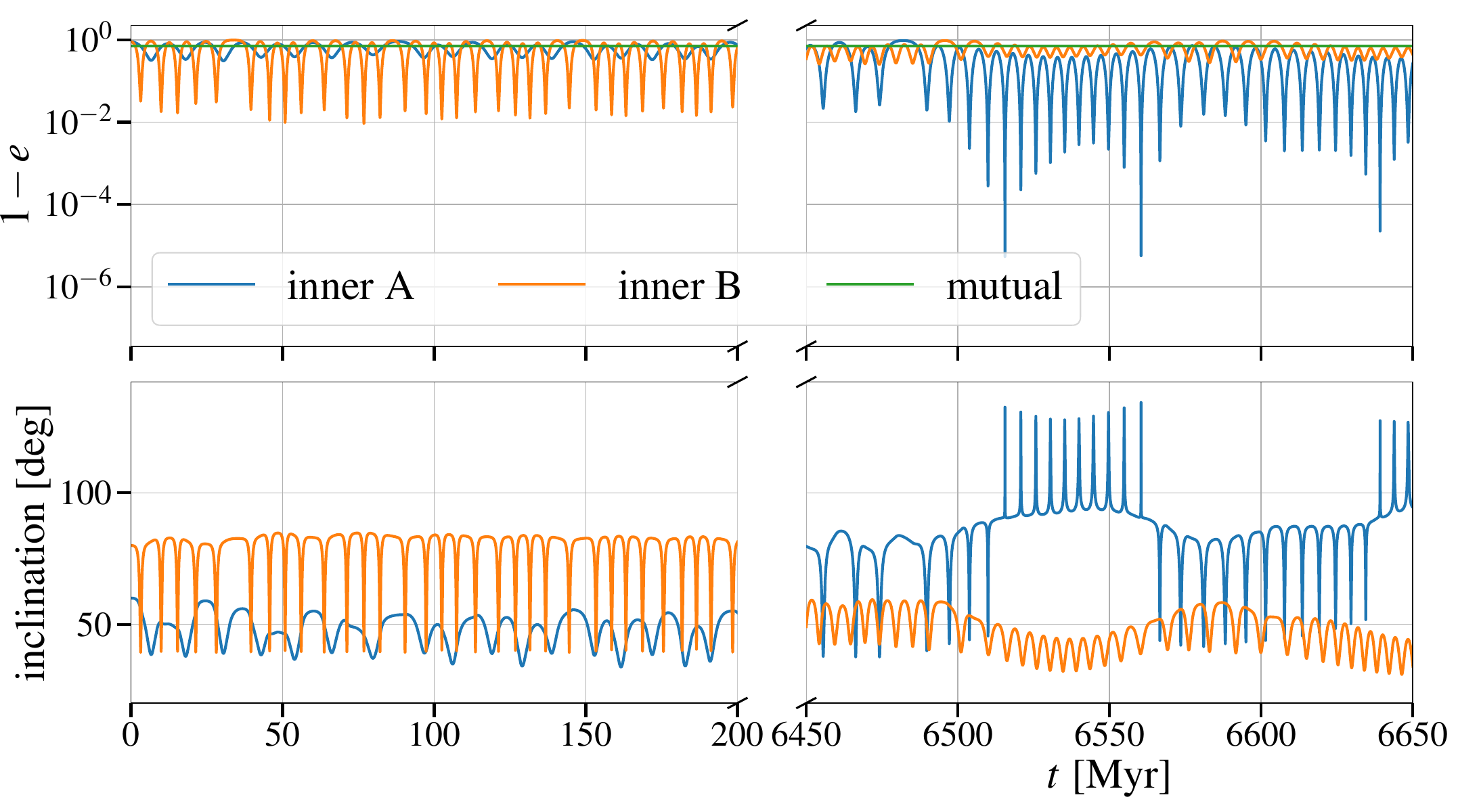}
\caption{The evolution of the ``4-star'' system discussed in \S\ref{subsubsec:4star}. The upper panel shows the eccentricities of the inner and outer orbits, while the lower panel shows the inclinations between the two inner orbits and the outer orbit. Both of the inner orbits evolve irregularly, and one of them reaches very high eccentricities that its equivalent triple counterpart system will not be able to reach with the same set of initial orbital elements. The initial orbital elements of this example system are listed in Table \ref{Tab:4star}. Note that the high eccentricity shown in the plot is significantly more than sufficient for the stars to collide.}
\label{fig:4star_e_inc}
\end{figure*}

The general ``4-Star'' systems behave rather chaotically. We assume the 4 stars are all solar-mass, and their initial elements are listed in Table \ref{Tab:4star}. The eccentricities of the two inner orbits and outer orbit and their mutual inclinations are shown in Figure \ref{fig:4star_e_inc}. It is interesting to see that in this example, one of the stellar binaries achieves very high eccentricity on a very long timescale, which is not possible for its equivalent triple system (\ie having a tertiary with mass 2M$_\odot$) with the same initial inclination. This opens the question of how much the fraction of systems evolving to high eccentricity is enhanced in quadruple systems relative to triples. We will explore the answer in the following subsections.

\begin{table}
\centering
\begin{tabular}{| c | c | c | c |}
 \hline
 \rule{0pt}{2.5ex} Elements & Inner Orbit A & Inner Orbit B & Mutual Orbit\\
 \hline
 $e$ & 0.1 & 0.1 & 0.3 \\
 $a$ & 10AU & 15AU & 1000AU \\
 $i$& 50$^\circ$ & 70$^\circ$ &10$^\circ$ \\
 $g$& 180$^\circ$ & 0 & 0  \\
 $h$& 0 & 0 & 180$^\circ$  \\
 \hline
\end{tabular}
\caption{The initial orbital elements of a ``4-star'' quadruple system, discussed in \S\ref{subsubsec:4star}. Both of the inner orbits consist of a pair of solar-mass stars.}
\label{Tab:4star}
\end{table}

\subsection{Enhanced high-$e$ fraction}
\label{subsec:enhanced_frac}
The fraction of systems that can reach high eccentricities is highly enhanced in the quadruple systems comparing to the triple systems.

For quadrupole order approximation of the test particle limit (TPL) of the inner companion on an initially circular orbit, the LK oscillation produces a maximal eccentricity of the inner orbit, given by \citep[\eg][]{1962P&SS....9..719L,1962AJ.....67..591K,1976CeMec..13..471L,1997AJ....113.1915I,1999ceme.symp..233K,2002ApJ...578..775B,2003ApJ...598..419W,2013MNRAS.431.2155N}
\begin{equation}
e_{\rm in,max} = \sqrt{1-\frac{5}{3}\cos^2 i_0}~,
\label{eq:triple_TPL}
\end{equation}
where $i_0$ is the initial inclination angle between the inner orbit and the outer orbit. In order to reach an eccentricity higher than $e_{\rm in,max}$, a high initial inclination is required, \ie $\cos^2 i_0\leq 3(1-e^2_{\rm in,max})/5$. Thus, the fraction to reach this in an ensemble of systems with randomly oriented inner and outer orbits is
\begin{equation}
f_{\rm triple, TPL} = \sqrt{\frac{3}{5}\left(1-e^2_{\rm in,max}\right)}~.
\label{eq:f_TPL}
\end{equation}
For non-zero initial $e_{\rm in}$ with $L_{\rm in}\ll L_{\rm out}$, the relation still holds\footnote{Here we denote $e_{\rm in},e_{\rm out}$ as the eccentricities of the inner and outer orbits of triple systems, $i_{\rm tot}$ as the mutual inclination angle. From the angular momentum conservation of $G_{\rm tot}$ and $G_{\rm out}$ and the energy conservation, at maximal $e_{\rm in}$ we have \citep{2013MNRAS.431.2155N}
\begin{align*}
&L_{\rm in}^2(1-e_{\rm in}^2)+2L_{\rm in}L_{\rm out}\sqrt{1-e_{\rm in}^2}\sqrt{1-e_{\rm out}^2}\cos i_{\rm tot} = {\rm const.}~{\rm and}\nonumber\\
&\cos^2 i_{\rm tot}\,(1+4e_{\rm in}^2)-3e_{\rm in}^2 = {\rm const.}~,
\end{align*}
where the first equation yields
\begin{equation*}
\cos^2 i_{\rm tot} = (1-e^2_{\rm in,init})\cos^2 i_0/(1-e_{\rm in}^2)
\end{equation*}
in the $L_{\rm in}\ll L_{\rm out}$ limit. Combining it with the second equation, we recover Eq.~(\ref{eq:triple_TPL}).
}. For Non-TPL triple systems, retrograde systems have a higher chance to reach high eccentricities \citep[\eg][]{1976CeMec..13..471L,2013MNRAS.431.2155N}.

When the inner binary stars have different masses, the octupole order term adds complexity. Some previous work has shown its effect on the eccentricity evolution \citep[\eg][]{2000ApJ...535..385F,2011PhRvL.107r1101K,2011ApJ...742...94L,2013MNRAS.431.2155N}, including the enhancement of the high-$e$ fraction.

In quadruple star systems, three (instead of two) orbits interact with each other, making it hard to derive a relation as simple as Eq.~(\ref{eq:triple_TPL}). It is not even clear whether there is an upper limit of $e_1$ less than 1, associated with any initial configuration. \cite{2013MNRAS.435..943P} has shown that the fraction of systems that reach high eccentricity is greatly enhanced in quadruple systems for several initial conditions with $N$-body simulations. However, full dynamical simulations are too computationally expensive to explore the huge range of the parameter space and for durations comparable to the age of the Universe.

As an example, we employ the secular code for $10^5$ systems with random orientations. The effects considered are the two quadrupole order perturbation terms (GR and tidal effects will be considered in \S\ref{sec:SN}). We take 4 main sequence stars with initial orbital configurations listed in Table \ref{Tab:4star_random}. Due to the isotropy of space, we can always choose an inertial frame with the coordinate axes in which the initial orientation of the mutual orbit (or one of the inner orbits) is fixed, thus reducing the parameter space of the initial conditions. Without loss of generality, we take $i=0.1$ rad and $g=h=0$. The initial values for $\cos i_1$ and $\cos i_2$ are drawn randomly from range $[-1,1]$, while $g_1$, $g_2$, $h_1$, and $h_2$ are drawn randomly from range $[0,2\pi]$. As a result, the cosines of the inclination angles $\cos i_A$ and $\cos i_B$ are also uniformly distributed in range $[-1,1]$. Each system runs for 10 Gyr if it is not stopped by meeting the criterion that the periastron distance of one inner orbit is less than 3 times the sum of two stars' radii, where the tidal effects could start to play an important role. This criterion is equivalent to setting a maximal eccentricity $e_{\rm in,max}=1-6R_{\odot}/a_{\rm in}$, \ie $e_{1,{\rm max}}=0.9972,e_{2,{\rm max}}=0.9981$. Note that due to the equal masses, octupole terms vanish. We consider unequal masses in \S\ref{subsec:mass_ratio}. Including the GR precession may detune the Kozai effect and lower the high-$e$ fraction. However, to make this effect substantial, the system would need to have GR precession timescale shorter than or comparable to the instantaneous Kozai timescales, which turns out not to be the case for the stellar systems considered in this section ($t_{\scriptscriptstyle\rm LK1}^{\rm (ins)}\sim 0.3\,$Myr, $t_{\scriptscriptstyle\rm pr1}^{\rm (1PN)}\sim 13\,$Myr, when the inner orbit A reaches $e_{1,{\rm max}}$, using Eqs.~\ref{eq:tLK1ins},\ref{eq:tpr1pn}).

\begin{table}
\centering
\begin{tabular}{| c | c | c | c |}
 \hline
 \rule{0pt}{2.5ex} Elements & Inner Orbit A & Inner Orbit B & Mutual Orbit\\
 \hline
 $m$ & 1+1M$_\odot$ & 1+1M$_\odot$ & -- \\
 $e$ & 0.1 & 0.1 & 0.3 \\
 $a$ & 10 AU & 15 AU & 1000 AU \\
 $\cos i$ & $[-1,1]$ & $[-1,1]$ & $\cos 0.1$ \\
 $g$ & $[0,2\pi]$ & $[0,2\pi]$ & 0  \\
 $h$ & $[0,2\pi]$ & $[0,2\pi]$ & 0 \\
 \hline
\end{tabular}
\caption{The initial orbital configurations of the ``4-star'' hierarchical quadruple systems, discussed in \S\ref{subsec:enhanced_frac}. The orbital sizes and shapes are fixed, while their orientations are randomly sampled. Because the physics is independent of the orientation of the coordinate system, we can reduce the degree-of-freedom of the system by fixing the initial orientation of one of the orbits (here the mutual orbit).}
\label{Tab:4star_random}
\end{table}

\begin{figure*}
\includegraphics[width=\textwidth]{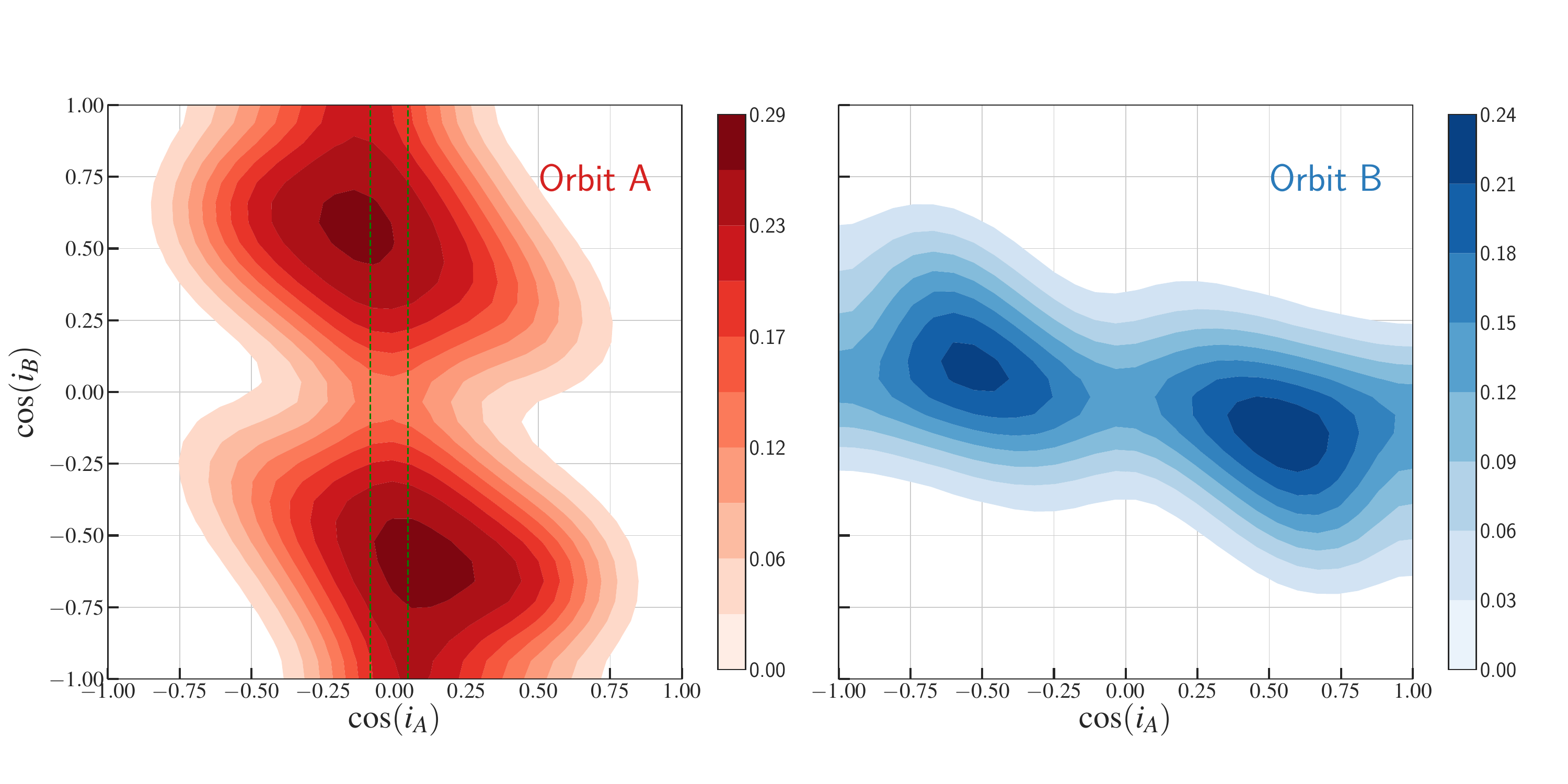}
\caption{The initial mutual inclination distributions of systems that reach high-$e$ before 10 Gyr, calculated from the $10^5$ randomly oriented ``4-star'' systems discussed in \S\ref{subsec:enhanced_frac}, whose initial orbital configurations are listed in Table \ref{Tab:4star_random}. The left panel shows systems whose inner orbits A reach high-$e$, while the right panel shows systems whose inner orbits B reach high-$e$. The higher density of the colour in each panel represents the higher fraction of high-$e$ systems, and it is normalized to the total high-$e$ fraction of each inner orbit. Between the two green dashed lines in the left panel shows the high-$e$ systems from the equivalent triple case, where the inner orbit B is replaced by a single star with mass $m_B=m_2+m_3=2M_{\odot}$.}
\label{fig:EnhancedFrac}
\end{figure*}

Figure \ref{fig:EnhancedFrac} shows the distribution of the systems on the $\cos i_A$-$\cos i_B$ plane that later reach the maximal high eccentricity. For equivalent triple systems (\ie the inner orbit B is replaced by a single star with mass $m_B=m_2+m_3=2M_{\odot}$), the corresponding region is very narrow and close to $\cos i_A=0$, as expected. At $t=10$ Gyr, the fraction of systems reaching the given maximal eccentricity is about 36.3\% in inner orbit A of quadruple systems, more than 6 times higher than that in triple systems ($\sim$5.8\% from the run, consistent with the analytical result calculated from Eq.~\ref{eq:f_TPL})! Note that about 19.8\% systems reach the given high eccentricity in their inner orbit B, so in total $\sim$56\% systems become dynamically interesting in this orbital configuration.

\subsection{Growing fraction over time}
\label{subsec:GrowingFrac}
In triple systems, the regular LK oscillation has a timescale $t_{\scriptscriptstyle \rm LK}\sim P_{\rm out}^2/P_{\rm in}$, which is longer than any of the orbital periods, but shorter than the evolution timescales of the main sequence stars with solar masses, for systems that are dynamically interesting. Especially for WD mergers to produce SNe Ia, we hope the mergers to be able to occur in a large range of timescales after the WDs are formed. At quadrupole order, once the mutual inclination is specified and, thus, the maximum eccentricity $e_{\rm in,max}$, all the systems that can reach $e_{\rm in,max}$ will do so within the first LK cycle on a timescale $t\sim t_{\scriptscriptstyle \rm LK}$, and after that there will be no more such events.

\begin{figure}
\centering
\includegraphics[width=\columnwidth]{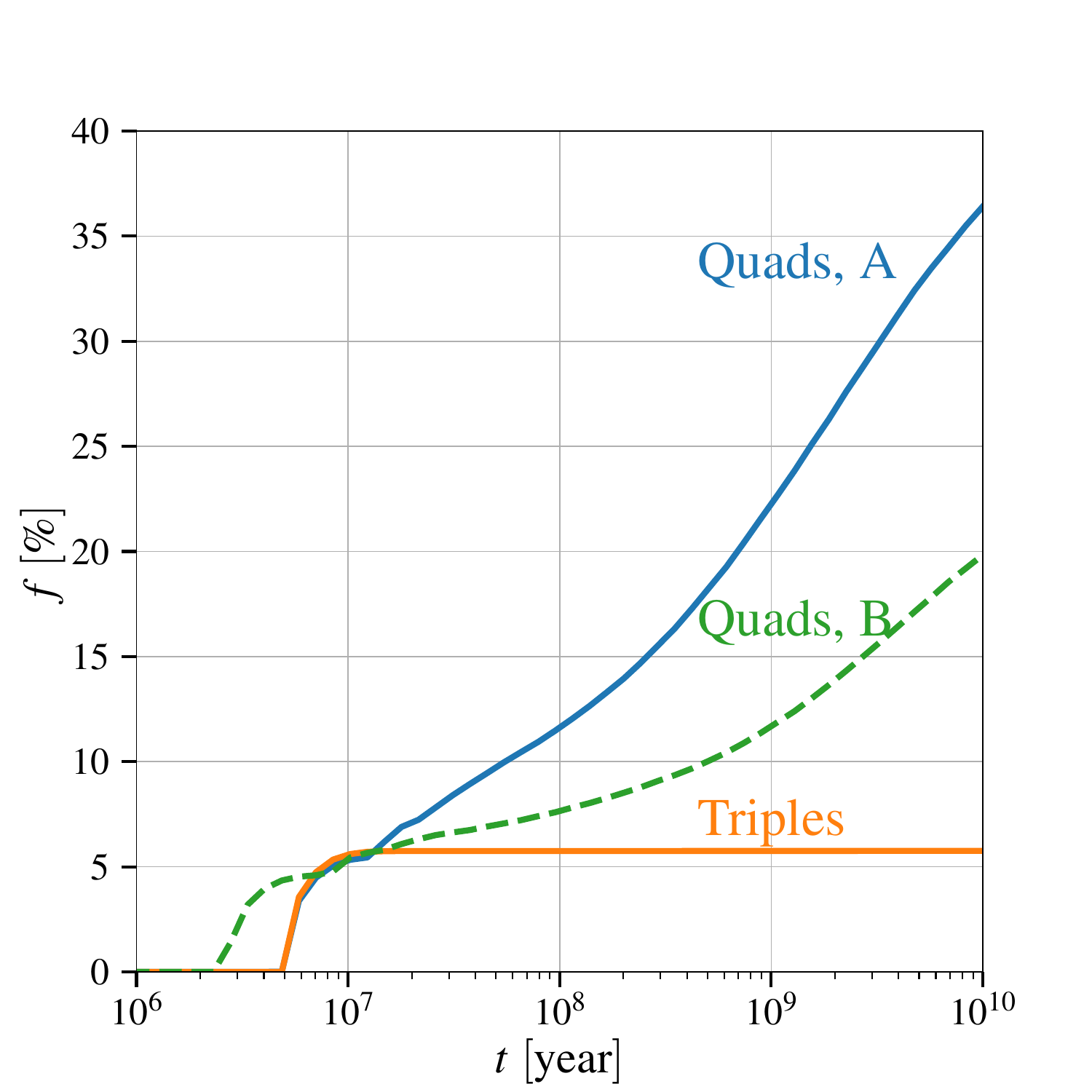}
\caption{The growing cumulative fractions of high-$e$ events from the $10^5$ randomly oriented ``4-star'' systems and their equivalent triple systems described in \S\ref{subsec:enhanced_frac}. At quadrupole order, the high-$e$ fraction of triples stops growing after the Kozai timescale ($\sim$6Myr), but for quadruples, the fraction keeps growing.}
\label{fig:GrowFrac}
\end{figure}

Quadruple systems, however, have the ability to produce high-$e$ events on timescale much longer than $t_{\scriptscriptstyle \rm LK}$, up to the age of the Universe. Figure \ref{fig:GrowFrac} shows the results from running the $10^5$ systems described in \S\ref{subsec:enhanced_frac}. The cumulative fraction $f_1,f_2$ (from the inner orbits A and B) grow with the logarithmic time, \ie about 10\% and 4\% per tenfold time. The fractional event rates $\Gamma_f$ in this orbital configuration are thus roughly given by
\begin{align}
\Gamma_{f1}&\equiv\dot{f}_1\sim\frac{0.10}{t\ln 10}~~{\rm for}~~t>t_{\scriptscriptstyle\rm LK1}~~{\rm and}
\nonumber \\
\Gamma_{f2}&\equiv\dot{f}_2\sim\frac{0.04}{t\ln 10}~~{\rm for}~~t>t_{\scriptscriptstyle\rm LK2}\,.
\end{align}
We find that $\Gamma_{f1}\sim 0.4$ Gyr$^{-1}$ at $t=0.1$ Gyr and $\Gamma_{f1}\sim 0.04\,$Gyr$^{-1}$ at $t=1\,$Gyr. We will discuss the implications for the stellar merger rate and the SN Ia rate in \S\ref{sec:conclusion}.

\subsection{Orbital size dependence}
\label{subsec:orbital_size}
Does the result we obtain in the previous subsections depend strongly on the orbital sizes? In this subsection we investigate the dependence of the size of the companion binary orbit. When $a_2$ (semi-major axis of orbit B) is very small, the orbital angular momentum of orbit B is small and the system reduces to the triple system limit (more precisely, if we treat the binary B as two point masses). Increasing $a_2$ enhances the high-$e$ fraction of the orbit A, as described in \S\ref{subsec:enhanced_frac},\ref{subsec:GrowingFrac}. However, when $a_2$ is very large, so that the LK timescale $t_{\scriptscriptstyle \rm LK2}$ is very small comparing to $t_{\scriptscriptstyle \rm LK1}$, the oscillatory perturbation exerted on the mutual orbit by the orbit B is rapid and is averaged out. At the TPL of the orbit A, the high-$e$ fraction of the orbit A should drop and approach the triple system limit, where the initial inclination should instead be estimated using the averaged angular momentum of the mutual orbit \citep[][]{2017MNRAS.470.1657H}. However, this is only true when the Kozai timescales are much longer than both inner and outer orbits, which sets an upper limit for the choice of $a_2$. For non-TPL systems, it is not clear to what extent the effect from the orbit B is averaged out and suppressed, so that it is likely that the triple limit may not be reached in the valid range of $a_1\ll a_2\ll a$.

\begin{figure}
\centering
\includegraphics[width=\columnwidth]{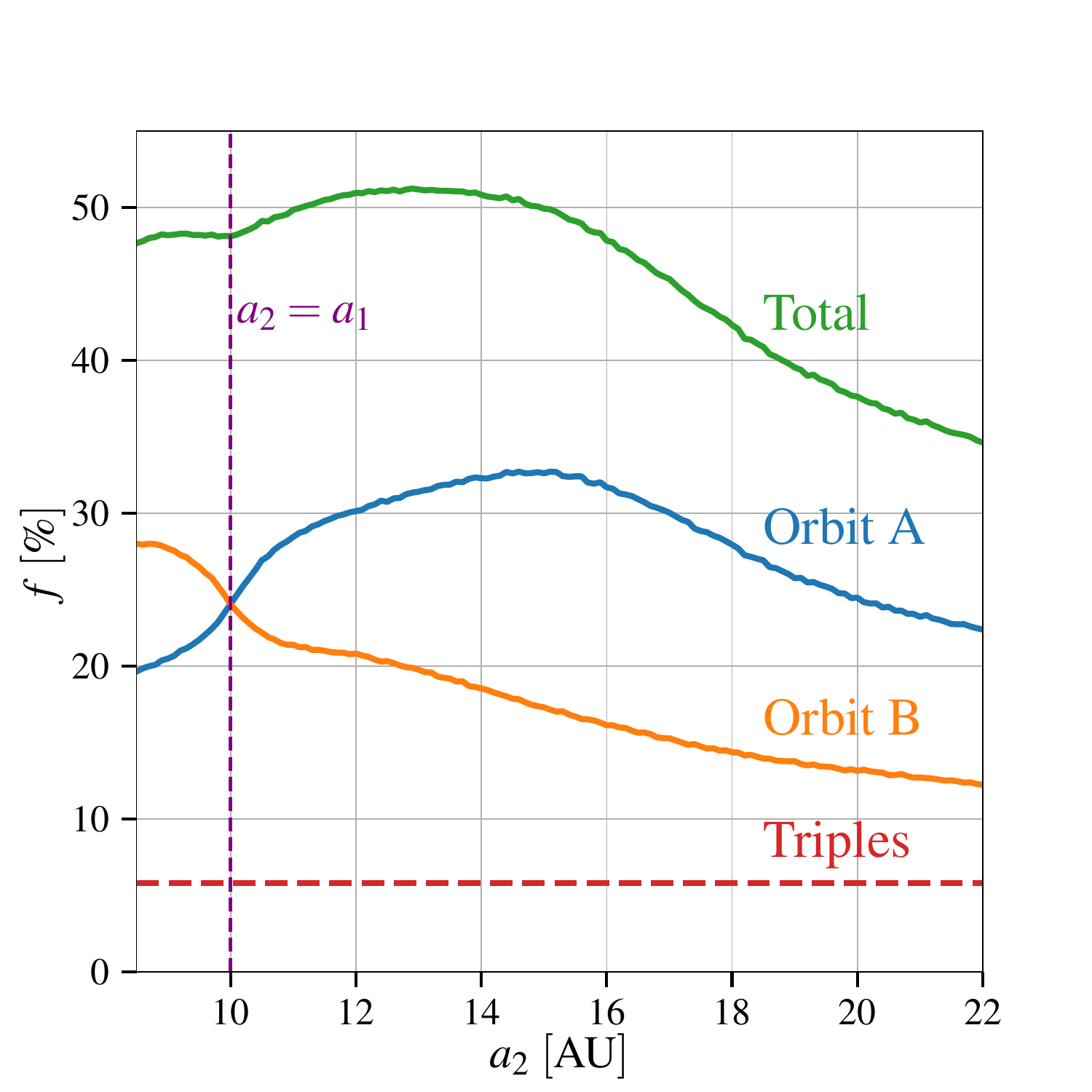}
\caption{The high-$e$ fractions from the inner orbit A, inner orbit B and the total fraction vary as functions of the semi-major axis of the inner orbit B, described in \S\ref{subsec:orbital_size}. $a_2$ is evenly sampled from 8.5AU to 22.0AU, with a stepsize 0.1AU. For each sampled $a_2$, we run $10^4$ systems up to $t=5\,$Gyr. The rest of initial orbital elements are listed in Table \ref{Tab:4star_random}.}
\label{fig:a2dependence}
\end{figure}

In Figure \ref{fig:a2dependence} we show how the percentages of systems whose ``A'' and ``B'' inner orbits reach the high-$e$, $f_1$ and $f_2$ respectively, change with different values of $a_2$ ranging from 8.5 AU to 22.0 AU with a step size of 0.1 AU (for the set of random oriented systems described in \S\ref{subsec:enhanced_frac} but with only $10^4$ systems for each $a_2$ and only up to 5 Gyr). The total fraction $f_1+f_2$  is also plotted. We can see that the high-$e$ fraction of the inner orbit A is much larger than the equivalent triple case ($\sim$5.8\%) for a large range of $a_2$. Thus, we confirm that our result that quadruple systems can largely enhance the high-$e$ fraction is true for a broad range of orbital size configurations.

To confirm the expectation of the ``triple system limit'' for large and small $a_2$ while avoiding large computational costs, we perform the following two tests: (1) $10^5$ systems with $a_2=1\,$AU; (2) $10^4$ systems with $a_2=100\,$AU. The second test satisfies the requirement $P\ll t_{\scriptscriptstyle \rm LK2}\ll t_{\scriptscriptstyle \rm LK1}$, where $P$ is the period of the mutual orbit. In each test, we run systems with random orientations up to 5 Gyr, and assume the inner binary B is composed of two point masses (\ie ignoring any high-$e$ event from the orbit B, $f_2\equiv 0$). In Test (1) we obtain $f_1 = 5846/100000\sim 5.8\%$, in agreement with our expectations, while in Test (2) we obtain $f_1 = 740/10000\sim 7.4\%$, confirming the descending trend of $f_1$ at large $a_2$. It is not surprising that $f_1$ does not reach the triple-limit because the system is not in the TPL.

\subsection{Mass ratio dependence}\label{subsec:mass_ratio}
In triple systems whose inner binary stars do not have equal masses, octupole order perturbations enhance the high-$e$ fraction on a much longer timescale than the LK timescale. However, in quadruple systems where the enhancement has been large, the contribution from octupole order terms becomes insignificant, because the systems that will reach high-$e$ under octupole order effect would likely have reached them under quadrupole order effect due to the second binary.

\begin{figure*}
% \centering
\includegraphics[width=\textwidth]{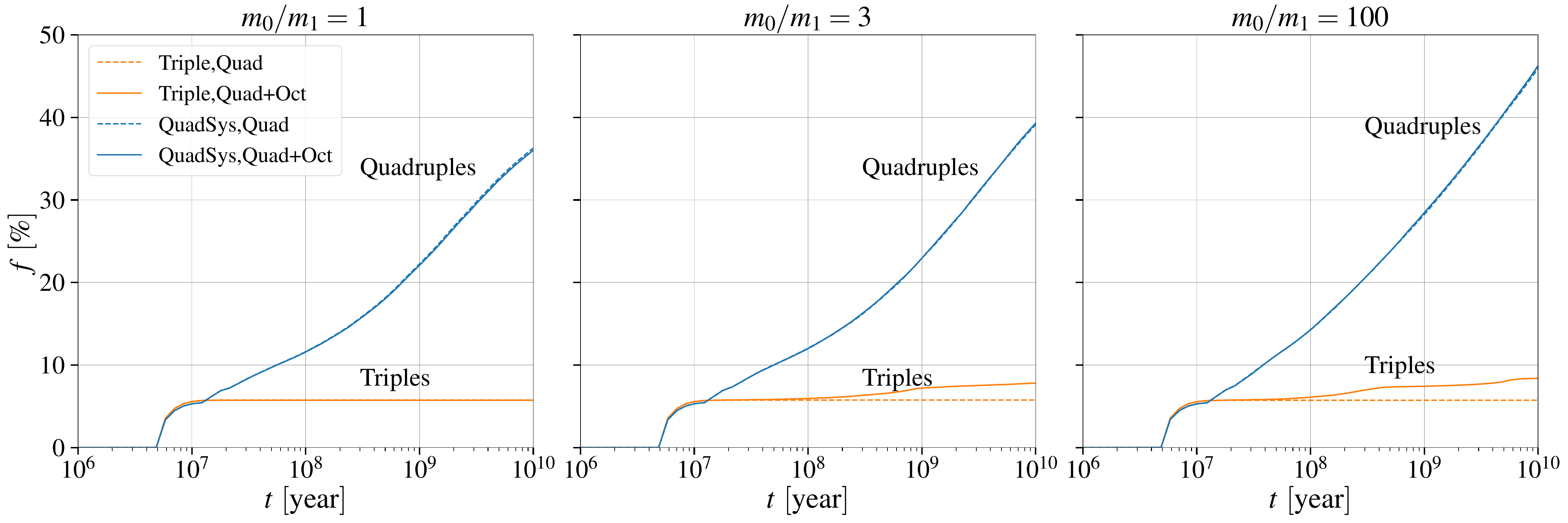}
\caption{The high-$e$ fractions of $10^5$ randomly oriented ``4-star'' systems with mass ratios 1 (\textit{left}), 3 (\textit{center}), 100 (\textit{right}) in the inner orbits A, discussed in \S\ref{subsec:mass_ratio}. The other initial orbital elements are listed in Table \ref{Tab:4star_random}. Their equivalent triple cases (\ie replacing the inner binary B with a single 2M$_\odot$ star) are plotted for comparison. The high-$e$ fraction enhancement for quadruples is remarkably robust against variations in $m_0/m_1$.}
\label{fig:massratio}
\end{figure*}

For a better comparison, we also plot the fractions from quadruple systems and their equivalent triples with different $m_0/m_1$ ratios but the same $m_A(=2M_\odot)$ and $e_{\rm in,max}$ values in Figure \ref{fig:massratio}, where for each mass ratio value, the plot shows the fraction growth curves for quadruple and triple systems with octupole order effects turned on or off. We can see quadruple systems produce much higher high-$e$ fractions than their equivalent triple cases, and octupole order contribution is negligible in these quadruple system configurations.

Note that the absolute values of $f$ have a strong dependence on the radii of stars since the $e_{\rm in,max}$ values also depend on the radii. Thus, for WD binaries the $f$ values are expected to be much smaller, as shown in \S\ref{sec:SN}.

\subsection{Possible ``safe'' regions}
\label{subsec:safe}
In triple systems, we can already conclude that the high-$e$ fraction $f_{\rm triple}$ is limited and the remaining fraction, $1-f_{\rm triple}$, is ``safe'' -- \ie at the order of approximations chosen they will not merge even on timescales longer than the age of the Universe. However, for quadruple systems, the results shown in \S\ref{subsec:GrowingFrac} seem to tell us that the event fraction keeps growing. But one might wonder whether there are ``safe'' regions in the initial parameter space where the system will never reach high eccentricity. In other words: does the fraction converge to some value $f_{\rm max} < 1$ as $t\rightarrow\infty$?

\begin{figure}
\centering
\includegraphics[width=\columnwidth]{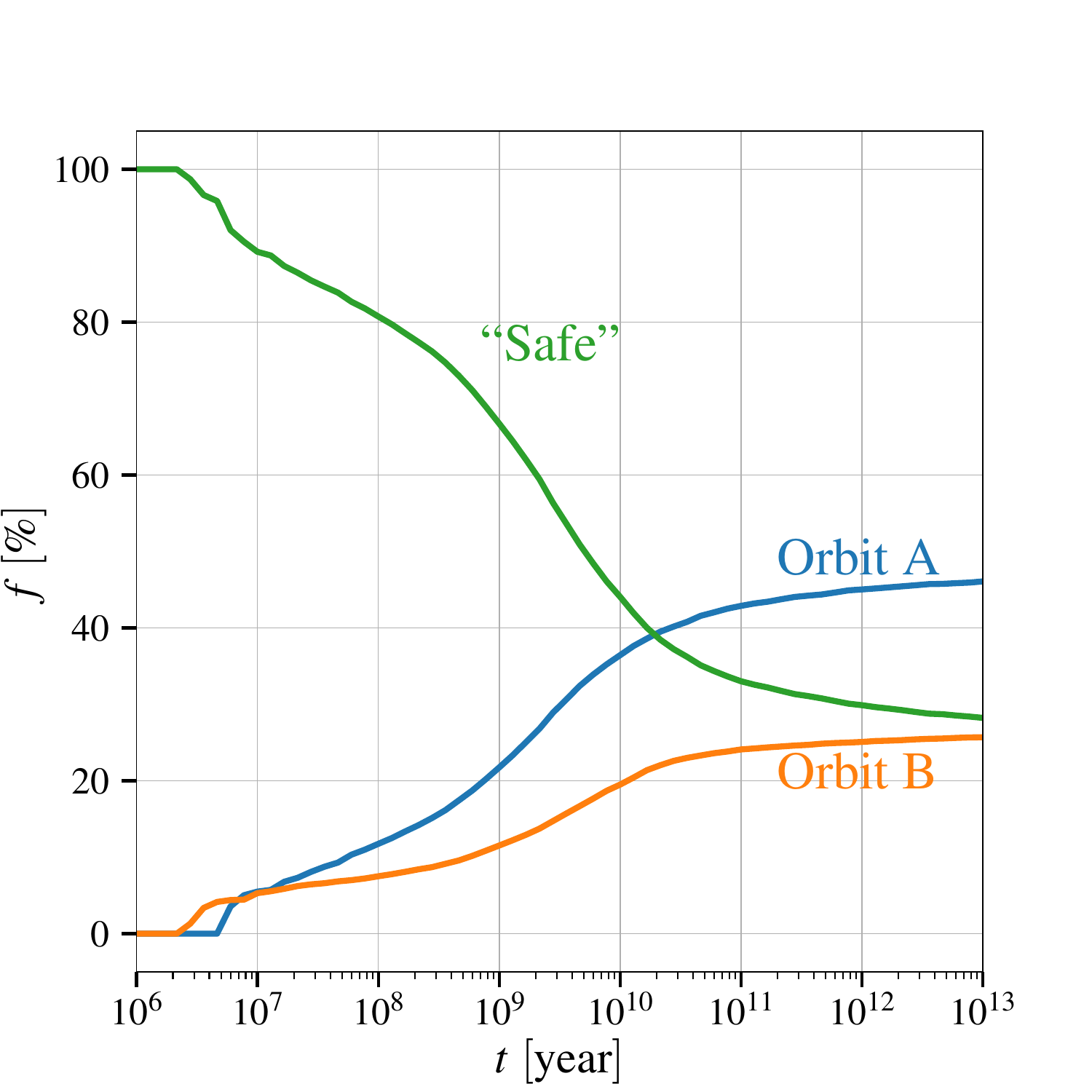}
\caption{The cumulative fractions of $10^4$ randomly oriented ``4-star'' systems whose inner orbits A and B reach high-$e$, shown in blue and orange solid lines, respectively. The rest of systems are ``safe'' and are shown in the green line. The initial orbital configurations are listed in Table \ref{Tab:4star_random}, and each system runs up to $10^{13}$ years, as discussed in \S\ref{subsec:safe}.}
\label{fig:frac_future}
\end{figure}

\begin{figure}
\centering
\includegraphics[width=\columnwidth]{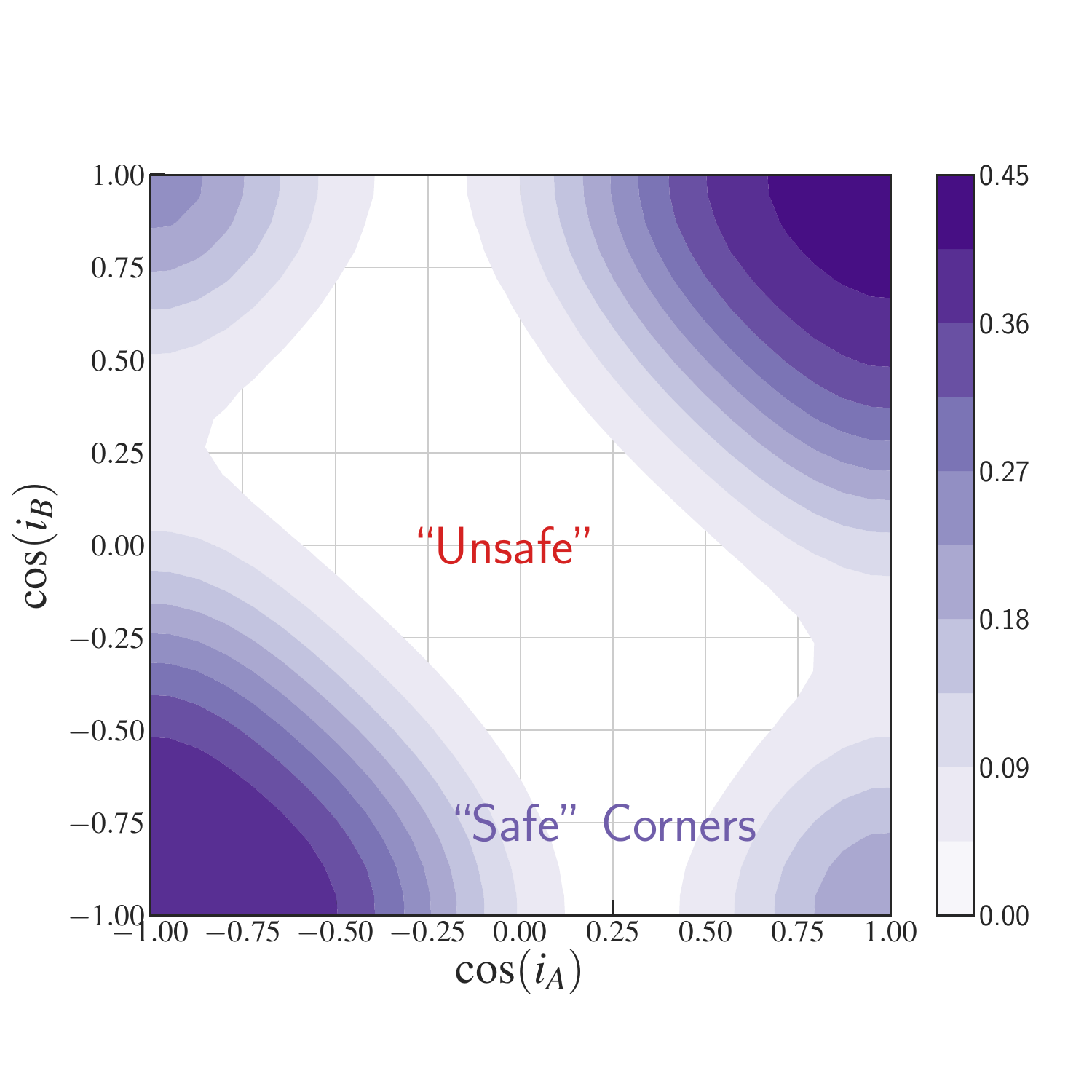}
\caption{The ``safe'' regions for the $10^4$ randomly oriented ``4-star'' systems with parameters from Table \ref{Tab:4star_random} running up to $10^{13}$ years, as discussed in \S\ref{subsec:safe}. All the systems that have never reached high-$e$ are initially coplanar, and the ``safe'' corners are larger for those systems whose two inner orbits are in the same direction. The density of the colour represents the fraction of ``safe'' systems in that region of $(\cos i_A,\cos i_B)$-space, and it is normalized to the total ``safe'' fraction.}
\label{fig:safe_initcosi}
\end{figure}

At quadrupole order of Hamiltonian and neglecting all other effects, we tested $10^4$ systems in the configuration described in \S\ref{subsec:enhanced_frac} up to $t=10^{13}$ years and confirmed the slow-down and convergence of the high-$e$ fraction growth. Figure \ref{fig:frac_future} shows that the fractions of reaching high-$e$ in the two inner orbits converge to $f_{\rm 1,max}\sim 47\%$, $f_{\rm 2,max}\sim 26\%$ for this case, leaving $\sim 27\%$ of systems ``safe'' (\ie never reaching high-$e$, at least on timescales of $10^{13}$ years). In Figure \ref{fig:safe_initcosi} we show that the ``safe'' regions are roughly at the corners of the $\cos i_A$-$\cos i_B$ plane, where the inner orbits and the mutual orbit are nearly coplanar. Also, we notice that the ``safe corners'' are much larger when the two inner orbits' angular momenta are in the same direction than when they are opposite.

\subsection{Quadruple systems of main sequence stars}
\label{subsec:4star_full}

We run $10^5$ systems for quadruple systems with four 1 $M_\odot$ main sequence stars and sample the values of their semi-major axes and eccentricities from given distributions. The eccentricities $e_1$, $e_2$, and $e$ are sampled from the thermal or uniform phase space density distribution \citep{1919MNRAS..79..408J}, \ie $e_1^2$, $e_2^2$, and $e^2$ are uniformly distributed in $[0,1]$. The semi-major axes $a_1$ and $a_2$ are sampled from a log-normal distribution; $\log_{10} a_1$ (in AU) is assigned a mean of 1.7038 and a standard deviation of 1.52, inferred from Figure 13 in \cite{2010ApJS..190....1R}. $a$ is then sampled assuming that $a/(a_1+a_2)$ is log-uniformly distributed in $[9,1900]$, based on observations of confirmed hierarchical multiple systems in \S 5.3.8 of \cite{2010ApJS..190....1R}, although the sample size is small. Also we impose the criteria that
\begin{enumerate}
	\item $a\leq 10^4$AU and $a(1-e)\geq 20$AU;
	\item the two inner orbits cannot be too close to each other, \ie $a(1-e)\geq 10a_1$ and $a(1-e)\geq 10a_2$; and
	\item the two inner orbits are not initially too small, \ie $a_1,a_2\geq 1$AU.
\end{enumerate}
Note that criterion (ii) is not based on observation, but is imposed since secular perturbation theory can break down for moderately hierarchical systems.

Other initial parameters, specifying the orientations of the orbits, are sampled randomly as described in \S\ref{sec:features}. The effect we include is quadrupole order only. The stopping criterion for the integrations is that any of the inner stellar binary reaches high-$e$ so that they are strongly impacted by tidal effects, where we set $r_{p,i}\equiv a_i(1-e_i)\leq 6R_\odot,(i=1,2)$. Such event is likely to produce close binaries or stellar mergers.

For comparison, we also run $10^5$ equivalent triple systems (\ie stellar binary with a tertiary 2$M_\odot$ star) and triple systems with tertiary mass 1$M_\odot$, with the same set of criteria adopted.

\begin{figure}
\centering
\includegraphics[width=\columnwidth]{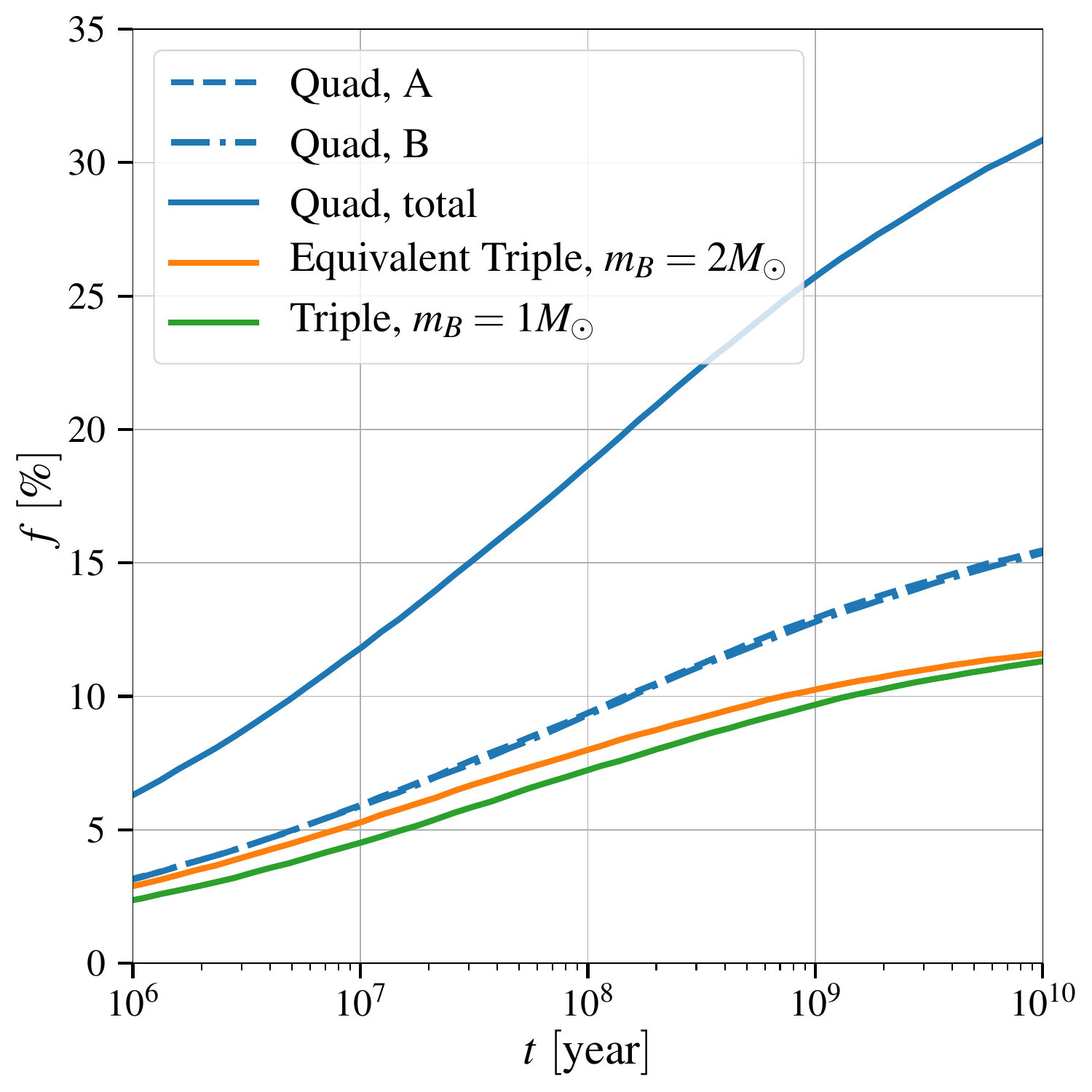}
\caption{The high-$e$ cumulative fractions in $10^5$ quadruple ``4-star'' systems versus in their ``equivalent triple'' systems (\textit{orange solid line}) and triples with solar-mass tertiary (\textit{green solid line}), as discussed in \S\ref{subsec:4star_full}. The dashed and dot-dashed blue lines are high-$e$ fractions from inner orbit A and B, respectively, which are almost the same because they are sampled from the same distributions, while the solid blue line is the sum of them, \ie total fraction.}
\label{fig:4star_general}
\end{figure}

Figure \ref{fig:4star_general} shows the total high-$e$ fraction in sampled quadruple systems is about 31\%, about 2.6 times higher than that from triples. Changing mass ratio of one inner binary is expected to increase the fraction, as shown in Figure \ref{fig:massratio}. We run $10^5$ quadruple systems with the same set of initial conditions except masses $[1+0.5]+[1+0.5]M_{\odot}$, and compare with both a set of octupole-order ``equivalent'' triple systems with 1.5\,M$_\odot$ tertiaries, and triple systems with 1\,M$_\odot$ tertiaries. The fraction of systems reaching high eccentricity increased by $\sim2-3\%$ in all cases.

\section{Implications for WD-WD Mergers}
\label{sec:SN}

We have seen from the last section that quadruple systems can largely enhance the probability of reaching the high eccentricities in a ``sustainable'' way up to the age of the Universe, with only quadrupole order terms considered. As we can see from \S\ref{sec:theory}, more interesting physical effects show up at high eccentricities, such as the GR effects and tidal effects. In this section, we will consider the hierarchical quadruple systems with a WD-WD binary and a main-sequence stellar binary, and discuss how the enhanced high-$e$ fraction can have implications for the WD merger rate.

\subsection{Merger rate}
\label{subsec:mergerrate}
In order to estimate the WD-WD merger rate, we need to run quadruple systems with different initial parameters, including their orbital orientations as well as their orbital sizes and shapes. For simplicity, we will only explore several configurations of the masses and show that the enhancement of the merger rate is generally true for all cases.

The majority of WDs are around $0.6-0.7M_\odot$ \citep[\eg][]{2007MNRAS.375.1315K,2017ASPC..509..421K}. We start by taking the WD-WD binary as equal mass with $m_0=m_1=0.7M_\odot$, and the companion binary as solar-like stars, $m_2=m_3=1M_\odot$. The radii of the WDs are $R_0=R_1=0.0084R_\odot$, estimated using the mass-radius relation \citep{1961ApJ...134..683H}. The initial orbital elements are sampled as described in \S\ref{subsec:4star_full}.
The effects we include are quadrupole order, the 1PN and tidal precession for both inner orbits, and the GW and tidal dissipation for the inner orbit A, \ie the WD-WD binary. The stopping criteria for the integrations are as follows:
\begin{enumerate}
\item the WD-WD binary collides, \ie $r_{p1}\leq R_0+R_1$, where $r_{p1}\equiv a_1(1-e_1)$ is the periastron distance of the WD-WD binary;
\item the orbital energy loss is of order unity per (inner) orbital period so that the orbital-averaged dissipation rate formulae are not valid and the WD-WD binary could collide directly, \ie $P_1\vert\dot{L}_1\vert\geq L_1$;
\item the WD-WD orbit shrinks significantly due to the GW and/or tidal dissipation, \ie $a_1<0.1$AU;
\item the stellar binary (\ie inner orbit B) reaches high-$e$ so that they are strongly impacted by tidal effects, where we set $r_{p2}\equiv a_2(1-e_2)\leq 3(R_2+R_3)$; or
\item the integrator has taken $10^7$ time-steps (\ie $4\times 10^7$ steps in RK4) so that we regard the system to be dynamically inert and uninteresting. It is likely that some of such systems would be unstable after an extremely long time (comparing to their LK timescales), but the chance should be small due to the existence of ``stable regions'' discussed in \S\ref{subsec:safe}.
\end{enumerate}
The stopping criteria (i), (ii), and (iii) contribute to WD-WD mergers, and we will call them channels (I), (II), and (III) respectively, while (iv) produces stellar mergers, which are interesting in their own right but lie outside the scope of this section.

\begin{figure*}
\centering
\includegraphics[width=\textwidth]{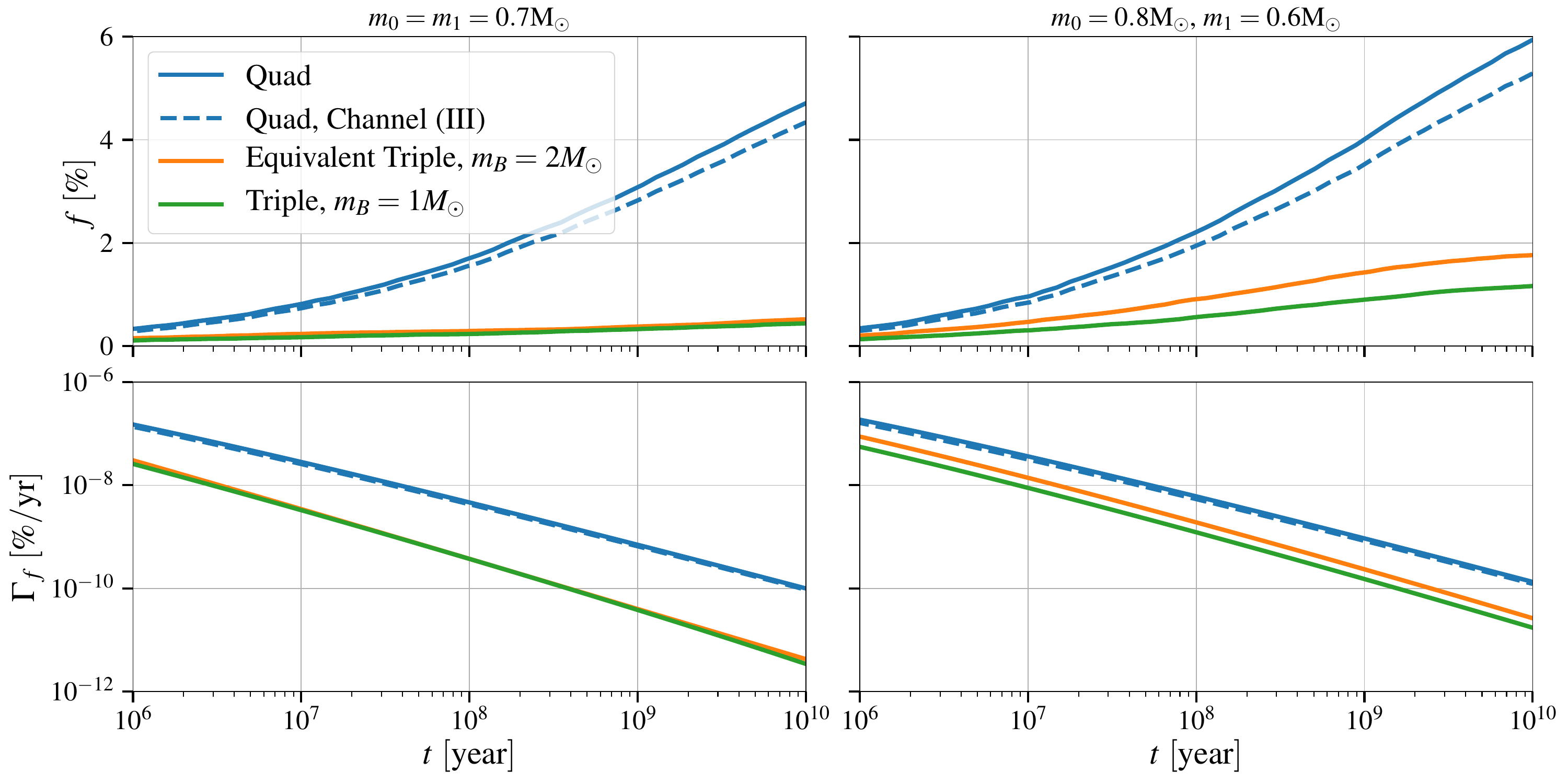}
\caption{The WD merger cumulative fractions (\textit{upper panels}) and rates (\textit{lower panels}) in $10^5$ quadruple systems (\ie [WD-WD]-[Star-Star]) (\textit{blue solid lines}) versus in their ``equivalent triple'' systems (\textit{orange solid lines}) and triples with solar-mass tertiary (\textit{green solid lines}), as discussed in \S\ref{subsec:mergerrate}. The left panel shows results from equal-mass WDs (both with 0.7$M_\odot$), while the right panel shows results from unequal-mass WDs ($0.8+0.6M_\odot$). The stellar masses in quadruple systems are both $1M_\odot$ and their ``equivalent triple'' systems have tertiary masses $2M_\odot$. The blue dashed lines are the fractions of Channel (III) mergers in quadruple systems. The rates $\Gamma_f\equiv\dot{f}$ are obtained by fitting $f$ to polynomial $f=A+B\tilde{t}+C\tilde{t}^2+D\tilde{t}^3+E\tilde{t}^4$, where $\tilde{t}\equiv\log_{10}t$ and $A,B,C,D,E$ are fitting parameters. Note that we show the early-time rates only for dynamical interest. Most of WDs form at late times depending on their progenitor masses, so we should only focus on the rate at late times.}
\label{fig:WD_secular}
\end{figure*}

We run $10^5$ such systems and, for comparison, we also run $10^5$ equivalent triple systems (\ie WD-WD binary with a tertiary 2$M_\odot$ star) and triple systems with tertiary mass 1$M_\odot$ (which is astrophysically more realistic) with the same set of criteria adopted. For unequal-mass WDs (\eg $0.8+0.6$M$_\odot$), octupole order effects are turned on, and the merger rates are expected to increase for both quadruple and triple systems. Figure \ref{fig:WD_secular} shows that the overall enhancement of merger rates from quadruple systems with respect to their equivalent triples (or with $1M_\odot$ tertiary) is $\sim$9 (or 10) for the equal-mass case, while it drops to $\sim$3.5 (or 5) for the unequal-mass case. We also find that the Channel (I) contribution is negligible (only 2 in $10^5$ for each run), while most of the mergers go through Channel (III), \ie the orbital shrinking, shown by the dashed lines in Figure \ref{fig:WD_secular}.

\subsection{Understanding the results}
From the overall results, there are two immediate questions:
\begin{enumerate}
\item Why is Channel (I) suppressed?
\item What role does each effect play?
\end{enumerate}
Let us first understand what the meanings of the three channels are. For the equal-mass case, Channel (I) is equivalent to having $a_1(1-e_1)\leq 2R_{\rm WD}$, \ie
\begin{equation}
1-e_1\leq 7.8\times 10^{-5}\left(\frac{a_1}{\rm AU}\right)^{-1}~.
\label{eq:I}
\end{equation}
Channel (II) is equivalent to
\begin{equation}
\frac{L_1}{\vert\dot{L}_1\vert}\leq \left(\frac{a_1}{\rm AU}\right)^{3/2}\left(\frac{m_A}{M_\odot}\right)^{-1/2}~,
\end{equation}
which is equivalent to
\begin{equation}
1-e_1\leq 3.6\times 10^{-6}\left(\frac{a_1}{\rm AU}\right)^{-5/7}
\label{eq:III_GW}
\end{equation}
if the dissipation is dominated by the GW emission. However, combining Eqs.~(\ref{eq:I}) and (\ref{eq:III_GW}) suggests that, in order to make Channel (II) more likely to happen than Channel (I), we need $a_1\gtrsim 5\times 10^{4}\,$AU, which does not explain the suppression of Channel (I).

In fact, Channel (II) is only made possible due to the tidal dissipation. For a simple order-of-magnitude estimation, we can use the tidal energy dissipation per (inner) orbit from \cite{1977ApJ...213..183P} (\textit{hereafter} PT), rewritten in our notation as
\begin{equation}
\Delta E = \frac{2Gm_0^2}{R_{\rm WD}}\left[\left(\frac{R_{\rm WD}}{r_{p1}}\right)^6T_2(\eta_1)+\left(\frac{R_{\rm WD}}{r_{p1}}\right)^8T_3(\eta_1)\right]~.
\end{equation}
In this expression, the dimensionless function $T_\ell$ corresponds to excitation of multipole-$\ell$ modes of the WDs, and $\eta_1$ is the ratio of the periastron passage timescale to the dynamical timescale of the WDs; for the equal-mass case,
\begin{equation}
\eta_1 = \frac{1}{\sqrt{2}}\left(\frac{r_{p1}}{R_{\rm WD}}\right)^{3/2}~.
\end{equation}

Since the $T_2$ term usually dominates\footnote{$T_2$ term dominates if $R_{\rm WD}/r_{p1}\ll \sqrt{T_2/T_3}$. From Fig. 1 in PT, $T_2/T_3\sim 1$ when $\eta_1$ approaches 2, and approaches $\sim 5/3$ at large $\eta_1$. However, $R_{\rm WD}/r_{p1}<1/2$, so that the $T_2$ contribution dominates.}, Channel (II) is equivalent to having
\begin{equation}
1-e_1\lesssim 2\left(\frac{R_{\rm WD}}{a_1}\right)^{5/6}[T_2(\eta_1)]^{1/6}\sim\left(\frac{R_{\rm WD}}{a_1}\right)^{5/6}~,
\end{equation}
where during the close passage, $T_2(\eta_1)\sim 0.01-0.1$. In order to make Channel (I) happen before Channel (II), we need $a_1\lesssim 4\times 10^{-3}$AU, impossible for our initial configurations. Thus, Channel (I) is largely suppressed.

Channel (III) assumes that when we detect a significant orbital shrinking (at least a factor of 10), the WD binary will merge on a short timescale. This is reasonable because orbital shrinking is only significant when the timescale of the energy dissipation is smaller than the LK timescale, which is typically of order $10^7$ years in our initial configurations.

There are 3 major scales: $a_1$, $a$, and $r_{p1}$, and they determine the timescales of all the effects we consider, hence their dominant regimes. Here we list all the relevant timescales using the ``AU, year, $M_{\odot}$'' unit system:
\begin{align}
&P = a^{3/2}M^{-1/2}~,\label{eq:P}\\
&P_1=a_1^{3/2}m_A^{-1/2}~\label{eq:P1},
\end{align}
\begin{align}
&t_{\scriptscriptstyle \rm LK,1} \simeq \frac{a^3}{a_1^{3/2}}(1-e^2)^{3/2}\sim 0.04\left(\frac{a}{a_1}\right)^3a_1^{3/2}~,\label{eq:tLK1}\\
&t_{\scriptscriptstyle \rm LK,1}^{\rm (ins)} \sim t_{\scriptscriptstyle \rm LK,1}\sqrt{1-e_1^2}\sim 0.07\left(\frac{a}{a_1}\right)^3a_1 r_{p1}^{1/2}~,\label{eq:tLK1ins}
\end{align}
\begin{align}
&t_{\rm pr}^{\rm (1PN)} \sim 4.1\times 10^7 a_1^{3/2} r_{p1}~,\label{eq:tpr1pn}\\
&t_{\rm pr}^{\rm (tide)}\sim 3.8\times 10^{23} a_1^{3/2} r_{p1}^5~,\label{eq:tprtide}
\end{align}
\begin{align}
&t_{\rm diss}^{\rm (GW)}\sim 9.5\times 10^{18} a_1^{1/2} r_{p1}^{7/2}~,~{\rm and}\label{eq:tdissGW}\\
&t_{\rm diss}^{\rm (tide,PT)}\sim 5.5\times 10^{21}[T_2(\eta_1)]^{-1} a_1^{1/2} r_{p1}^6~,
\end{align}
where $P$ and $P_1$ are the orbital periods of mutual and inner orbit A, $t_{\scriptscriptstyle \rm LK,1}^{\rm (ins)}$ stands for the instantaneous LK timescale of the inner orbit A \citep[][]{2014MNRAS.438..573B}, ``pr'' stands for precession and ``diss'' stands for dissipation. The tidal dissipation used here is from PT (see \S\ref{subsec:TidalDissip}), which is much simpler than but at order-of-magnitude level consistent with \cite{1986AcA....36..181G}, and its analytic form shows that it is negligible at most of separation scales, but may take over when the periastron is small. We have estimated the mutual orbit eccentricity $e$ as $1/\sqrt{2}$ due to its thermal distribution. $T_2(\eta_1)$ can be estimated as a power law
\begin{equation}
T_2(\eta_1)\sim 0.4\left(\frac{\eta_1}{2}\right)^{-2.47}
\end{equation}
for $\eta_1\gg 2$. Although this expression overestimates $T_2$ when $\eta_1$ is approaching 2, it is still good at the order-of-magnitude level.

\begin{figure}
\centering
\includegraphics[width=\columnwidth]{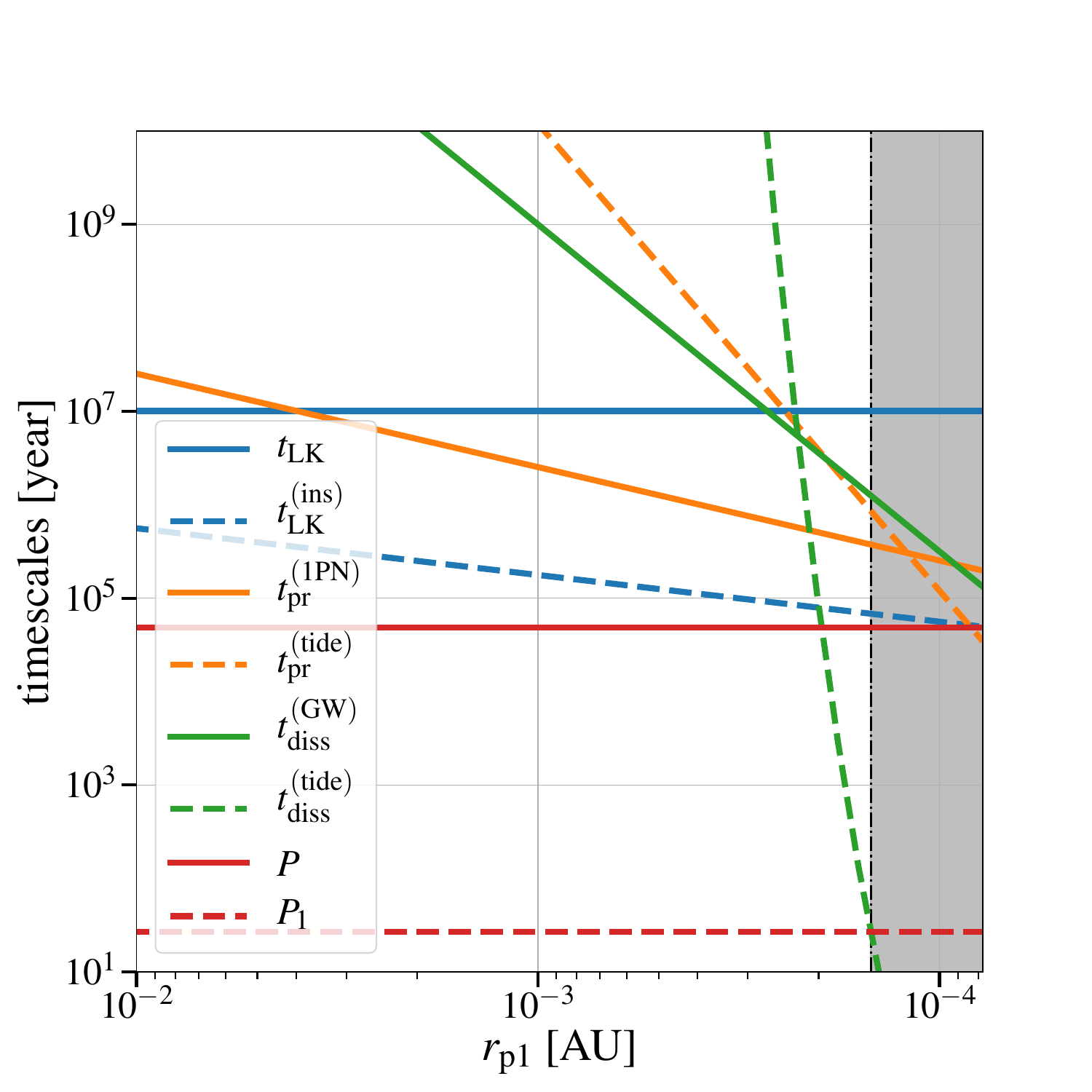}
\caption{The timescales versus the periastron of the inner orbit A, $r_{p1}$, for a system with $a=2000$AU, $a_1=10$AU. All the timescales except the tidal dissipation are calculated using Eqs.(\ref{eq:P}-\ref{eq:tdissGW}), while the tidal dissipation timescale is calculated from \protect\cite{1986AcA....36..181G} as we use in our code. The shaded region is Channel (II) region.}
\label{fig:timescales}
\end{figure}

For $a=2000$AU and $a_1=10$AU, we plot the timescales versus the periastron of the inner orbit A, $r_{p1}$, in Figure \ref{fig:timescales}, where the tidal dissipation is calculated from \cite{1986AcA....36..181G} as we use in our code. The minimal $r_{p1}$ shown in the figure is $2R_{\rm WD}=7.8\times 10^{-5}$AU, below which we assume a collision (Channel (I) event) occurs. The shaded region represents Channel (II) region, and is determined by the intersection between the tidal dissipation timescale $t_{\rm diss}^{\rm (tide)}$ and the WD-WD binary orbital period $P_1$.

\subsection{Classification of orbital shrinking}
\label{subsec:classificatioon}
Channel (III), \ie undergoing orbital shrinking, is divided into 3 categories, Type-IIIL, Type-IIIC and Type-IIIS, based on how they fall into the shrinking phase. The normal Kozai motion of triple systems has two types of trajectories: libration and circularization. In quadruple systems, the inner orbits can switch between these two types. The Type-IIIL mergers undergo rapid orbital shrinking when they are on the libration trajectory, while the Type-IIIC mergers shrink on the circularization trajectory. The Type-IIIS mergers are initially at the orbital shrinking phase, which are of less interest. We also identify a subtype in each category, \ie those systems that show eccentricity oscillations on the GR precession timescale during their beginning phase of orbital shrinking. We denote those ``wiggled'' systems with ``w'', \ie Type-IIILw, Type-IIICw and Type-IIISw.

We examine a set of $10^4$ [WD-WD]-[Star-Star] systems and find 491 (\ie $\sim$4.91\%) systems undergoing orbital shrinking (\ie Channel III). Among these systems, 209 (49.9\%) systems are Type-IIIL with 27 (6.4\%) in Type-IIILw, 207 (1.4\%) systems are Type-IIIC with 6 in Type-IIICw, the rest of 3 (0.7\%) systems are Type-IIIS with 1 (0.24\%) in Type-IIISw. About 8\% of the orbital shrinking systems experience the ``wiggled'' phase during shrinking, which we will call ``precession oscillation'' phase from now on.

\begin{figure*}
\centering
        \subfigure[Type-IIIL: Orbital shrinking from the libration trajectory]{
                \centering
                \includegraphics[width=.5\linewidth]{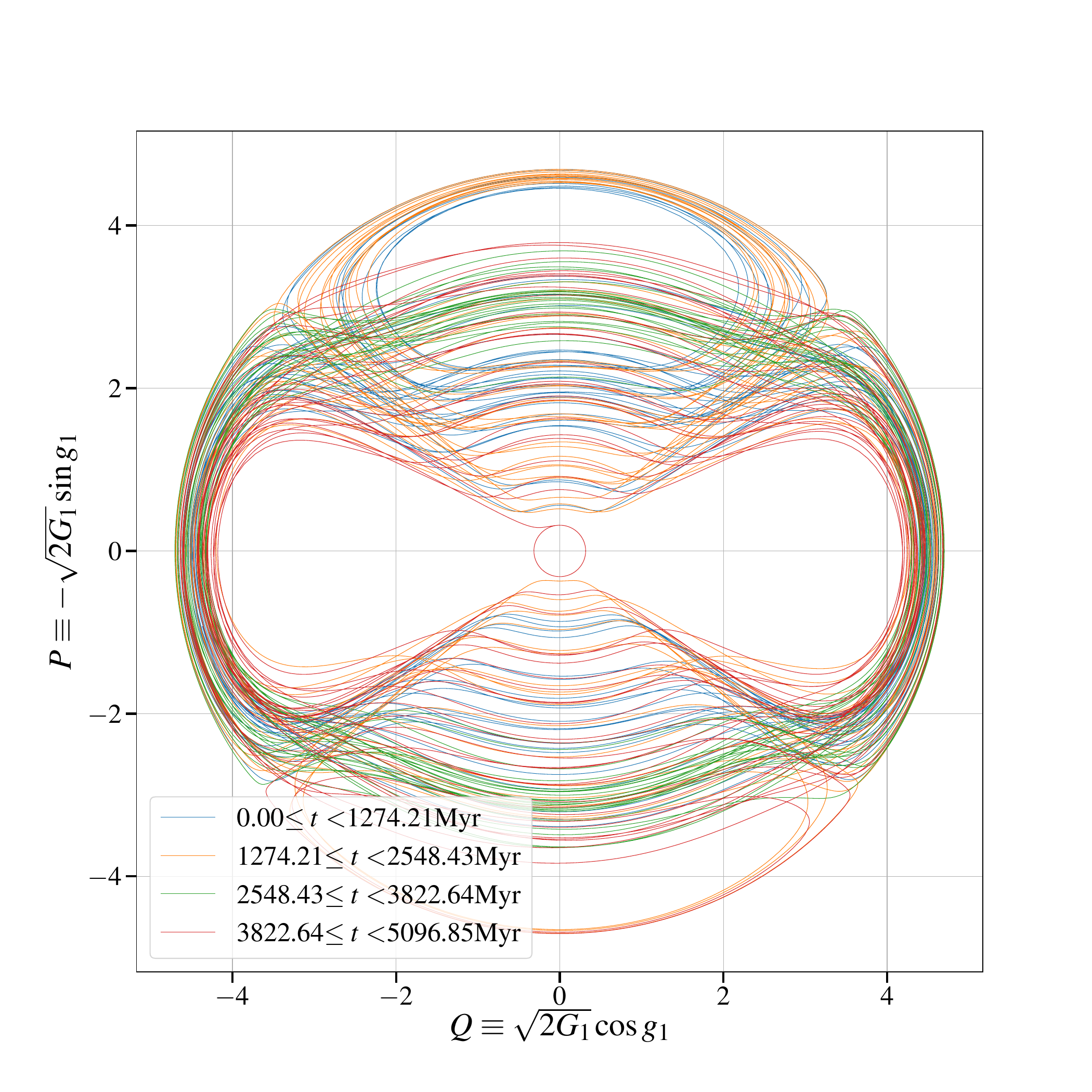}
        }%
        \subfigure[Type-IIIC: Orbital shrinking from the circularization trajectory]{
                \centering
                \includegraphics[width=.5\linewidth]{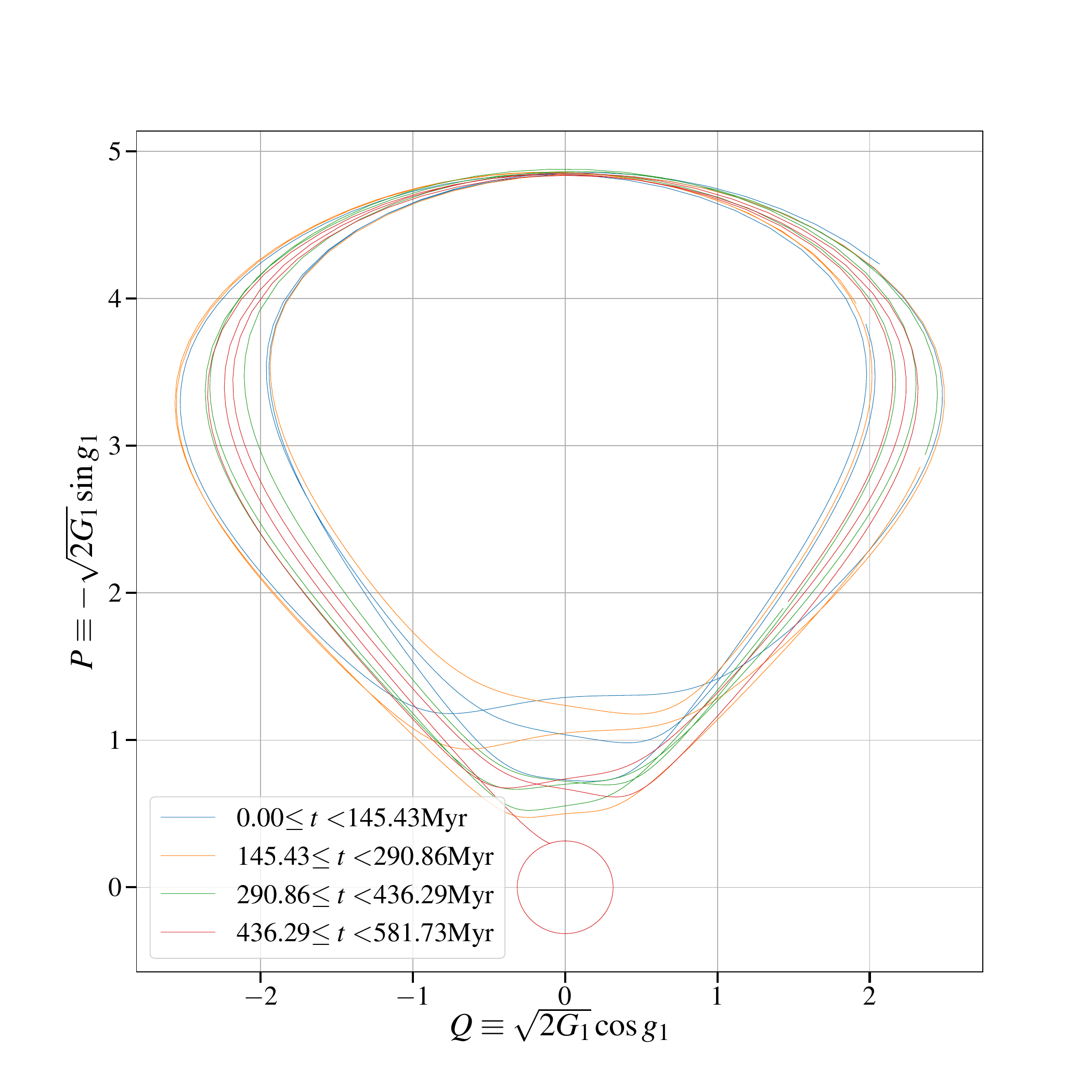}
        }%
        \\
        \subfigure[Type-IIILw: Type-IIIL with ``precession oscillation'' phase]{
                \centering
                \includegraphics[width=.5\linewidth]{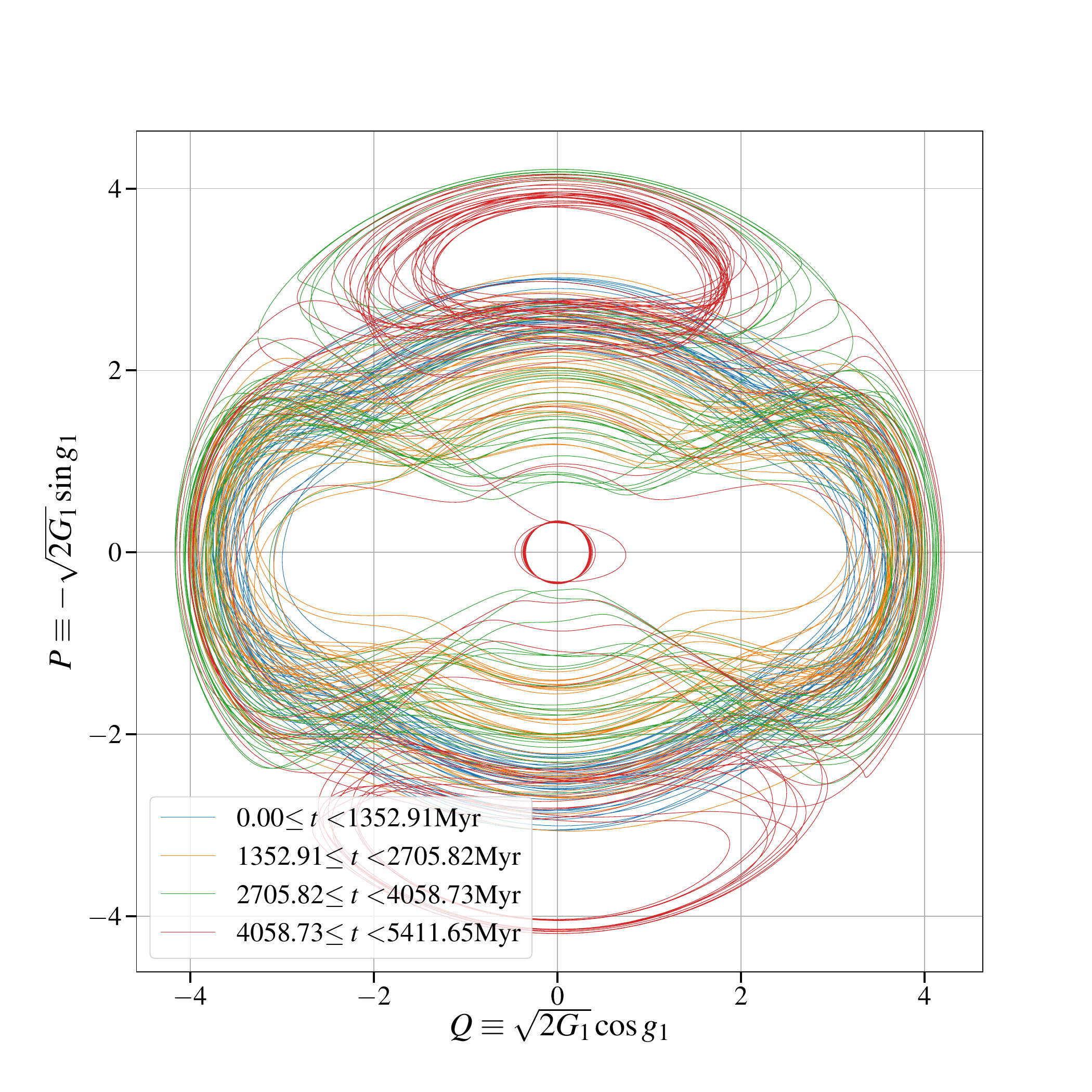}
        }%
        \subfigure[Type-IIICw: Type-IIIC with ``precession oscillation'' phase]{
                \centering
                \includegraphics[width=.5\linewidth]{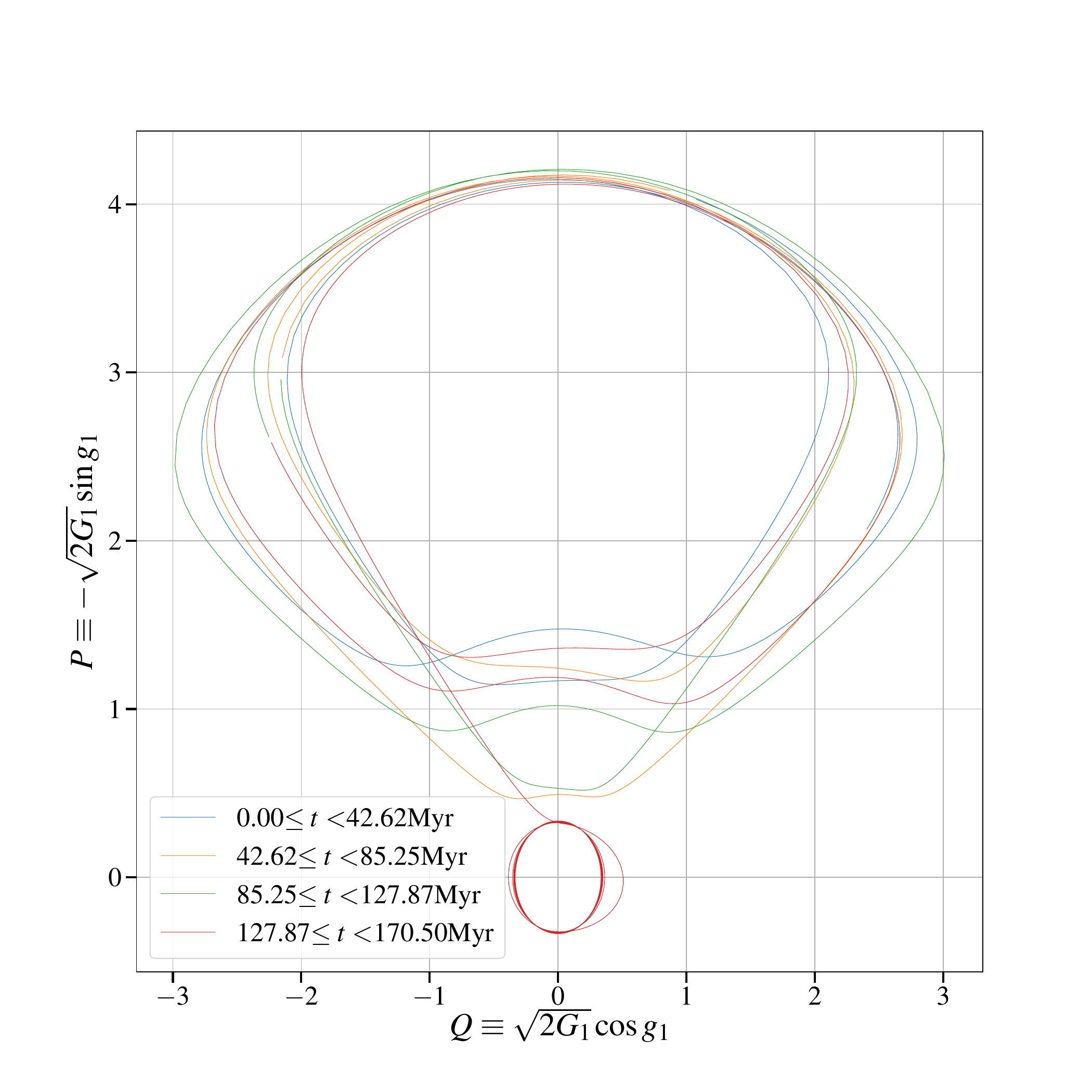}
        }
        \caption{Classification of orbital shrinking WD mergers.}\label{fig:classification}
\end{figure*}

In Figure \ref{fig:classification} we show the example phase diagrams of Type-IIIL, Type-IIIC and their ``wiggled'' subtypes. The ``non-wiggled'' subtypes (\textit{upper panel}) show that at the final stage the WD binaries go on extremely high eccentricities and then rapidly shrink their orbits, so that they decouple from their companions and directly enter the small circular trajectories on the phase diagram, where the angular momentum is approximately conserved. The ``wiggled'' subtypes (\textit{lower panel}) show that the inner orbit angular momentum $G_1$ oscillates several times before decoupling from the companions (also see Figure \ref{fig:equalSN_Quad_NoEvec_part0_sys9469_5plots_zoom}, \ref{fig:equalSN_Quad_NoEvec_part0_sys9469_G_vs_g_zoom}). We show the underlying physics of this ``precession oscillation'' in detail in Appendix \ref{app:rapidwiggle}.

\section{Nonsecular effects: evection}
\label{sec:evection}

The ``double-averaging'' procedure neglects non-secular effects, including the ``rapid eccentricity oscillations'' of the inner orbits on the timescale of the outer period. This effect was discovered in the motion of the Moon by Ptolemy, known as the Moon's ``second inequality'' and much later as ``evection'' \citep{1515Ptolemy,1896itlt.book.....B,1984ptal.book.....T}. In this section, we discuss how evection affects the eccentricities, and show that our conclusion still holds that quadruples are more efficient in producing mergers than triples.

\begin{figure}
\centering
\includegraphics[width=\columnwidth]{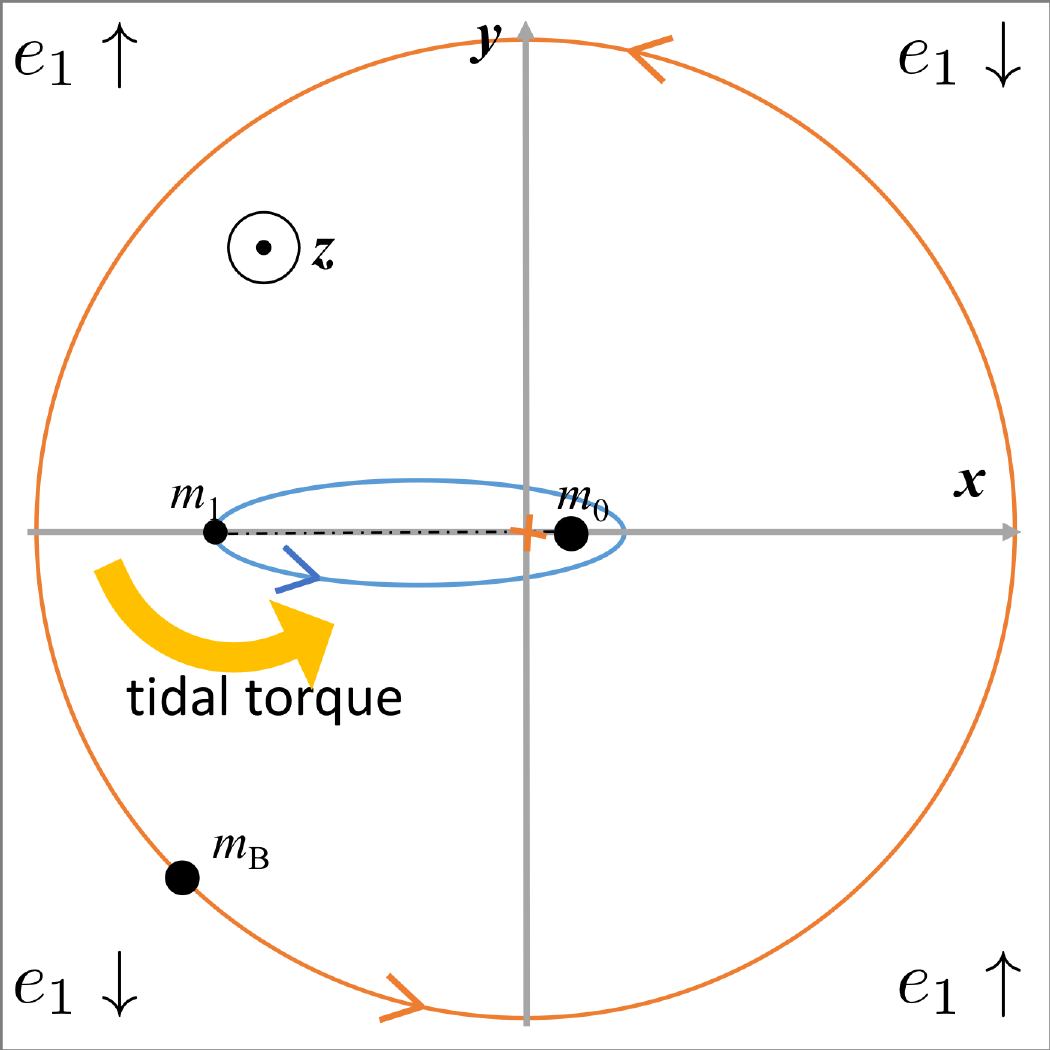}
\caption{An illustration of evection. The inner orbit (in \textit{blue}) and the mutual orbit (in \textit{orange}) are both in the $x-y$ plane, with their angular momenta along the $+z$ direction. The perturber $m_B$ at the position shown in the figure exerts a tidal torque on the inner orbit, as shown by the yellow arrows, which decreases the inner eccentricity $e_1$. As the perturber moves around, the tidal torque will change its direction according to the quadrant the perturber is in, and the change of $e_1$ is shown at the four corners. During one period of the mutual orbit, the eccentricity $e_1$ goes up and down twice.}
\label{fig:evection}
\end{figure}

The nature of evection is that the tidal torque on the inner orbit exerted by the outer perturber varies and changes its sign four times during the period of the outer orbit, as illustrated in Figure \ref{fig:evection}. The oscillations of the orbital elements on the timescale of the outer period in the context of triple star systems were discussed by \cite{1975A&A....42..229S}. Assuming a circular outer orbit, the amplitude of the oscillation of the inner binary angular momentum was derived by \cite{2005MNRAS.358.1361I} in the high eccentricity limit of the inner orbit (\ie $e_{\rm in}\rightarrow 1$). Later this phenomenon was observed in simulations by \cite{2014MNRAS.438..573B} in the test particle limit and by \cite{2012ApJ...757...27A} in the equal-mass inner binary case, and discussed by \cite{2012arXiv1211.4584K} in the WD-WD context. Its impact on GW observations has also been discussed \eg by \cite{2013PhRvL.111f1106S}. In the presence of the eccentric LK mechanism (from octupole order) and the non-secular evection, the merger times can be orders of magnitude shorter than predicted by double-averaged secular calculations in triple systems with low outer orbit eccentricities \citep{2014MNRAS.439.1079A}.

We generalize the derivation in \cite{2005MNRAS.358.1361I} (assuming circular outer orbit) to triple systems with eccentric outer orbits in Appendix \ref{App:Evec}. We find that in addition to the ``$\frac12P_{\rm out}$'' periodic eccentricity oscillations that we have seen in the circular outer orbit case, now we also have ``$P_{\rm out}$''-periodic and ``$\frac{1}{3}P_{\rm out}$''-periodic oscillations, due to the modulation of the tidal field (with frequency $2n$) by the varying distance of the perturber (with frequency $n$), as shown in Eqs.~(\ref{eq:DeltaG1z_prime1},\ref{eq:DeltaG1z_prime2}).

Since evection can cause rapid mergers and/or cause mergers/collisions in otherwise non-merger systems, we need to assess its importance in increasing merger rates for quadruple and triple systems. We first calculate the ``upper bound'' of the eccentricities of the inner orbits at each time-step using the equations derived in Appendix \ref{App:evec_bound}:
\begin{equation}
e_1^{(\rm bound)} = \sqrt{1-\left[\sqrt{1-e_1^2}-\frac{15}{8}\sqrt{\frac{m_B^2a_1^3}{m_A (m_A+m_B)a^3}}\frac{F(e,i_A)}{(1-e^2)^{3/2}}\right]^2}
\end{equation}
and similarly for $e_2^{(\rm bound)}$, where the functions $F(e,i_{A/B})$ are defined in Eq.~(\ref{eq:F-ei}). Then we use these values to check whether some of the stopping criteria are satisfied. If the upper bound is high enough to make a collision, then it could mean that there is a collision due to evection, or that the evection amplitude is overestimated, so we will switch to a tighter estimation as discussed in Appendix \ref{app:evec_TEE}. That is, we calculate the ``true evection envelope'' (\textit{hereafter} TEE) using Eqs.~(\ref{eq:e_TEE},\ref{eq:DeltaG1z_prime1}), which are more computationally expensive but do not contain any inequality.

We reran the $10^5$ [WD-WD]-[Star-Star] systems discussed in \S\ref{sec:SN}, but this time we use the highest evection eccentricities to determine whether the mergers occur. The merger fraction estimated in this way is expected to be an upper limit, since in a finite number of periastron passages the maximum of the evection envelope may not be sampled. When the dissipation effects are turned on, we use the secular eccentricities to calculate the dissipation rates. We also test the cases where we use the TEE eccentricities for the dissipation rate estimation, which does not affect our results significantly. Note that the expressions are derived for highly eccentric inner orbits, so we only turn on the evection calculation if either inner orbit when its eccentricity is very large (\eg $e_1,e_2>0.9$ in our code).

\begin{figure*}
\centering
\includegraphics[width=\textwidth]{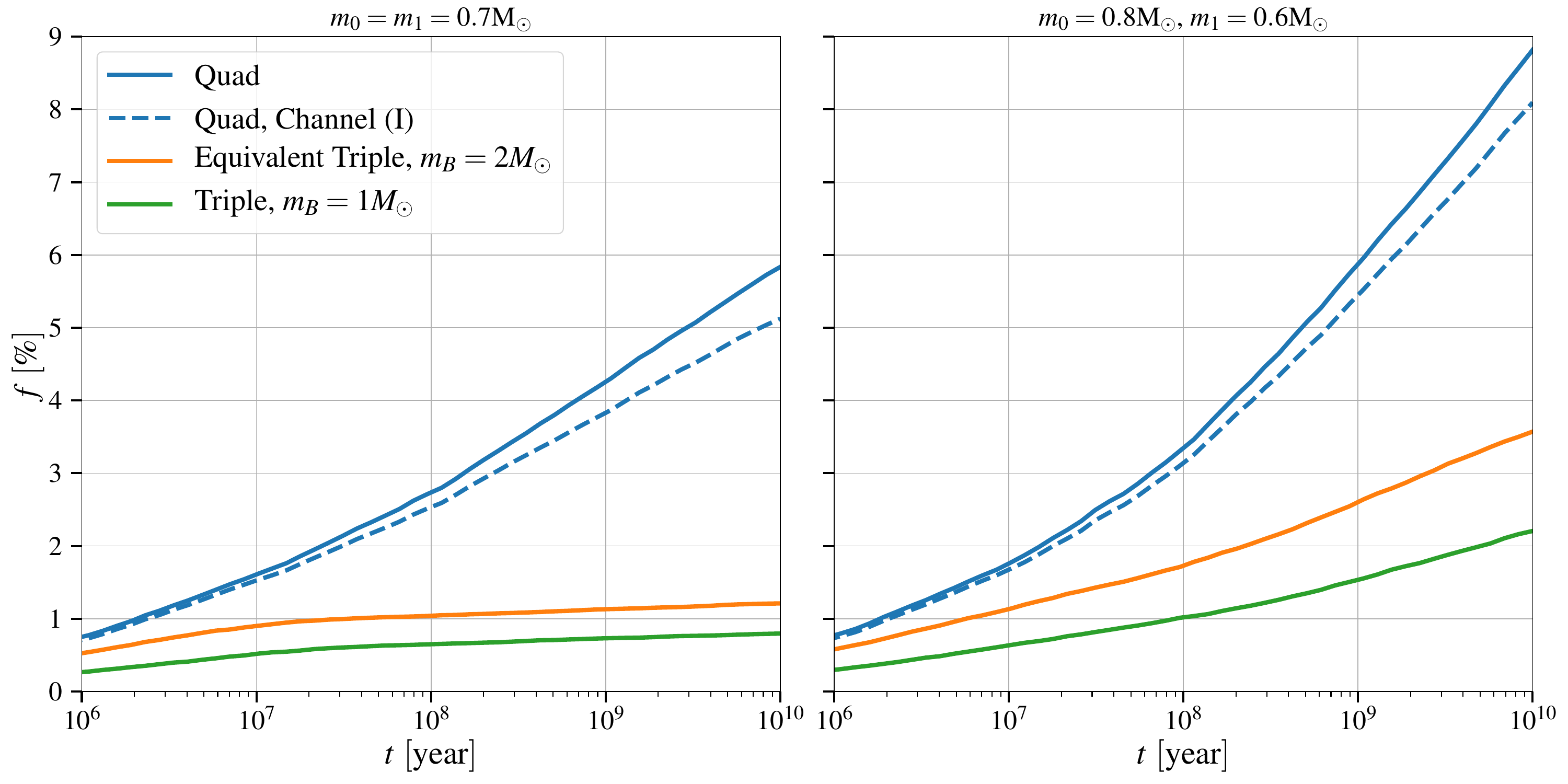}
\caption{The merger fractions from $10^5$ random [WD-WD]-[Star-Star] systems (\textit{blue solid lines}) and their equivalent triple cases (\textit{orange solid lines}), with evection included, as described in \S\ref{sec:evection}. Results for triples with solar-mass tertiary are shown with green solid lines. With evection, the merger fractions are enhanced for both quadruple systems and triples, but the fraction from quadruples remains much larger than that from triples. Mergers from Channel (I) (\textit{blue dashed lines}) now dominates over Channel (III), and Channel (II) becomes negligible. The runs for this figure is equivalent to those for Fig.~\ref{fig:WD_secular} except for including evection.}
\label{fig:WD_evec}
\end{figure*}

Figure \ref{fig:WD_evec} shows the merger rates from quadruple systems and their equivalent triple cases, for different WD-WD masses. Comparing to the secular results in Figure \ref{fig:WD_secular}, the merger fractions increase by small fractions for both quadruples and triples, and the enhancement from quadruple systems with respect to their equivalent triples (or with 1$M_\odot$ tertiary) are still large, \ie $\sim$5 (or 7) times for the equal-mass WDs and $\sim$2.5 (4) times for the unequal-mass case. Note that the mergers from evection runs are mostly from Channel (I), and the rest are almost from Channel (III). Mergers from Channel (II) become very rare. This makes sense because systems that undergo orbital shrinking must be at very high eccentricities (\ie with small orbital angular momentum), where the torque from the companion binary can more easily extract most of the inner orbital angular momentum and result in a Channel (I) collision. However, we must note that the dominance of Channel (I) mergers does not mean that we have more ``direct collision'' events, because TEE is just a possible eccentricity maximum: in reality the eccentricity may not reach TEE value and the ``direct collision'' may be avoided. Although the exact merging channel is important to the outcome (\eg whether the interaction results in a SN Ia), we leave that discussion for future work.

Finally, we must emphasize that our runs are based on the validity of secular calculations as a representation of the mean evolution, which limits our explorations to the ``highly hierarchical'' cases, as we have assumed in the sampling of initial parameters (\ie $a_1,a_2\leq r_p/10$). In this regime, we find that evection can modestly enhance the merger rates by a factor of $\sim1.5$ (compare the right panels of Figures ~\ref{fig:WD_secular} and \ref{fig:WD_evec}). In more moderately hierarchical systems, evection could be much more important, as suggested by \eg \cite{2012arXiv1211.4584K} and \cite{2014MNRAS.439.1079A}. However, to assess evection fully in less hierarchical systems, one would have to correct the double-averaging equations or drop the outer orbit average entirely \citep[\eg][]{2016MNRAS.458.3060L}.

\section{Discussion and Conclusion}
\label{sec:conclusion}

Hierarchical quadruple systems are more complex than triple systems and can exhibit qualitatively different behaviour on long timescales. Most interestingly, the fraction of systems that can reach high eccentricities is significantly enhanced in quadruple systems, with a correspondingly higher probability of producing WD-WD and stellar mergers.

We have derived the secular equations for general hierarchical quadruple systems up to octupole order, and shown that the fraction of reaching high-$e$ is enhanced even at quadrupole order. We have run the systems up to the age of the Universe and found the event rate of reaching high eccentricities goes approximately as $1/t$ (Figure \ref{fig:GrowFrac}), consistent with current observations of the delay-time distribution of SNe Ia \citep[\eg][]{2012PASA...29..447M}. We also found that for a given initial configuration, running through an extremely long time, a fraction of systems will never reach high eccentricities, and that these are initially mostly coplanar (Figures \ref{fig:frac_future} and \ref{fig:safe_initcosi}).

We have calculated the amplitude of eccentricity oscillations due to evection when the inner orbits are at high eccentricities (\S\ref{sec:evection} and Appendix \ref{App:Evec}), and used it to estimate the enhancement. This method is much faster than full $N$-body simulations but accurately describes the eccentricity ``envelope'' systems attained during successive eccentricity maxima.

\subsection{Stellar quadruples}

We have investigated the fraction of systems reaching high eccentricity in quadruple systems consisting of two pairs of solar-mass stars and compared our results to stellar triple systems over a large portion of parameter space.

With only quadrupole order effects turned on and our baseline distribution of initial orbital elements, about 31\% of quadruple systems reach high eccentricities within the age of universe (Figure \ref{fig:4star_general}), about 2.6 times higher than that from triples, indicating a high probability of producing close binaries, stellar mergers, or even blue stragglers \citep[\eg][]{1979A&A....77..145M,2001ApJ...562.1012E,2006A&A...450..681T,2009ApJ...697.1048P,2013ApJ...766...64S,2014MNRAS.439.1079A,2014ApJ...793..137N} in quadruple systems on long timescales. We also ran quadruple systems with $[1+0.5]+[1+0.5]$\,M$_{\odot}$ including octupole-order terms, and compared with both a set of octupole-order ``equivalent'' triple systems with 1.5\,M$_\odot$ tertiaries, and triple systems with 1\,M$_\odot$ tertiaries. The fraction of systems reaching high eccentricity increased by $\sim2-3\%$ in all cases. 

Our results suggest a dynamical explanation for the observation that the ratio of quadruples to triples, as well as the ratio of triples to binaries, seems to be in excess among young stars \citep[\eg][]{2006A&A...459..909C,2013ApJ...768..110C}, and that a large fraction of triples and quadruples are in tight binaries \citep[\eg][]{2009AJ....137.3646P}. The anomalously high number of (relatively) high-mass stars in the thick disk observed by APOGEE could also be evidence of stellar mergers \citep{2017arXiv170905237I}. 

A simple estimate for the role of quadruple systems in producing tight stellar binaries, mergers, or collisions proceeds as follows. The rate at which systems reach high eccentricities can be estimated as 
$$\dot{N}_{\rm high-e}(t) = \int_0^t{\rm SFR}(t')\bar M_{\rm sys}^{-1}f_{\rm Quad}f_{\rm cut}\dot{f}(t-t')\,dt',$$ where SFR is the star formation rate (units: $M_\odot$ yr$^{-1}$), $\bar M_{\rm sys}\sim 0.6 M_\odot$ is the mean mass of a stellar system\footnote{The number fractions of single, binary, triple and quadruple systems are about 56\%, 33\%, 8\% and 3\% \citep{2010ApJS..190....1R}, giving an average of 1.58 stars per system, despite the dependence of multiplicity on masses \citep[\eg][]{2016AJ....152...40G,2017ApJS..230...15M}. Using the initial mass function Eq.~(2) in \cite{2001MNRAS.322..231K}, we obtain the average mass of stars $\langle M_*\rangle \sim 0.38M_\odot$.}, $f_{\rm Quad}\sim 0.03$ is the fraction of systems that are quadruples \citep[\eg][]{2010ApJS..190....1R}; and $f_{\rm cut}\sim 1$ is the fraction of quadruple systems of interest. Assuming a constant Milky Way star formation rate, the event rate is simply $\dot{N}_{\rm high-e}(t) \sim {\rm SFR}\times f_{\rm Quad}f(t<10^{10}{\rm yr})/\bar M_{\rm sys}$. Recent estimates of the Milky Way SFR range from around $0.68-4$\,M$_\odot\,{\rm yr}^{-1}$ \citep[\eg][]{2006Natur.439...45D,2006A&A...459..113M,2010ApJ...709..424M,2010ApJ...710L..11R,2012ARA&A..50..531K}, which combined with our estimate of $f(t<10^{10}{\rm yr})=0.31-0.33$ gives an event rate from quadruple systems of $\sim 0.01-0.06\,{\rm yr}^{-1}$. Accounting for the relative frequency of stellar quadruples and triples, the same calculation for triples gives an additional $\sim 0.01-0.06\,{\rm yr}^{-1}$, very similar to the quadruple contribution. In total, the rate is about $0.02-0.12\,{\rm yr}^{-1}$, around $2-50\%$ of the observed Galactic rate of bright stellar mergers ($M_V\geq -3$), which is $0.24-1.1\,{\rm yr}^{-1}$ \citep{2014MNRAS.443.1319K}. Considering that the SFR was higher in the past, and that we have only used $f$ from two fixed-mass configurations, our estimates for the quadruple and triple contributions could be conservative. We conclude that it is plausible, but by no means certain, that triple + quadruple systems are an important channel for stellar mergers.

\subsection{WD-WD binaries and Type Ia supernovae}

We propose a new channel for producing Type Ia SNe: WD-WD mergers in hierarchical quadruple systems. Although selection effects make the observation of WD-WD binaries in triple or quadruple systems difficult, the high multiplicity of A-type stars indicates that it is common for WD-WD binaries to live in triple and quadruple systems \citep[\eg][]{2012MNRAS.422.2765D,2014MNRAS.437.1216D}. In the [WD+WD]+[star+star] case, we have added GR and tidal precession and dissipation effects. We have sampled the systems from a distribution of sizes and shapes, as well as random orientations. We find a significantly enhanced merger rate, a factor of $3.5-10$ higher than that in triple systems (Figure \ref{fig:WD_secular}), and a $\sim 1/t$ delay-time distribution for both quadruples and triples, consistent with that from observations \citep[\eg][]{2010ApJ...723..329H,2014ARA&A..52..107M}. We classify the major type of mergers, \ie those undergoing rapid orbital shrinking, into 3 categories by their evolution patterns in phase space (Figure \ref{fig:classification}), and identify $\sim 8\%$ of orbital shrinking mergers that experience a ``precession oscillation'' phase, whose underlying physics is explained in Appendix \ref{app:rapidwiggle} and Figure \ref{fig:wiggle}.

The secular merger rate from quadruple systems inferred from Figure \ref{fig:WD_secular} is $\sim 10^{-12}\,{\rm yr}^{-1}\,{\rm Quad}^{-1}$ at $t=10^{10}$\,yr, which corresponds to a merger rate per unit of initial stellar mass of $\sim f_{\rm Quad}f_{\rm cut}\dot{f}(t=10{\rm Gyr})/\bar M_{\rm sys}.$ We impose a cut based on a Kroupa initial mass function for the primary in each of the inner binaries, and uniform distribution for secondary-to-primary mass ratio, and take the WD progenitor mass range to be $1-8\,M_\odot$, leading to $f_{\rm cut}\sim 4.5\%$\footnote{Variations in the secondary mass distribution can result in a factor of $\sim2$ lower \citep[\eg][]{2017MNRAS.465L..44K}.}. Considering the rates from the ``equal-'' and ``unequal-mass'' cases presented in \S\ref{sec:SN} and \ref{sec:evection}, we obtain a WD-WD merger rate from quadruple systems of $(2.7-5.3)\times 10^{-15}\,{\rm yr}^{-1}\,{M_\odot}^{-1}$. Including the contribution of triples, we have a total merger rate $(3-8)\times 10^{-15}\,{\rm yr}^{-1}\,{M_\odot}^{-1}$. This can be compared to the observed SNe Ia rate at 10 Gyr, $(1-5)\times 10^{-14}\,{\rm yr}^{-1}\,{M_\odot}^{-1}$ \citep[\eg][]{2014ARA&A..52..107M}. We conclude that in an optimistic scenario, the quadruples + triples provide enough merging white dwarf pairs to explain of order half of the SN Ia rate at long delay times. Indeed, our rate estimate is close to those from traditional binary stellar synthesis \citep[\eg][]{2009ApJ...699.2026R}. On the other hand, it is unclear whether all of these mergers result in Ia supernovae, and at the more pessimistic end of the rate calculation the rate is only 6\%\ of the observed SN Ia rate. Additionally, the lowest mass stars we consider are still on the main sequence and hence unavailable for WD-WD mergers at short delay times. If $f_{\rm cut}$ is an increasing function of $t$, then this could spoil the $\propto 1/t$ delay-time distribution derived from dynamics. Also we have not taken into account the possible production of stellar mergers before WDs are formed, which could lower $f_{\rm cut}$ by $\sim 30\%$ for quadruples and $\sim 10\%$ for triples (estimated from Figure \ref{fig:4star_general}). In any case, it is noteworthy that the quadruples dominate over the triples in our calculation.

We found that evection enhances the overall merger rate in quadruple systems by a modest factor of $\sim1.5$ (compare Figs.\ \ref{fig:WD_secular} and \ref{fig:WD_evec}). However, it could play an important role in determining the branching ratio of different merging/collision channels, which may affect whether or not a SN Ia occurs and its observed properties. Importantly, the estimate of the importance of evection in this paper involves bounds rather than a calculation of the full probability distribution of outcomes. A more careful and thorough treatment of evection is required to determine the exact final pathways of WD-WD mergers in quadruple and triple systems.

\subsection{Future directions and outlook}

Several problems in both dynamics and stellar astrophysics are left for future work. For the case of main sequence star mergers, we have only studied two fixed-mass configurations and only consider secular quadrupole + octupole order effects. An exploration of the stellar mass distribution and inclusion of tidal effects will give a more accurate prediction of the high-$e$ rate, while mass loss and mass transfer may be important when stars are close to each other and stellar merger rate is concerned. For the WD-WD merger case, in terms of dynamics, we need a more detailed treatment of non-secular effects, particularly evection, to explore the final stages of WD-WD mergers driven by quadruple dynamics. The role of higher order effects, such as hexadecapole-monopole \citep[see][for recent discussion in context of triple systems]{2017PhRvD..96b3017W} and quadrupole-quadrupole interactions, is also unclear. In terms of astrophysics, this paper did not consider effects such as mass loss \citep[which causes the dynamical characteristics of the system to change, see][]{2012ApJ...760...99P,2013ApJ...766...64S} and interactions and common envelope evolution (important if $a_1$ is small). A potential complication that involves both dynamics and astrophysics is that mass loss will cause the orbital periods to change and many quadruple systems will sweep over resonances between the orbital periods of the two inner binaries before producing white dwarfs. Finally, these results may differ when changing the initial distribution of orbital elements, \eg exploring the moderately hierarchical regime ($a/a_1\lesssim 10$) where secular codes such as ours are least applicable. Many of these same considerations are also relevant for stellar binaries.

There are several observational signatures that could test for a ``quadruple channel'' of Type Ia supernovae. Most are related to similar signatures for triples. For example, in historical SN Ia remnants, one can look for low-mass binaries inside, exhibiting blue-shifted absorption in their spectra or anomalous abundances. Similarly, the discovery of a pre-explosion main-sequence binary at the position of a nearby SN Ia would provide confirmation of the quadruple nature of the system \citep[see \eg][]{2011ApJ...741...82T,2009ApJ...707.1578K}. For newly-discovered SNe Ia, one can look for two soft X-ray flashes $\sim 10^5\,{\rm s}\,(a/10\,{\rm AU})$ after the explosion, with a time separation $\sim 10^4\,{\rm s}\,(a_2/{\rm AU})$, as the shock wave overtakes the binary companion \citep[\eg][]{2010ApJ...708.1025K,2011ApJ...741...82T}. Very small changes to early-time optical/UV light curve might also signal the existence of a companion binary.  

Due to the colour selection of WD-WD searches \citep[\eg][]{2001AN....322..411N,2009ApJ...707..971B,2010ApJ...723.1072B}, it remains difficult to make a census of WD binaries in triple or quadruple systems \citep[\eg][]{2014arXiv1402.7083K}. In the future, the population of Galactic WD-WD binaries driven to high eccentricities by their companion stars or binaries may be detectable in \textit{LISA} \citep{2011ApJ...741...82T,2011ApJ...729L..23G,2017arXiv170200786A}, or in microlensing searches \citep[\eg \textit{WFIRST},][]{2015arXiv150303757S}. For now, radial velocity surveys of stars to identify the signal of a massive, but unseen companion, may be a robust way to explore the population of old and massive WDs in triple and quadruple systems \citep[][]{2011ApJ...741...82T}, and thus alleviate a primary uncertainty in the determination of the role of few-body systems in driving WD-WD mergers.

\section*{Acknowledgments}
XF is supported by the Simons Foundation and NSF 1313252, and is grateful to Joseph McEwen, Paulo Montero-Camacho, Shirley Li, John Beacom, Annika Peter, Joe Antognini, Ben Wibking and Daniel Fabrycky for useful discussions and comments. TAT is supported in part by NSF 1313252 and thanks Ondrej Pejcha and Joe Antognini for conversations and collaborations. CMH is supported by the Simons Foundation, the US Department of Energy, the Packard Foundation, NASA, and the NSF. Some of the simulations in this paper made use of the \texttt{REBOUND} code which can be downloaded freely at \texttt{http://github.com/hannorein/rebound}. Many computations in this paper were run on the CCAPP condo of the Ruby Cluster at the Ohio Supercomputer Center \citep{OhioSupercomputerCenter1987}. We
are also grateful for suggestions from an anonymous referee which improved the paper.

%%%%%%%%%%%%%%%%%%%%%%%%%%%%%%%%%%%%%%%%%%%%%%%%%%



\bibliographystyle{mnras}
\bibliography{references} % if your bibtex file is called example.bib

%%%%%%%%%%%%%%%%%%%%%%%%%%%%%%%%%%%%%%%%%%%%%%%%%%

%%%%%%%%%%%%%%%%% APPENDICES %%%%%%%%%%%%%%%%%%%%%

\appendix

\section{Coefficients in Octupole Order Hamiltonian}
\label{App:Octu}
The coefficients with non-negative $m$ values in Eq.~(\ref{eq:oct_avg}) are listed as follows,
\begin{align}
\mathcal{A}_1^{(3)} =&\left[6+29e_1^2-(6+e_1^2)\cos 2i_1 + 7e_1^2\cos 2g_1(7+\cos 2i_1)\right]\nonumber\\
& \times 40 e_1\sin g_1\cos i_1~,\\
\mathcal{B}_1^{(3)} =&\left[6-13e_1^2-3(2+5e_1^2)\cos 2i_1 + 7e_1^2\cos 2g_1(5+3\cos 2i_1)\right]\nonumber\\
& \times 40 e_1\cos g_1~,
\end{align}
\begin{align}
\mathcal{A}_1^{(2)} =&\left[-2+9e_1^2-(6+e_1^2)\cos 2i_1 + 7e_1^2\cos 2g_1(3+\cos 2i_1)\right]\nonumber\\
& \times 20 e_1\sin g_1\sin i_1~,\\
\mathcal{B}_1^{(2)} =&\left(-2-5e_1^2 + 7e_1^2\cos 2g_1\right) 40 e_1\cos g_1\sin 2i_1~,
\end{align}
\begin{align}
\mathcal{A}_1^{(1)} =&\left[5(6+e_1^2)\cos 2i_1 + 7(-2+e_1^2+10e_1^2\cos 2g_1\sin^2 i_1)\right]\nonumber\\
& \times 2 e_1\sin g_1\cos i_1~,\\
\mathcal{B}_1^{(1)} =&\left[6-13e_1^2+5(2+5e_1^2)\cos 2i_1 + 70e_1^2\cos 2g_1\sin^2 i_1)\right]\nonumber\\
& \times 2 e_1\cos g_1~,\\
\mathcal{A}_1^{(0)} =&\left[18+31e_1^2+5(6+e_1^2)\cos 2i_1 + 70e_1^2\cos 2g_1\sin^2 i_1\right]\nonumber\\
& \times e_1\sin g_1\sin i_1~,
\end{align}
\begin{align}
\mathcal{A}^{(3)} =&\sin g\sin^2 i\cos i~,~~~\mathcal{B}^{(3)} =\cos g\sin^2 i~,\\
\mathcal{A}^{(2)} =&\sin g(\sin i - 3\sin 3i)~,~~~\mathcal{B}^{(2)} =-4 \cos g\sin 2i~,\\
\mathcal{A}^{(1)} =&\sin g(\cos i+15\cos 3i)~,~~~\mathcal{B}^{(1)} =2\cos g(3+5\cos 2i)~,
\end{align}
and
\begin{align}
\mathcal{A}^{(0)} =&4\sin g(\sin i+5\sin 3i)~.
\end{align}
For negative $m$'s, coefficients can be obtained using the relation (\ref{eq:coeff_relation}).

\section{Code Description and Tests}
\label{App:code}
Our secular code is written in {\tt C}. It uses the fourth-order Runge-Kutta (RK4) integrator with the adaptive time step size determined by
\begin{equation}
\Delta t = \epsilon \min_X \left|\frac{X}{\dot X}\right|,
\label{eq:epsdef}
\end{equation}
where $X$ can be $g_i$, $h_i$, $G_i+H_i$, $G_i-H_i$, $L_i-G_i$, $G_i$, or $L_i$ for any of the orbits ($i=1$, 2, or mutual).\footnote{We exclude the mutual $L$ since this does not evolve. We also exclude the mutual $G$: because the mutual orbit carries most of the angular momentum, its relative rate of change $|\dot G|/G$ is much smaller than the relative rates for the inner orbits.} For $X=g_i$ or $h_i$, we use $1$ in the numerator in place of $X$, since nothing special happens when an angle passes through zero. These reduces the step size when the system approaches any singularity of the equations of motion. Some of the singularities are real (\eg $G_i\rightarrow 0$, which corresponds to a plunging orbit) and others are merely coordinate artifacts (\eg $G_i - H_i\rightarrow 0$, which corresponds to a zero-inclination orbit). The default setting is $\epsilon=0.05$.

The code takes a list of \texttt{effect\_flag} specified by the user. The element of the flag list is either 1 or 0, turning on or off any of the secular effects in either one of the inner orbits discussed in \S\ref{sec:theory}, \ie the secular quadrupole and octupole order effects, the GR 1PN precession and 2.5PN dissipation, and the tidal precession and dissipation at the parabolic-orbit limit. The stopping criteria are also defined by the user and highly depend on the specific tasks.

In this appendix, we present some test results of the code, including the convergence tests of the integrator, the tests of total energy and angular momentum conservation, the tests of the Kozai constant for a triple system at test particle limit and quadrupole order.

\subsection{Convergence with step size}
Although many quadruple systems are very chaotic and the time we integrate over is generally much longer than the Lyapunov time, the ensemble properties, such as the high-$e$ fraction and the merger fraction, are expected to be well defined and not sensitive to the accuracy of the integrator.

\begin{figure}
\centering
\includegraphics[width=\columnwidth]{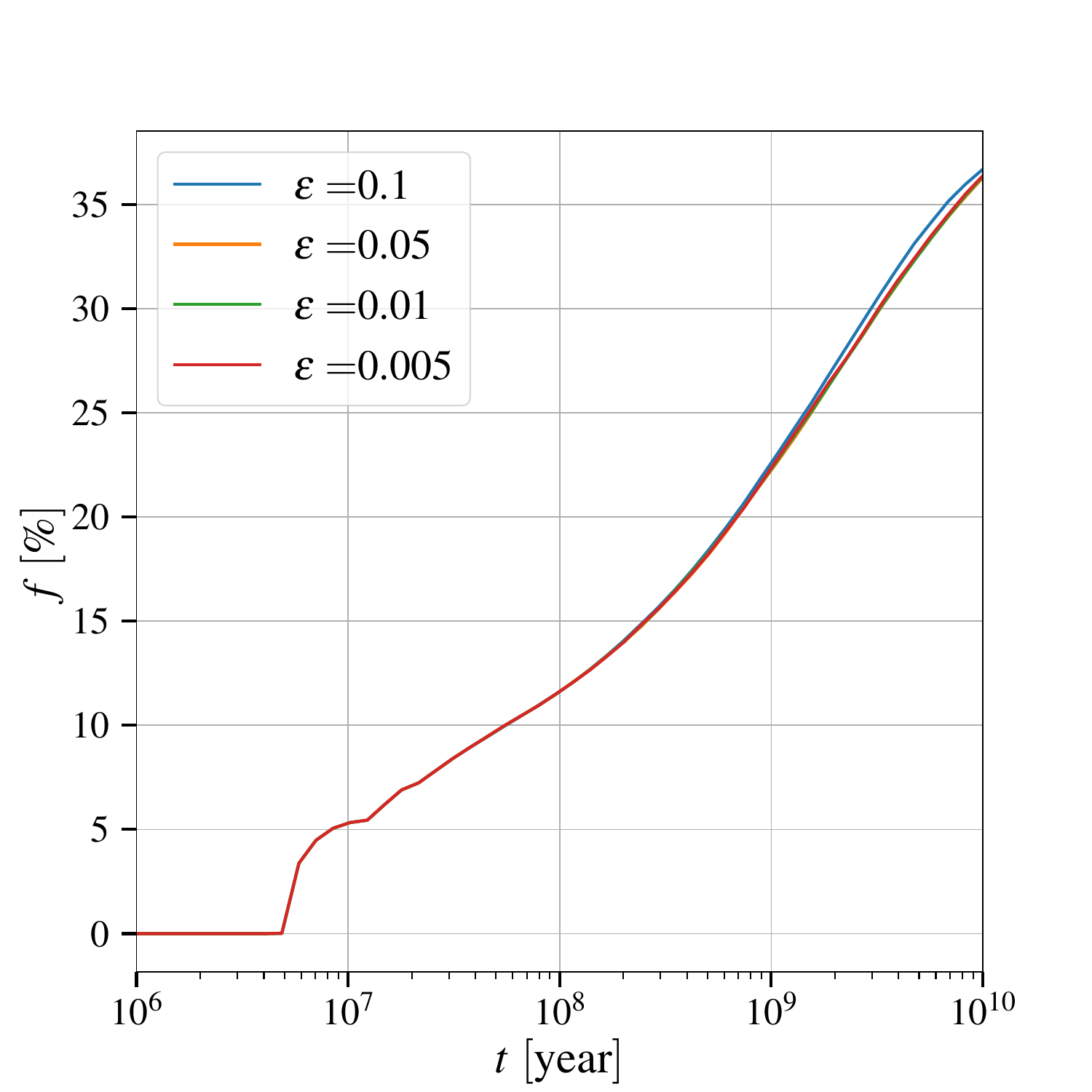}
\caption{Convergence test for $10^5$ randomly oriented ``4-star'' systems discussed in \S\ref{subsec:GrowingFrac}. Only quadrupole order effect is turned on. The stepsize prefactor $\epsilon$ of Eq.~(\ref{eq:epsdef}) is chosen to be 0.1, 0.05 (default setting), 0.01, 0.005, respectively, and the high-$e$ fractions converge very well.}
\label{fig:convergence_test_4star}
\end{figure}

\begin{figure}
\centering
\includegraphics[width=\columnwidth]{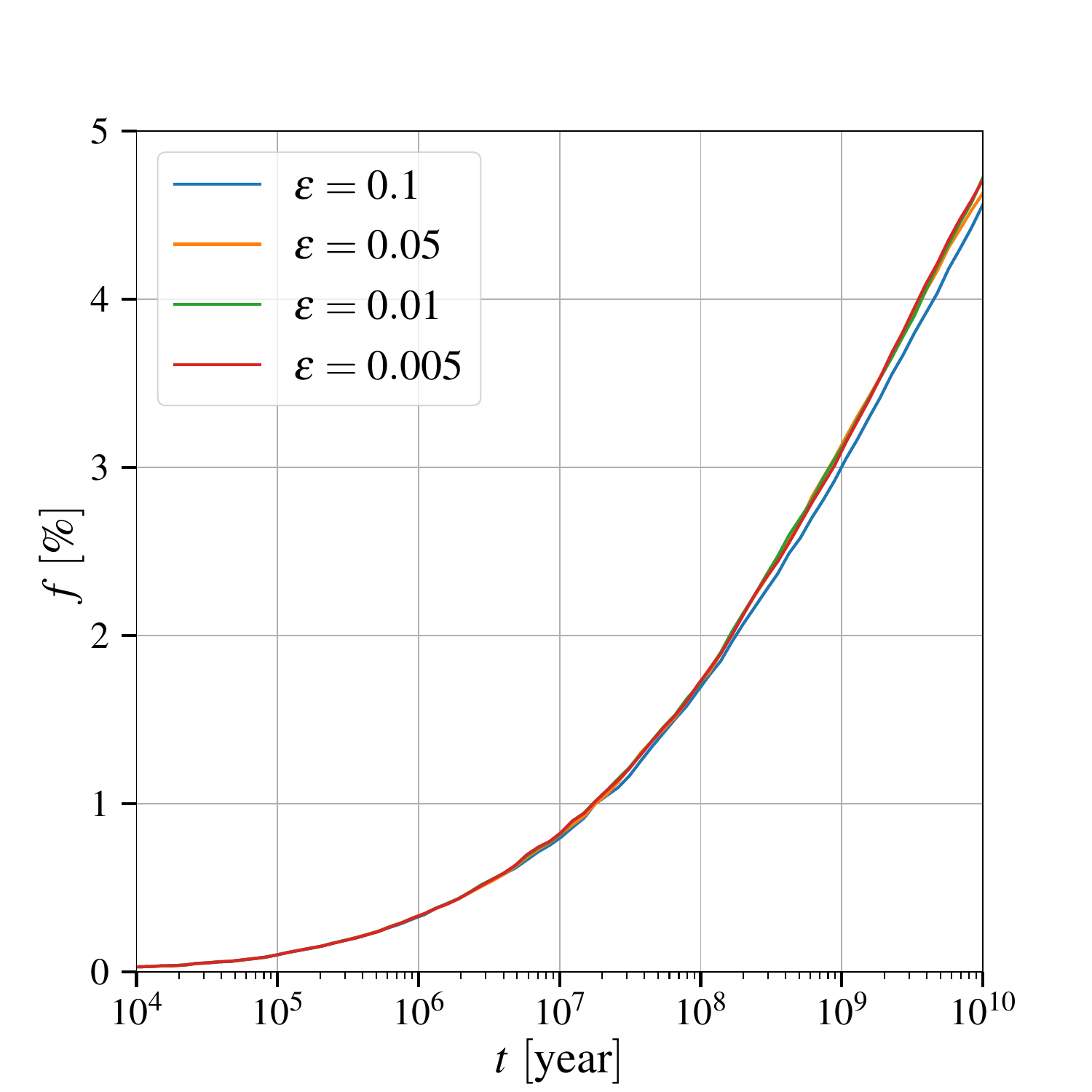}
\caption{Convergence test for $10^5$ randomly oriented ``[WD-WD]-[Star-Star]'' systems with equal-mass (0.7M$_\odot$) WDs, discussed in \S\ref{sec:SN}. Only quadrupole order effect is turned on. The stepsize prefactor $\epsilon$ of Eq.~(\ref{eq:epsdef}) is chosen to be 0.1, 0.05 (default setting), 0.01, 0.005, respectively, and the high-$e$ fractions converge very well.}
\label{fig:convergence_test_WD}
\end{figure}

Here we perform the convergence test by changing the adaptive time step size to larger and smaller values, and test for the ``4-star'' case in \S\ref{subsec:GrowingFrac} and the equal-mass ``[WD-WD]-[Star-Star]'' case in \S\ref{sec:SN}, respectively. The results are shown in Figure \ref{fig:convergence_test_4star} and Figure \ref{fig:convergence_test_WD}.


\subsection{Conservation of energy and angular momentum}
\label{app:code_erg_ang}

\begin{figure}
\centering
\includegraphics[width=\columnwidth]{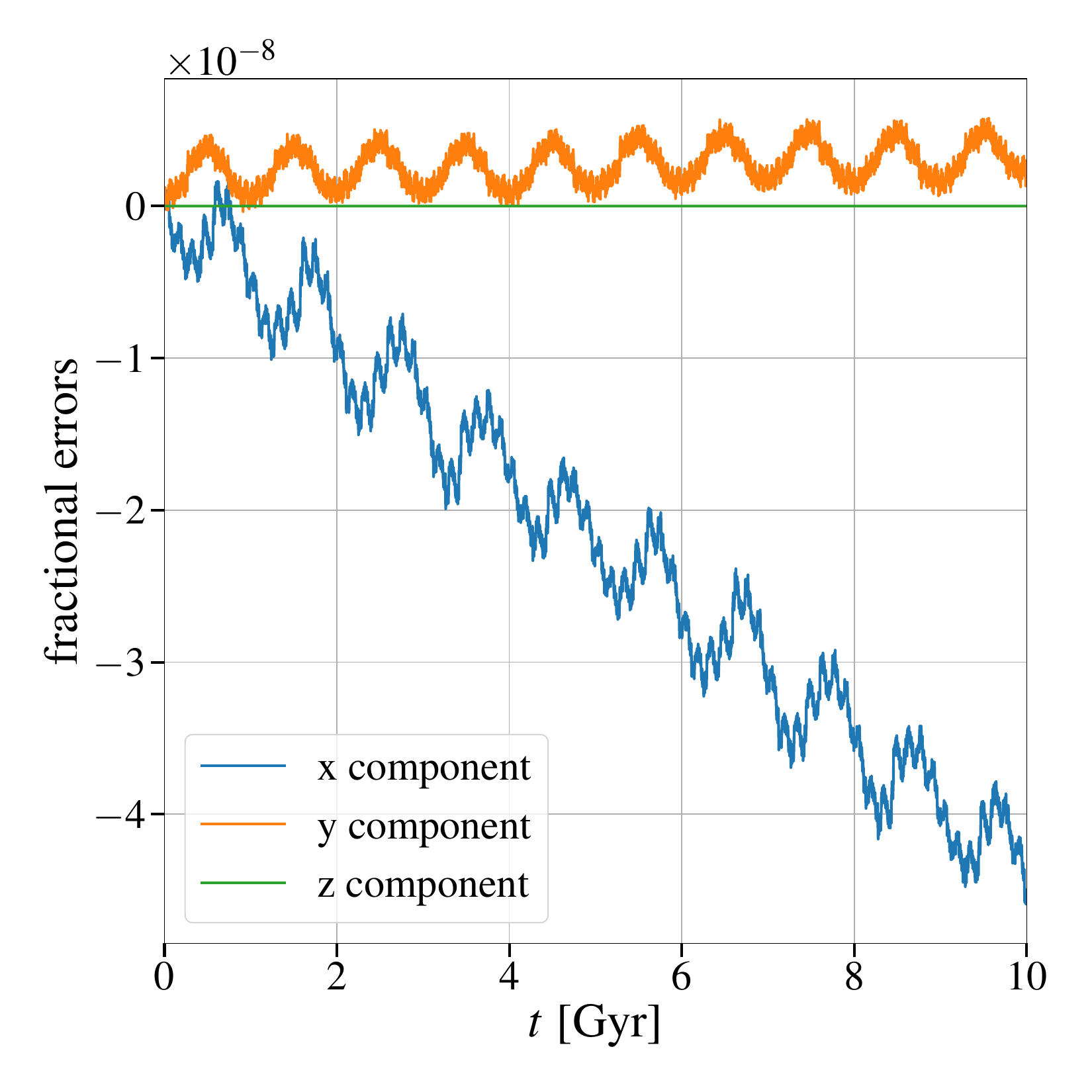}
\caption{Angular momentum conservation test for a ``[WD-WD]-[Star-Star]'' system with its initial orbital parameters listed in Table \ref{Tab:WD_test_erg_ang}. The plot shows the deviations of 3 components of the total angular momentum from their initial values, \ie $(G_{{\rm tot},(x,y,z)}-G_{{\rm tot},(x,y,z)}^{\rm initial})/G_{\rm tot}$, up to 10Gyr. Note that the RK4 integrator exactly conserves the total $z$-angular momentum, but not the $x$ or $y$ components.}
\label{fig:quad_angmom_test}
\end{figure}

\begin{figure}
\centering
\includegraphics[width=\columnwidth]{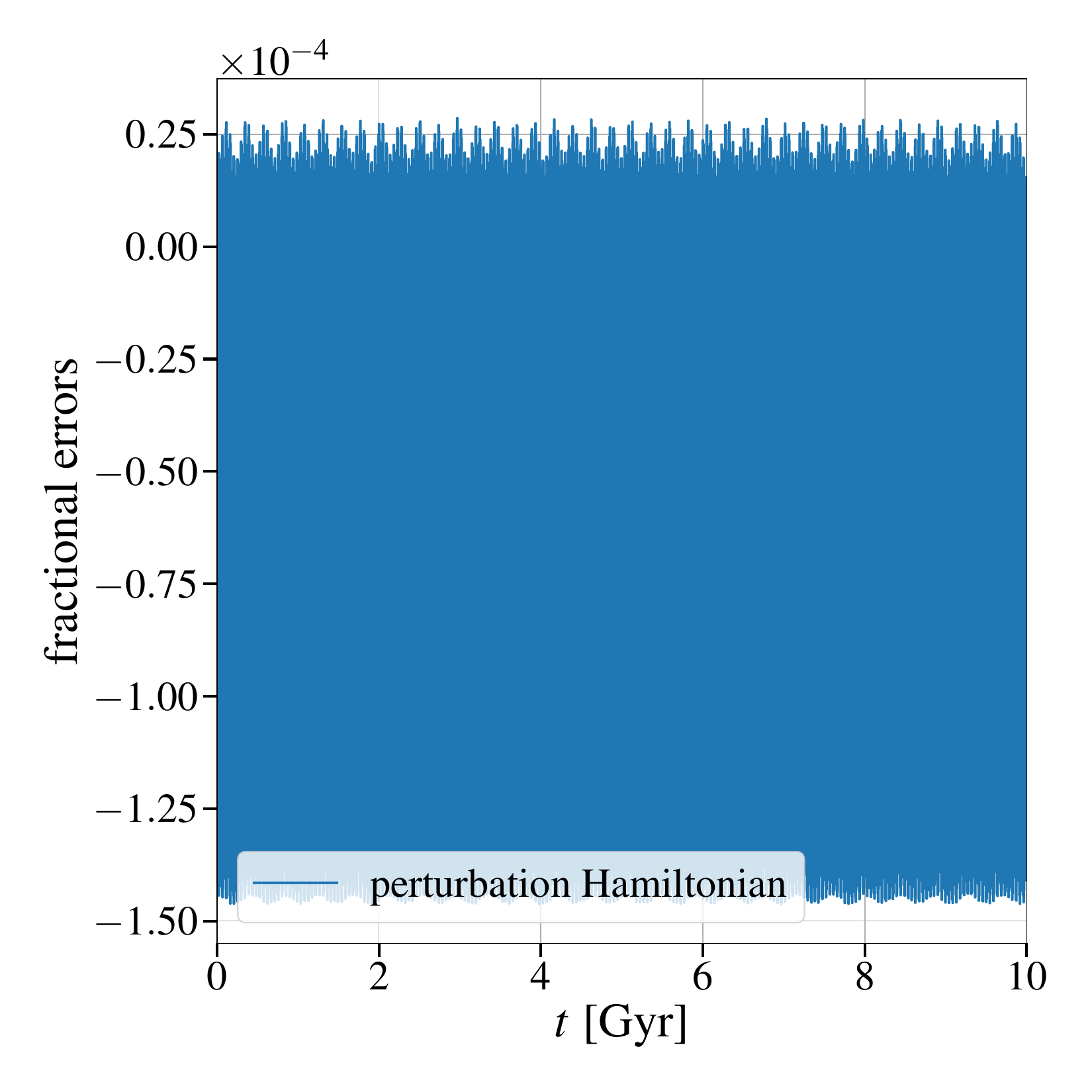}
\caption{Energy conservation test for a ``[WD-WD]-[Star-Star]'' system with its initial orbital parameters listed in Table \ref{Tab:WD_test_erg_ang}. A conservative system preserves its Hamiltonian, and this plot shows the fractional deviation of the perturbation Hamiltonian (only including quadrupole and octupole order terms and the GR and tidal precession terms, but {\em excluding} the Kepler parts) from its initial value. The errors are small and remain bounded up to 10Gyr.}
\label{fig:quad_erg_test}
\end{figure}

The RK4 integrator is not symplectic, phase-space conserving, or time-reversible. Thus one concern is that it may not preserve the energy and angular momentum, especially when integrating over a long time. However, for secular equations, the situation can be much better. We test the conserved quantities of the ``[WD-WD]-[Star-Star]'' case (with masses 0.8+0.6+1+1 M$_\odot$), where all the secular non-dissipative effects in \S\ref{sec:theory} are taken into account. The initial orbital elements are listed in Table \ref{Tab:WD_test_erg_ang}. The results are shown in Figures \ref{fig:quad_angmom_test} and \ref{fig:quad_erg_test} using the default setting $\epsilon=0.05$, which are quite good: over the lifetime of the Universe, errors are of order $10^{-8}$ for the angular momentum and $10^{-4}$ for the {\em perturbation} to the energy.

\begin{table}
\centering
\begin{tabular}{| c | c | c | c |}
 \hline
 \rule{0pt}{2.5ex} Elements & Inner Orbit A & Inner Orbit B & Mutual Orbit\\
 \hline
 $m$ & 0.8+0.6M$_\odot$ & 1+1M$_\odot$ & -- \\
 $e$ & 0.6 & 0.2 & 0.1 \\
 $a$ & 10AU & 15AU & 1000AU \\
 $i$ & 20$^\circ$ & 10$^\circ$ & 0.1$^\circ$ \\
 $g$ & 0 & 0 & 0  \\
 $h$ & 0 & 0 & 0 \\
 \hline
\end{tabular}
\caption{The initial orbital elements of the example ``[WD-WD]-[Star-Star]'' system for energy and angular momentum conservation tests in Appendix \ref{app:code_erg_ang}.}
\label{Tab:WD_test_erg_ang}
\end{table}

\subsection{Kozai constant}
\label{app:code_kozai}
\begin{figure}
\centering
\includegraphics[width=\columnwidth]{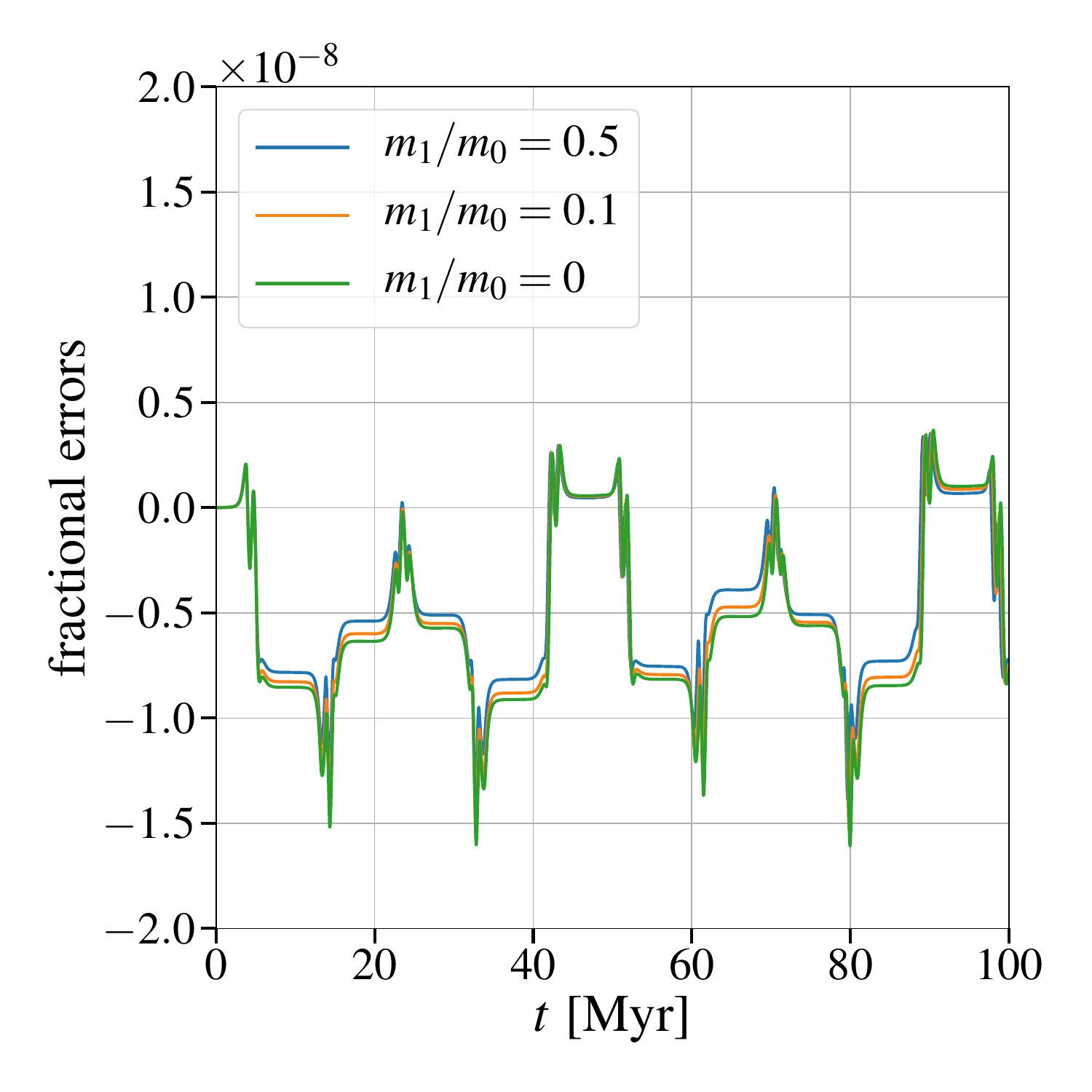}
\caption{The fractional deviations of Kozai constant at quadrupole order for triples with initial orbital parameters listed in Table \ref{Tab:triple_test_kozai} and discussed in Appendix \ref{app:code_kozai}. The constant is preserved at $10^{-8}$ level for all inner binary mass ratios $m_1/m_0$.}
\label{fig:Kozai_test}
\end{figure}

A triple system at TPL and quadrupole order exhibits the standard LK oscillation, with the Kozai constant, \ie $\sqrt{1-e_{\rm in}^2}\cos i={\rm const}$. For non-TPL cases at quadrupole order, the Kozai constant is generalized as (see \eg \citealt{2013MNRAS.431.2155N}, Eq. 23\footnote{A similar conservation law arises as a result of \citet{2013MNRAS.431.2155N}, Eq.~(23). That conservation law is slightly different because their inclination is measured relative to the fixed total angular momentum, not the outer orbit.}; \citealt{2012arXiv1211.4584K}, Eq. 14)
\begin{equation}
K = \sqrt{1-e_{\rm in}^2}\,\cos i + \frac{L_{\rm in}}{2G_{\rm out}}(1-e_{\rm in}^2)
\label{eq:K-const}
\end{equation}
using the fact that angular momenta $G_{\rm out}$, $G_{\rm tot}$ and Delaunay's variable $L_{\rm in}$ are conserved, and applying the law of cosines to the vector sum $\bm{G}_{\rm tot}=\bm{G}_{\rm in}+\bm{G}_{\rm out}$. We test this property with our code for triple systems with different mass ratios $m_1/m_0$ but with the same $m_0+m_1=1$M$_\odot$, $m_2=2$M$_\odot$, and the same initial orbital elements listed in Table \ref{Tab:triple_test_kozai}. We only turn on their quadrupole order effects. The result is shown in Figure \ref{fig:Kozai_test} using the default setting $\epsilon=0.05$. The Kozai constant is well-conserved in all cases.

\begin{table}
\centering
\begin{tabular}{| c | c | c |}
 \hline
 \rule{0pt}{2.5ex} Elements & Inner Orbit& Outer Orbit\\
 \hline
 $e$ & 0.1 & 0.3 \\
 $a$ & 10AU  & 1000AU \\
 $i$ & 50$^\circ$ & 10$^\circ$ \\
 $g$ & 0 & 0  \\
 $h$ & 0 & 180$^\circ$ \\
 \hline
\end{tabular}
\caption{The initial orbital elements of the triple systems for the Kozai constant test in Appendix \ref{app:code_kozai}.}
\label{Tab:triple_test_kozai}
\end{table}

\subsection{Comparison with few-body calculations}
\label{app:code_rebound}
\begin{figure*}
\centering
\includegraphics[width=\textwidth]{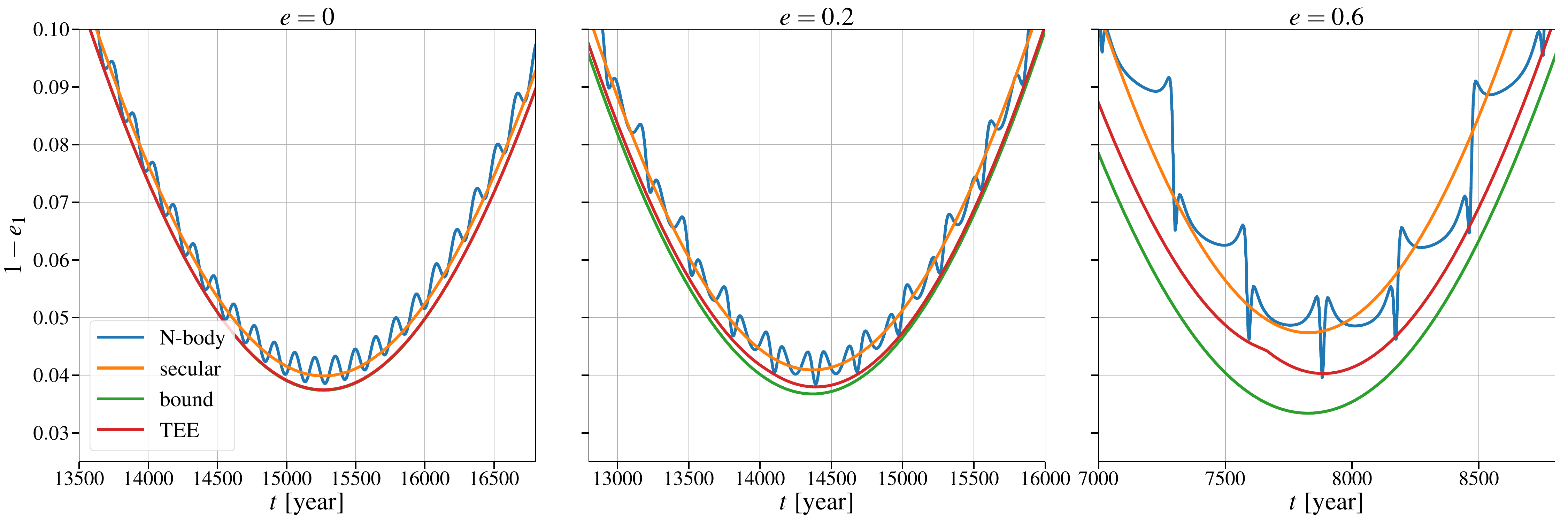}
\caption{The eccentricities of the inner orbits at their local maximum. The three triple systems are described in Appendix \ref{app:code_rebound} and Table \ref{Tab:triple_test_rebound} and their initial parameters only differ in the outer eccentricities ($e=0$(\textit{left}), 0.2 (\textit{middle}), and 0.6 (\textit{right})). The blue lines are results from \texttt{REBOUND} few-body simulations, while the orange lines are produced by our secular code. The green and red lines are the estimated ``upper bound'' and the ``true evection envelope'', calculated based on the secular results. The details of evection calculations are discussed in \S\ref{sec:evection} and Appendix \ref{App:Evec}.}
\label{fig:rebound_triple151}
\end{figure*}

We compare our secular results (including quadrupole and octupole order effects) of triple systems with the results from the fully dynamical few-body code \texttt{REBOUND} \citep[][]{2012A&A...537A.128R,2015MNRAS.446.1424R}. We consider inner binary masses 1+0.5M$_\odot$ and tertiary mass 1M$_\odot$, and a hierarchy ratio $a_{\rm out}/a_{\rm in} = 20$; the initial orbital elements are listed in Table \ref{Tab:triple_test_rebound}. We allow the initial outer orbit eccentricity to change and fix other parameters. In Figure \ref{fig:rebound_triple151} we show that our secular results match with the few-body results, and our ``upper bound'' (bound) and ``true evection envelope'' (TEE) results estimate the amplitude of the eccentricity oscillations very well (see Appendix \ref{App:Evec} for details). TEE results work better than the ``bound'' especially when the outer eccentricities are large. The test is done using the default setting $\epsilon=0.05$.

\begin{table}
\centering
\begin{tabular}{| c | c | c |}
 \hline
 \rule{0pt}{2.5ex} Elements & Inner Orbit& Outer Orbit\\
 \hline
 $m$ & 1+0.5M$_\odot$ & 1M$_\odot$ \\
 $e$ & 0.1 & 0, 0.2, 0.6 \\
 $a$ & 3AU  & 60AU \\
 $i$ & 80$^\circ$ & 0 \\
 $g$ & 0 & 0  \\
 $h$ & 0 & 180$^\circ$ \\
 \hline
\end{tabular}
\caption{The initial orbital elements of the triple systems for comparison with the \texttt{REBOUND} few-body simulation, discussed in Appendix \ref{app:code_rebound}. We test for different outer orbit eccentricities.}
\label{Tab:triple_test_rebound}
\end{table}

\subsection{Coordinate independence test}
\label{app:CItest}

\begin{figure}
\centering
\includegraphics[width=\columnwidth]{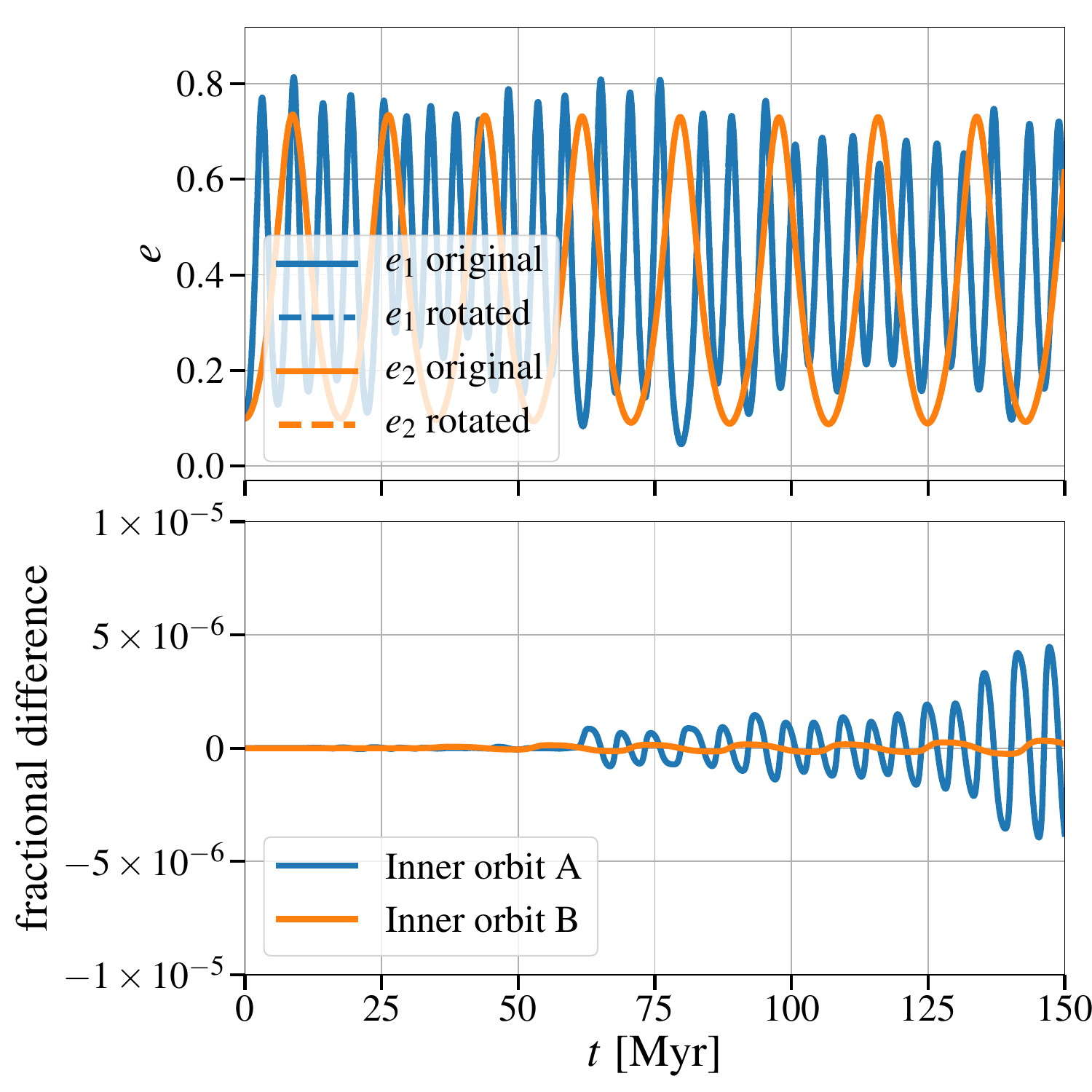}
\caption{The eccentricity evolution of the [Star-Planet]-[Star-Star] quadruple system in Table \ref{Tab:coordinate_rotation}. In the upper panel, solid lines and dashed lines are calculated from the original and the rotated coordinate system, respectively. The lower panel shows the fractional difference between the results from the two coordinate systems. The Star-Planet binary starts to deviate earlier than the Star-Star binary because it is more chaotic, as we have discussed in \S\ref{subsubsec:planet}, although the deviations are very tiny at least in the first 150Myr.}
\label{fig:coordinate_test}
\end{figure}

Finally, we test the behaviour of our code under the rotation of the coordinate system. Here we test the [Star-Planet]-[Star-Star] quadruple system in Table \ref{Tab:coordinate_rotation}. This system differs from the one in \S\ref{subsubsec:planet} in the masses of the stellar binary. We make them unequal to fully test both octupole order terms. We rotate the coordinate system such that $\hat{\bm{x}}'=\hat{\bm{y}}$, $\hat{\bm{y}}'=\hat{\bm{z}}$, $\hat{\bm{z}}'=\hat{\bm{x}}$. The comparison of the eccentricity evolutions between the original and the basis-rotated is shown in Figure \ref{fig:coordinate_test}, and it shows the two systems evolve in the same way at least in the first 150 Myr with high accuracy. The long term chaotic motion will show up later but has no significance in terms of the ensemble properties we care about in this paper. The test is done using the default setting $\epsilon=0.05$.

\begin{table}
\centering
\begin{tabular}{| c | c | c | c |}
 \hline
 \rule{0pt}{2.5ex} Elements & Inner Orbit A & Inner Orbit B & Mutual Orbit\\
 \hline
 $m$ & 1+0.001M$_\odot$ & 1+2M$_\odot$ & -- \\
 $e$ & 0.3 & 0.1 & 0.1 \\
 $a$ & 10AU & 15AU & 1000AU \\
 $i$ & 50$^\circ$ & 50$^\circ$ & 10$^\circ$ \\
 $g$ & 0 & 0 & 0  \\
 $h$ & 0 & 0 & 180$^\circ$ \\
 \hline
\end{tabular}
\caption{The initial orbital elements of the example ``[Star-Planet]-[Star-Star]'' system for the coordinate-independence test in Appendix \ref{app:CItest}.}
\label{Tab:coordinate_rotation}
\end{table}

\section{``Precession oscillation'' phase}
\label{app:rapidwiggle}
\begin{figure*}
\centering
\includegraphics[width=\textwidth]{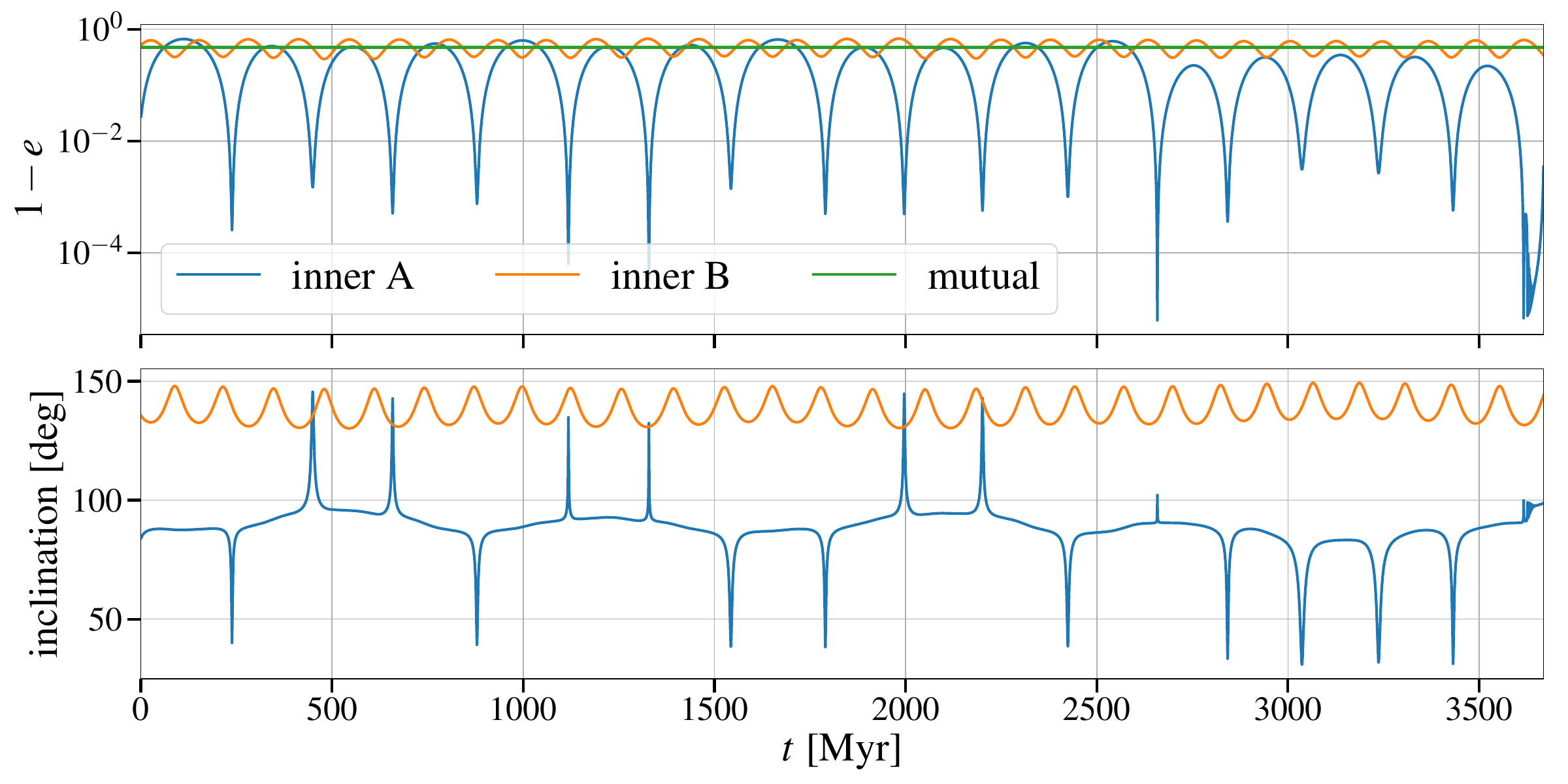}
\caption{An example quadruple system with a WD binary ($0.7+0.7$M$_\odot$) and a stellar binary ($1+1$M$_\odot$). We include the Newtonian secular effects up to octupole order, and the 1PN and tidal precession for both inner orbits, as well as the 2.5PN and tidal dissipation for the WD binary. The upper panel shows the eccentricities of both inner and mutual orbits evolve with time and the lower panel shows the inclinations between two inner orbits and the mutual orbit. At the end the eccentricity $e_1$ shows the ``precession oscillation'' phase while circularizing.}
\label{fig:equalSN_Quad_NoEvec_part0_sys9469_e_inc}
\end{figure*}

\begin{figure*}
\centering
\includegraphics[width=\textwidth]{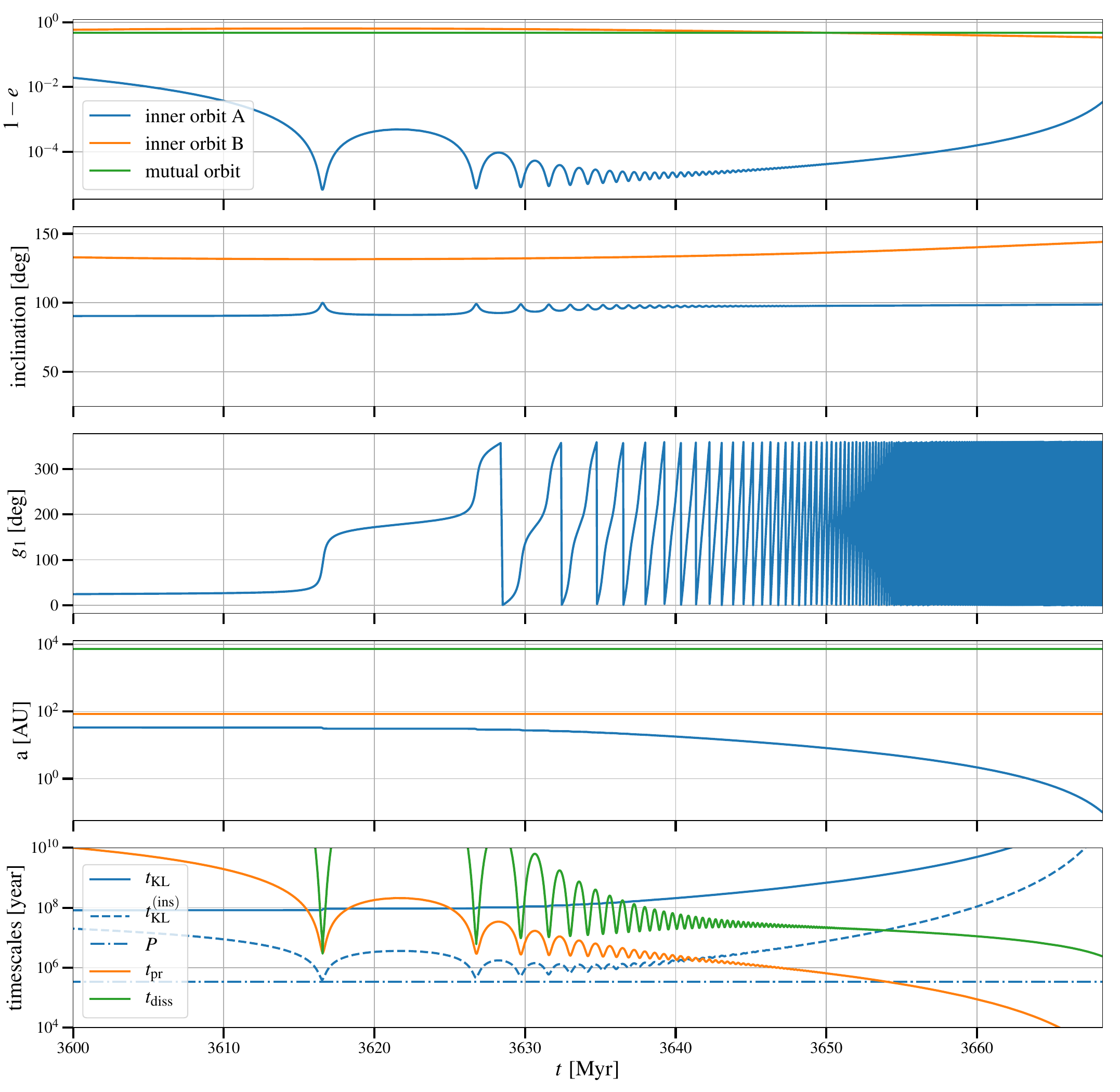}
\caption{Zoom-in of the final stage of the system shown in Figure \ref{fig:equalSN_Quad_NoEvec_part0_sys9469_e_inc}. Here we show the evolution of eccentricities, inclinations, argument of periastron of WD binary ($g_1$), semi-major axis of WD binary ($a_1$), and the timescales relevant to the WD binary, respectively.}
\label{fig:equalSN_Quad_NoEvec_part0_sys9469_5plots_zoom}
\end{figure*}

It is intriguing to investigate the origin of the ``precession oscillation'' phase. Figure \ref{fig:equalSN_Quad_NoEvec_part0_sys9469_e_inc} shows the evolution of a quadruple system with a pair of WDs ($0.7+0.7$ M$_\odot$) and a pair of solar-mass stars. At $t\sim 3620$ Myr, the WD binary experiences a rapid orbital decay, as shown in the zoomed Figure \ref{fig:equalSN_Quad_NoEvec_part0_sys9469_5plots_zoom}. During this time, the eccentricity and inclination under oscillations (the ``precession oscillation'' phase) on the timescale of the GR precession, rather than the much longer LK timescale.

We can understand the precession oscillations using a simple dynamical model, including the 1PN precession and the perturbing quadrupole of the outer orbit. We further take the inclination to be nearly constant (\ie $\sim 90^\circ$, $H_1\sim 0$) so that we can construct a one degree-of-freedom model. In this case, we can simplify the Hamiltonian of the WD binary in this phase as
\begin{align}
\overline{\mathcal{H}}^{(\rm simp)} &= \overline{\mathcal{H}}_1^{(\rm quad)} + \overline{\mathcal{H}}_1^{(\rm 1PN)} \nonumber\\
&= -\frac{15B_1L_1^4}{G^3L^3}e_1^2\cos 2g_1 \sin^2 i_1 -\frac{3\mathcal{G}^2\mu_1 m_A^2 L_1}{c^2a_1^2 G_1} \nonumber\\
&\equiv -K_{\rm quad}\cos 2g_1 - \frac{K_{\rm prec}}{G_1}~,
\end{align}
where we have dropped the constant terms in the first equality because they have no effect on the motion, and defined the parameters
\begin{equation}
K_{\rm quad}\equiv \frac{15B_1L_1^4}{G^3L^3}e_1^2\sin^2 i_1~~{\rm and}~~K_{\rm prec}\equiv \frac{3\mathcal{G}^2\mu_1 m_A^2 L_1}{c^2a_1^2}~.
\end{equation}
Now we do a canonical transformation to the variables $Q\equiv \sqrt{2G_1}\cos g_1$ and $P\equiv-\sqrt{2G_1}\sin g_1$, and rewrite the Hamiltonian in terms of them as
\begin{equation}
\overline{\mathcal{H}}^{(\rm simp)} = \frac{2K_{\rm quad} P^2 - 2K_{\rm prec}}{Q^2+P^2}~,
\end{equation}
where a constant term ``$-K_{\rm quad}$'' is dropped.

In the $(Q,P)$-plane, $\overline{\mathcal{H}}^{(\rm simp)}$ has a separatrix at $\overline{\mathcal{H}}^{(\rm simp)}=0$, \ie
\begin{equation}
P_{\rm separatrix}=\pm\sqrt{ \frac{K_{\rm prec}}{K_{\rm quad}} } = \pm\sqrt{\frac{\mathcal{G}^4\mu_1^5 m_A^4 G^3L^3}{5B_1c^2 e_1^2\sin^2 i_1L_1^7} }~,
\end{equation}
where we have used $L_1=\mu_1\sqrt{\mathcal{G}m_A a_1}$. If $\vert P\vert>\sqrt{K_{\rm prec}/K_{\rm quad}}$, the phase diagram trajectory is hyperbolic, while if $\vert P\vert<\sqrt{K_{\rm prec}/K_{\rm quad}}$, the trajectory is elliptic and trapped. When the system is close enough to the separatrix, a small energy dissipation (\ie decrease in $L_1$) can lead to an outward expansion of the separatrix and may cause the system to jump from the hyperbolic trajectory to the trapped trajectory.

\begin{figure}
\centering
\includegraphics[width=\columnwidth]{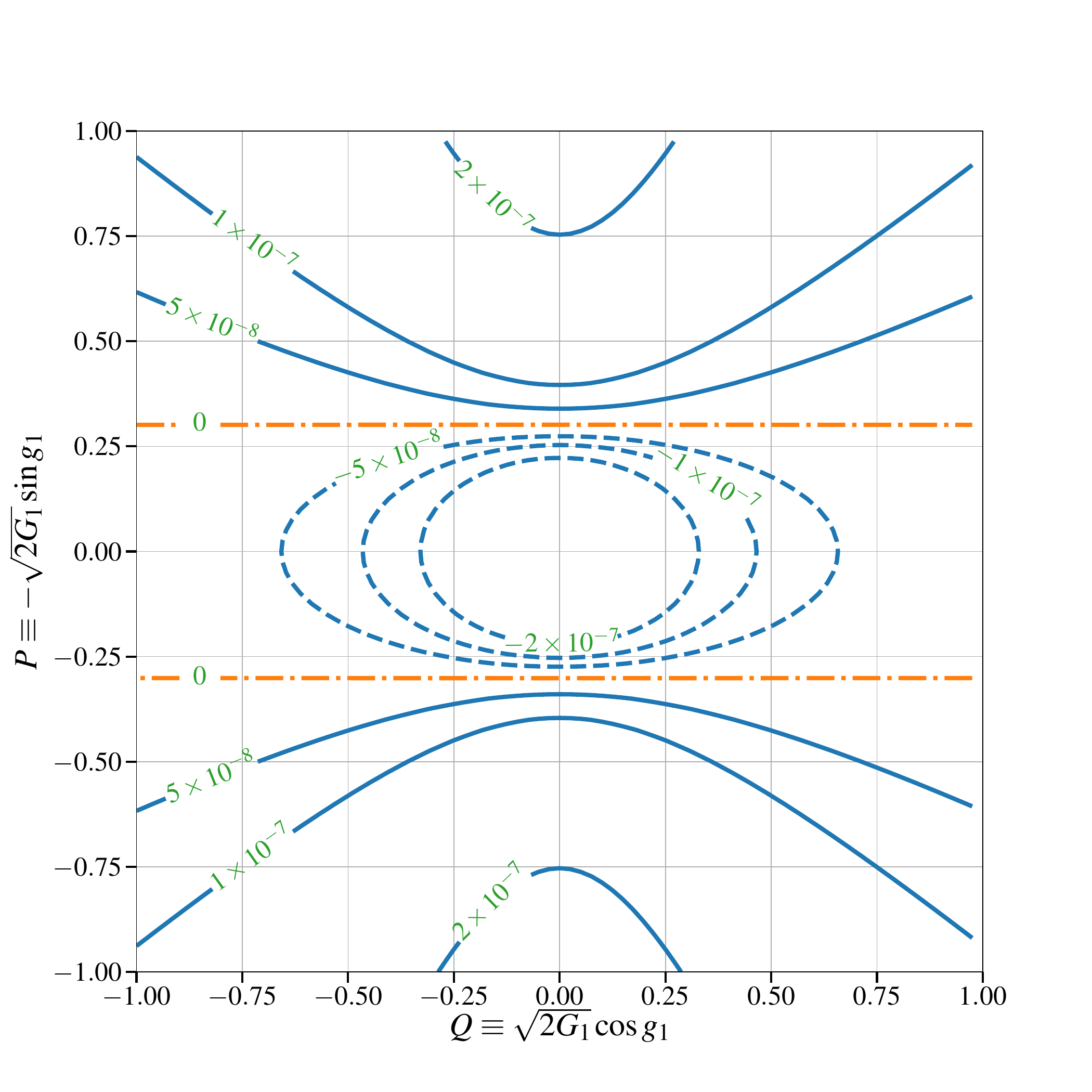}
\caption{The equal-$\overline{\mathcal{H}}^{(\rm simp)}$ contours of the example system in Figure \ref{fig:equalSN_Quad_NoEvec_part0_sys9469_e_inc} as its eccentricity $e_1$ approaches the maximal eccentricity. The orange dot-dashed lines are the separatrix between the hyperbolic trajectories and the trapped elliptic trajectories.}
\label{fig:equalSN_Quad_NoEvec_part0_sys9469_G_vs_g_contour}
\end{figure}

\begin{figure}
\centering
\includegraphics[width=\columnwidth]{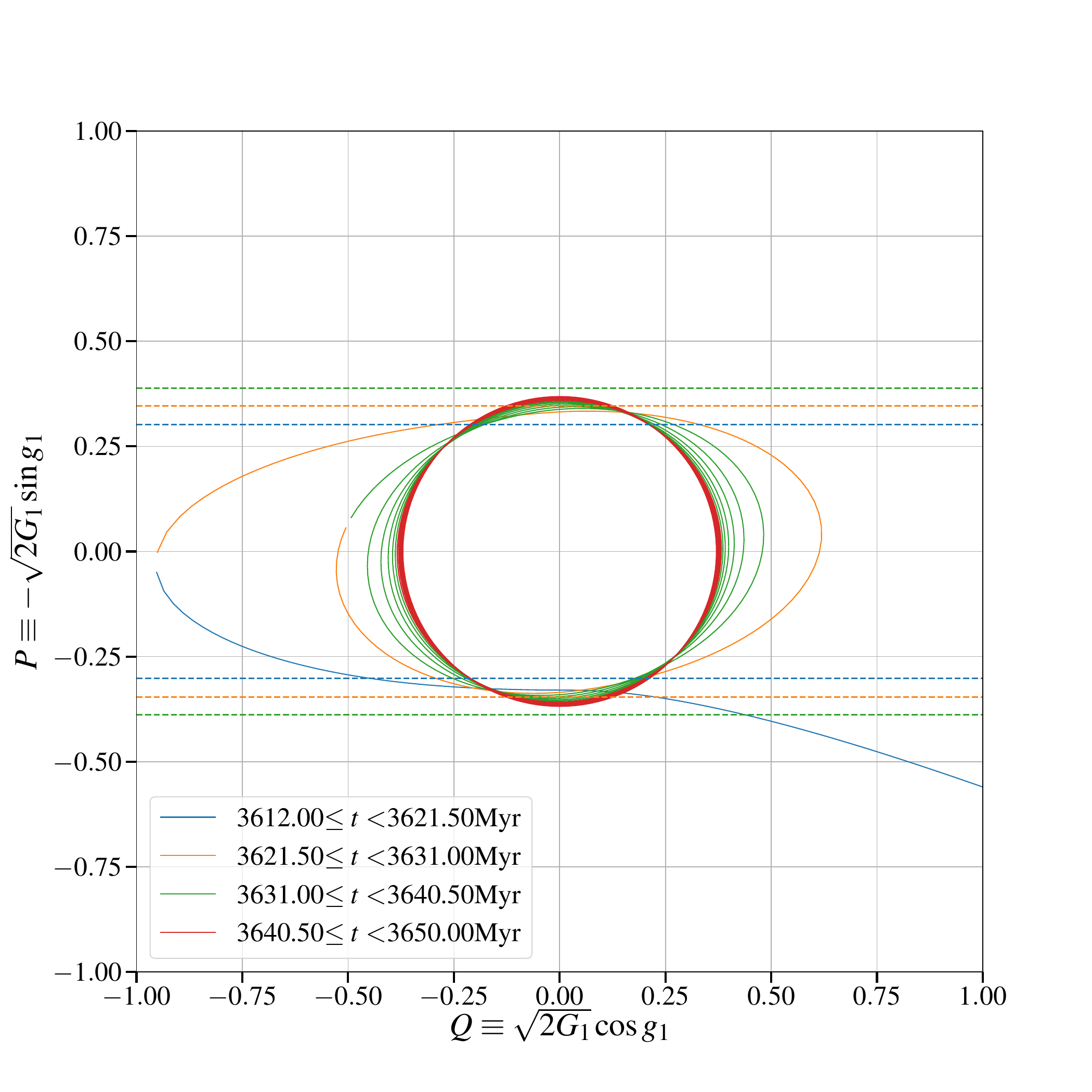}
\caption{The phase diagram of the example system in Figure \ref{fig:equalSN_Quad_NoEvec_part0_sys9469_e_inc} between $t=3612$Myr and $t=3650$Myr. Before and after the first eccentricity peak, the separatrix lines move from the blue dashed lines to the orange dashed lines, and the second eccentricity peak moves the separatrix lines further out to the green dashed lines. Meanwhile, the trajectory transits from the hyperbolic to the trapped elliptic.}
\label{fig:equalSN_Quad_NoEvec_part0_sys9469_G_vs_g_zoom}
\end{figure}

Figure \ref{fig:equalSN_Quad_NoEvec_part0_sys9469_G_vs_g_contour} shows the equal-$\overline{\mathcal{H}}^{(\rm simp)}$ contours of the example system as its eccentricity $e_1$ is approaching the maximal eccentricity, where the separatrix lines are the orange dot-dashed lines. Figure \ref{fig:equalSN_Quad_NoEvec_part0_sys9469_G_vs_g_zoom} shows the evolution of the system between $t=3612$ Myr and $t=3650$ Myr. Passing through the first and second eccentricity peaks, the separatrix lines move from the blue dashed lines to the orange dashed lines and further out to the green dashed lines, and keep moving outwards due to the energy dissipation. When $L_1$ (hence $a_1$) is small enough, the separatrix moves to $|P_{\rm separatrix}|\gg \sqrt{G_1}$, and the trajectory becomes a circle, \ie $G_1={\rm constant}$.

\begin{figure}
\centering
\includegraphics[width=\columnwidth]{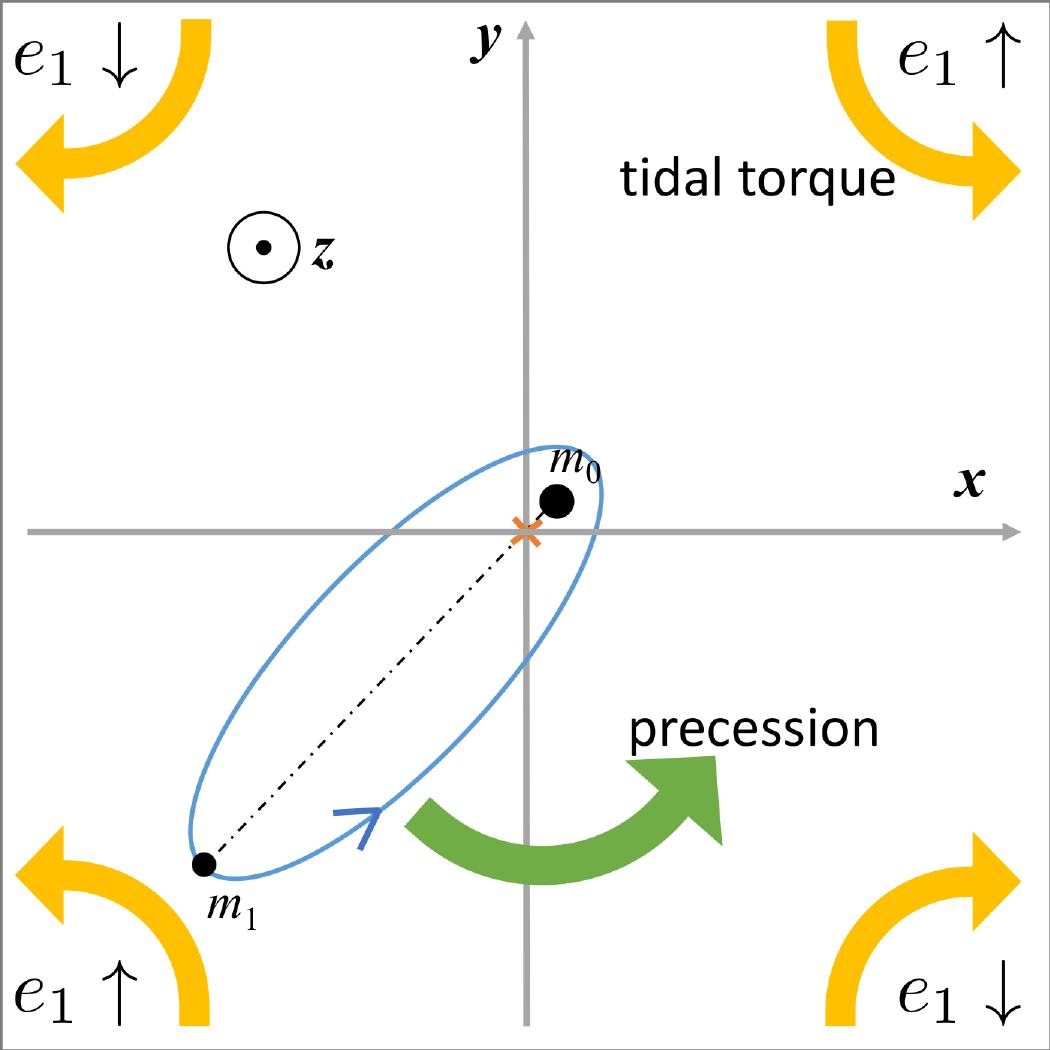}
\caption{A sketch of the system during its ``precession oscillation'' phase. The inner orbit (in \textit{blue}) is in the $x-y$ plane, with its angular momentum along the $+z$ direction. The mutual orbit is in the $x-z$ orbit and exerts an average tidal torque on the inner orbit, as shown by the yellow arrows. The torque has different directions in each quadrant of the $x-y$ plane, hence increasing or decreasing $e_1$ depending on which quadrant the apastron of the inner orbit is in. As the inner orbit swings around the $z$-axis for one cycle, the eccentricity goes up and down twice.}
\label{fig:wiggle}
\end{figure}

The wiggles in the eccentricity evolution come from the interaction of GR precession and the quadrupole-order perturbation from the other binary. For simplicity, consider the case where the inclination between the WD binary and the mutual orbit is close to 90$^\circ$, as shown in Figure \ref{fig:wiggle}. The mutual orbit is in the $x-z$ plane and the WD binary is in the $x-y$ plane with its angular momentum along the $+z$ direction. In the $x-y$ plane, the average tidal field from the companion is sketched by the curved arrows. Due to the rapid GR precession, the inner orbit swings counterclockwise in the $x-y$ plane, and the tidal field increases its eccentricity when its apastron is in the $xy>0$ quadrants and decreases otherwise. Thus, the wiggles are on the timescale of half-period of the 1PN precession.

\section{Short-term eccentricity changes due to evection}
\label{App:Evec}

In this appendix, we consider how evection can change the maximum eccentricity of an inner binary. We work in the limit where the inner orbit is at high eccentricity, \ie $e_1\rightarrow 1$, but unlike some past works we allow general outer eccentricity $e$.

It is convenient here to work with two right-handed frames. We define an unprimed frame $(X,Y,Z)$ aligned with the outer orbit: the outer orbit is on the $XY$-plane, $\hat{\bm{X}}$ is pointing from centre of mass of the inner orbit $m_A$ to the periastron of the outer orbit, and $\hat{\bm{Z}}$ is along the direction of the angular momentum of the outer orbit. At any time, we define the vector from $m_A$ to the tertiary $m_B$ as $\bm{D}$, and its true anomaly as $f$. We also define a primed coordinate frame $(X',Y',Z')$ aligned with the mean inner orbit: $\hat{\bm{X}}'$ points to the apastron\footnote{We recognize that this is an unusual convention, since anomalies are measured from periastron. However, in the $e_1\rightarrow 1$ limit, the separation vector is almost always in the direction of apastron, so this choice seems more natural.} of the inner orbit from the centre of mass $m_A$, and $\hat{\bm{Z}}'$ along the direction of the angular momentum of the inner orbit. The tertiary (or the centre of mass of the companion binary), with mass $m_B$, is orbiting around the inner system on a much larger orbit. The inclination angle between the outer and the inner orbit is $i$. We define the argument of apastron of the inner orbit as $\alpha_1$ and the argument of periastron of the outer orbit (relative to the line of nodes) as $g$; due to our choice to put $\hat{\bm{X}}$ in the direction of periastron of the outer orbit, $-g$ is the longitude of the ascending node of the inner orbit in the $(X,Y,Z)$ frame.

First we average over the inner orbit. The torque exerted on the inner orbit by the tertiary at $\bm{D}$ is given by
\begin{align}
\bm{\tau} &= \frac{\mathcal{G}\mu_1m_B}{D^3}\left\langle r_1^2\right\rangle 3\sin\theta\cos\theta\, \widehat{\hat{\bm{X}}'\times\hat{\bm{D}}}\nonumber\\
&= \frac{15}{2}\mathcal{G}\mu_1m_B\frac{a_1^2}{D^3}\sin\theta\cos\theta\,\widehat{\hat{\bm{X}}'\times\hat{\bm{D}}}~,
\end{align}
where $\mu_1\equiv m_0m_1/m_A$ is the reduced mass of inner binary. $\theta$ is the angle between $\bm{D}$ and $\hat{\bm{X}}'$ and $r_1$ is the separation between the inner binary stars. The rate of the angular momentum of the inner orbit, written in the Delaunay variables, is thus
\begin{equation}
\dot{\bm{G}}_1 = \bm{\tau} = \frac{15}{2}\mathcal{G}\mu_1m_B\frac{a_1^2}{D^3}\sin\theta\cos\theta\,\widehat{\hat{\bm{X}}'\times\hat{\bm{D}}}~.
\end{equation}
The components in the $(X',Y',Z')$ coordinates are
\begin{align}
\dot{G}_{1X'} &= 0~, \nonumber\\
\dot{G}_{1Y'} &= -\frac{15}{2}\mathcal{G}\mu_1m_B a_1^2 \frac{\hat{D}_{X'}\hat{D}_{Z'}}{D^3}~, {\rm ~and}\nonumber\\
\dot{G}_{1Z'} &= \frac{15}{2}\mathcal{G}\mu_1m_B a_1^2 \frac{\hat{D}_{X'}\hat{D}_{Y'}}{D^3}~,
\end{align}
where $\hat{D}_{X',Y',Z'}$ are components of $\hat{\bm{D}}$.

In order to obtain the angular momentum oscillation amplitude during one period of the outer orbit assuming the inner orbit dissipation is not important, we are interested in the integrals
\begin{equation*}
I_{X'Z'}\equiv\int\frac{\hat{D}_{X'}\hat{D}_{Z'}}{D^3}dt~,~{\rm and}~~I_{X'Y'}\equiv\int\frac{\hat{D}_{X'}\hat{D}_{Y'}}{D^3}dt~.
\end{equation*}
It is then convenient to express the variables in the $(X,Y,Z)$ coordinates, where $\hat{\bm{D}}=(\cos f,\sin f,0)^{\rm T}$. Using the rotation matrices, the components are given by
\begin{align}
\left(\begin{array}{c}
\hat{D}_{X'} \\ \hat{D}_{Y'} \\ \hat{D}_{Z'} \end{array}\right)
&= R_Z(-\alpha_1) R_X(-i) R_Z(g) \left(\begin{array}{c}
\cos f \\ \sin f \\ 0 \end{array}\right) \nonumber\\
&= \left(\begin{array}{cc}
a_{X'} & b_{X'} \\
a_{Y'} & b_{Y'} \\
a_{Z'} & b_{Z'} \end{array}\right)
\left(\begin{array}{c}
\cos f \\ \sin f \end{array}\right)~,
\end{align}
where the elements are given by
\begin{align}
a_{X'}&\equiv \sin\alpha_1 \cos i \sin g+\cos\alpha_1\cos g~, \nonumber\\
b_{X'}&\equiv \sin\alpha_1\cos i\cos g-\cos\alpha_1\sin g~, \nonumber\\
a_{Y'}&\equiv \cos\alpha_1\cos i \sin g -\sin\alpha_1\cos g~, \nonumber\\
b_{Y'}&\equiv \cos\alpha_1\cos i\cos g+\sin\alpha_1\sin g~, \nonumber\\
a_{Z'}&\equiv -\sin i \sin g~,~~{\rm and} \nonumber\\
b_{Z'}&\equiv -\sin i\cos g~.
\label{eq:coeff_apbp}
\end{align}

The integrals of interest are just combinations of the following integrals
\begin{align}
&I_{XX}\equiv\int\frac{\cos^2 f}{D^3}dt~,~I_{XY}\equiv\int\frac{\sin f\cos f}{D^3}dt~,{\rm ~and}
\nonumber \\
&~I_{YY}\equiv\int\frac{\sin^2 f}{D^3}dt~.
\end{align}
The first integral is evaluated as follows
\begin{align}
I_{XX} &=\frac{1}{na^3}\int\cos^2 f\frac{(1+e\cos f)^3}{(1-e^2)^3}\frac{(1-e^2)^{3/2}}{(1+e\cos f)^2}df \nonumber\\
&=\frac{2f+\sin 2f + e\left(3\sin f+\frac{1}{3}\sin 3f\right)}{4n a^3(1-e^2)^{3/2}}~.
\end{align}
Similarly, we have
\begin{equation}
I_{XY} =\frac{1+\frac{4}{3}e-\cos 2f - e\left(\cos f+\frac{1}{3}\cos 3f\right)}{4n a^3(1-e^2)^{3/2}}
\end{equation}
and
\begin{equation}
I_{YY} =\frac{2f-\sin 2f + e\left(\sin f-\frac{1}{3}\sin 3f\right)}{4n a^3(1-e^2)^{3/2}}~,
\end{equation}
where the $\sin 2f$ and $\cos 2f$ terms give evection with a frequency twice the outer orbital frequency, and $n$ is the mean motion of the outer orbit. The integrals of interest are thus
\begin{align}
I_{X'Z'} &= a_{X'}a_{Z'}I_{XX}+(a_{X'}b_{Z'}+a_{Z'}b_{X'})I_{XY}+b_{X'}b_{Z'}I_{YY}~~{\rm and}\\
I_{X'Y'} &= a_{X'}a_{Y'}I_{XX}+(a_{X'}b_{Y'}+a_{Y'}b_{X'})I_{XY}+b_{X'}b_{Y'}I_{YY}~,
\end{align}
and the angular momentum changes are given by
\begin{equation}
\Delta G_{1Y'} = -\frac{15}{2}\mathcal{G}\mu_1m_B a_1^2 I_{X'Z'}~~{\rm and}~~
\Delta G_{1Z'} = \frac{15}{2}\mathcal{G}\mu_1m_B a_1^2 I_{X'Y'}~.
\label{eq:DG1yz}
\end{equation}
Since we are only interested in the oscillation of the angular momentum around its mean (``orbital averaged'') evolution and the angular momentum components are linear combinations of $I_{XX},I_{XY},I_{YY}$, the oscillatory parts depend only on the oscillatory parts of the $I$'s. Subtracting the mean parts, we obtain
\begin{align}
I_{XX}^{(\rm osc)} =&\frac{2(f-M)+\sin 2f + e\left(3\sin f+\frac{1}{3}\sin 3f\right)}{4n a^3(1-e^2)^{3/2}}~,\nonumber\\
I_{YY}^{(\rm osc)} =&\frac{2(f-M)-\sin 2f + e\left(\sin f-\frac{1}{3}\sin 3f\right)}{4n a^3(1-e^2)^{3/2}}~,~~{\rm and}\nonumber\\
I_{XY}^{(\rm osc)} =&-\frac{\cos 2f-\langle\cos 2f\rangle_{\scriptscriptstyle M}}{4n a^3(1-e^2)^{3/2}} \nonumber\\ 
&-\frac{e\left[\cos f-\langle\cos f\rangle_{\scriptscriptstyle M}+\frac{1}{3}\left(\cos 3f-\langle\cos 3f\rangle_{\scriptscriptstyle M}\right)\right]}{4n a^3(1-e^2)^{3/2}}~,
\end{align}
where $\langle\cdots\rangle_{\scriptscriptstyle M}$ represents the time-averaged mean value, or equivalently the averaged value over the mean anomaly $M$. The mean values of the oscillatory parts are zero. Utilizing the Keplerian relation $dM = (1-e^2)^{3/2}(1+e\cos f)^{-2}\,df$, we can evaluate the orbit averages
\begin{align}
&\langle\cos f\rangle_{\scriptscriptstyle M} \equiv \oint\frac{dM}{2\pi}\cos f = -e~,\nonumber\\
&\langle\cos 2f\rangle_{\scriptscriptstyle M} = 3-2\sqrt{1-e^2}-\frac{2}{e^2}\left(1-\sqrt{1-e^2}\right)~,~~{\rm and}\nonumber\\
&\langle\cos 3f\rangle_{\scriptscriptstyle M} = 3e-\frac{4}{e}\left(3-2\sqrt{1-e^2}\right)+\frac{8}{e^3}\left(1-\sqrt{1-e^2}\right)~,\nonumber\\
\end{align}
so that $I_{XY}^{(\rm osc)}$ simplifies to
\begin{align}
I_{XY}^{(\rm osc)} =& -\frac{1}{4n a^3(1-e^2)^{3/2}}
\Bigl[\cos 2f + e\left(\cos f+\frac{1}{3}\cos 3f\right) +\frac{1}{3}
\nonumber \\
&-\frac{2(1-e^2)}{3e^2}\left(1-\sqrt{1-e^2}\right)\Bigr].
\end{align}
Assuming that the angular momentum oscillation is small, the eccentricity oscillation is mainly contributed by $\Delta G_{1Z'}$. The oscillatory part is
\begin{align}
\Delta G_{1Z'}^{(\rm osc)} =& \frac{15\mathcal{G}\mu_1m_Ba_1^2}{8na^3(1-e^2)^{3/2}}\nonumber\\
& \times\left\lbrace(a_{X'}a_{Y'}-b_{X'}b_{Y'})\sin 2f - (a_{X'}b_{Y'}+b_{X'}a_{Y'})\cos 2f \right.\nonumber\\
& + e(3a_{X'}a_{Y'}+b_{X'}b_{Y'})\sin f - e(a_{X'}b_{Y'}+b_{X'}a_{Y'})\cos f \nonumber\\
& +\frac{e}{3}(a_{X'}a_{Y'}-b_{X'}b_{Y'})\sin 3f
\nonumber \\ &
- \frac{e}{3}(a_{X'}b_{Y'}+b_{X'}a_{Y'})\cos 3f \nonumber\\
& +2(a_{X'}a_{Y'}+b_{X'}b_{Y'})(f-M) \nonumber\\
& -\left.(a_{X'}b_{Y'}+b_{X'}a_{Y'})\left[\frac{1}{3}-\frac{2(1-e^2)}{3e^2}\left(1-\sqrt{1-e^2}\right)\right] \right\rbrace~.
\label{eq:DeltaG1z_prime1}
\end{align}
This can be simplified using the expressions in Eq.~(\ref{eq:coeff_apbp}) and trigonometric identities:
\begin{align}
\Delta G_{1Z'}^{(\rm osc)} =& \frac{15\mathcal{G}\mu_1m_Ba_1^2}{8na^3(1-e^2)^{3/2}}\left\lbrace-\frac{\cos^2 i+1}{2}\sin 2\alpha_1\right.\nonumber\\
& \times\left[\sin(2f+2g)+e\sin(f+2g)+\frac{e}{3}\sin(3f+2g)\right] \nonumber\\
&- \cos i\cos 2\alpha_1\nonumber\\
& \times\left[\cos(2f+2g)+e\cos(f+2g)+\frac{e}{3}\cos(3f+2g)\right] \nonumber\\
& - (f-M+e\sin f)\sin^2 i\sin 2\alpha_1 \nonumber\\
&-\left.(a_{X'}b_{Y'}+b_{X'}a_{Y'})\left[\frac{1}{3}-\frac{2(1-e^2)}{3e^2}\left(1-\sqrt{1-e^2}\right)\right] \right\rbrace~.
\label{eq:DeltaG1z_prime2}
\end{align}

\subsection{Upper Bound}
\label{App:evec_bound}
Equation~(\ref{eq:DeltaG1z_prime2}) shows that the inner orbit angular momentum oscillates around its secular value on the timescale comparable to the outer orbit period, where $f$ is the fastest-varying quantity indicating the position of the tertiary. Since usually the timescales of the precession of both orbits are longer than the variation of the mutual inclination, we can obtain a rough estimation of the highest possible eccentricity of the inner orbit using some inequalities to eliminate $g$ and $\alpha_1$ in order to simplify the calculation.

For the $e=0$ case, $f-M=0$, we can choose $g=0$ without loss of generality, then Eq.~(\ref{eq:DeltaG1z_prime2}) reduces to
\begin{align}
\Delta G_{1Z'}^{(\rm osc)} =& \frac{15\mathcal{G}\mu_1m_Ba_1^2}{8na^3}\times\nonumber\\
&\left(-\frac{\cos^2 i+1}{2}\sin 2\alpha_1\sin 2f- \cos i\cos 2\alpha_1\cos 2f\right)~.
\end{align}
The amplitude is constrained by
\begin{align}
\left\vert\Delta G_{1Z'}^{(\rm osc)}\right\vert&\leq\frac{15\mathcal{G}\mu_1m_Ba_1^2}{8na^3}
\nonumber \\ & \times
\sqrt{\left(\frac{\cos^2 i+1}{2}\right)^2\sin^2 2\alpha_1 +\cos^2 2\alpha_1\cos^2i} \nonumber\\
&\leq\frac{15\mathcal{G}\mu_1m_Ba_1^2}{8na^3}\left(\frac{1+\cos^2i}{2}\right) ~.
\end{align}
Note that $n=\sqrt{\mathcal{G}(m_A+m_B)/a^3}$, we obtain the upper limit of the amplitude
\begin{equation}
\frac{\left\vert\Delta G_{1Z'}^{(\rm osc)}\right\vert}{\mu_1} \leq \frac{15}{8}\left(\frac{a_1}{a}\right)^2\left(\frac{1+\cos^2i}{2}\right)\sqrt{\mathcal{G}(m_A+m_B)a}\frac{m_B}{m_A+m_B}~,
\end{equation}
which is consistent with \cite{2005MNRAS.358.1361I}.

For $e\neq 0$, we can roughly estimate the upper limit of the amplitude as
\begin{align}
\left\vert\Delta G_{1Z'}^{(\rm osc)}\right\vert<&\frac{15\mathcal{G}\mu_1m_Ba_1^2}{8na^3(1-e^2)^{3/2}}\nonumber\\
&\times \left[ \left(1+e+\frac{e}{3}\right)\frac{1+\cos^2i}{2} + (\vert e\sin f\vert + \vert f-M\vert)\sin^2 i\right. \nonumber\\
&+\left.\left\vert a_{X'}b_{Y'}+b_{X'}a_{Y'}\right\vert \left\vert\frac{1}{3}-\frac{2(1-e^2)}{3e^2}\left(1-\sqrt{1-e^2}\right)\right\vert\right]\nonumber\\
<&\frac{15\mathcal{G}\mu_1m_Ba_1^2}{8na^3(1-e^2)^{3/2}}\nonumber\\
&\times\left\lbrace\left[\frac{4}{3}+\frac{4e}{3}-\frac{2(1-e^2)}{3e^2}\left(1-\sqrt{1-e^2}\right)\right]\frac{1+\cos^2i}{2} \right. \nonumber\\
&\left.+\left(e+\vert f-M\vert\right)\sin^2 i \right\rbrace ~,
\end{align}
where $\vert f-M\vert<\pi$. The accurate upper limit of $\vert f-M\vert$ can be derived as follows
\begin{align}
f-M=&f-(1-e^2)^{3/2}\int_0^f\frac{df'}{(1+e\cos f')^2}\nonumber\\
=&f-2\arctan\left(\sqrt{\frac{1-e}{1+e}}\tan\frac{f}{2}\right)+e\sqrt{1-e^2}\frac{\sin f}{1+e\cos f}\nonumber\\
=&2\arctan y-2\arctan\left(\sqrt{\frac{1-e}{1+e}}y\right)+\frac{2e\sqrt{1-e^2}y}{1+e+(1-e)y^2}\nonumber\\
&
\end{align}
(for $|f|\le\pi$) where $y\equiv\tan(f/2)$ and the maximum is reached when
\begin{equation}
y=y_0\equiv\sqrt{\frac{e+1-(1-e^2)^{3/4}}{e-1+(1-e^2)^{3/4}}}~,
\end{equation}
or equivalently, when $e\cos f+1=(1-e^2)^{3/4}$. In general, we have
\begin{align}
\frac{\left\vert\Delta G_{1Z'}^{(\rm osc)}\right\vert}{\mu_1}<\frac{15}{8}\left(\frac{a_1}{a}\right)^2\sqrt{\mathcal{G}(m_A+m_B)a}\frac{m_B}{m_A+m_B} \frac{F(e,i)}{(1-e^2)^{3/2}} ~,
\end{align}
where $F(e,i)$ is defined as
\begin{align}
F(e,i)\equiv &\left(e+\vert f-M\vert_{y=y_0}\right)\sin^2 i\nonumber\\
 &+ \left[\frac{4}{3}+\frac{4e}{3}-\frac{2(1-e^2)}{3e^2}\left(1-\sqrt{1-e^2}\right)\right]\frac{1+\cos^2i}{2}~.
 \label{eq:F-ei}
\end{align}
Assuming that the energy dissipation is negligible during the orbital period of the outer orbit, we can estimate the local maximum of the eccentricity of the inner orbit due to evection as
\begin{align}
e_1^{(\rm bound)} &= \sqrt{1-\frac{(G_1-\vert\Delta G_{1Z'}\vert)^2}{L_1^2}} \nonumber\\
& \leq \sqrt{1-\left[\sqrt{1-e_1^2}-\frac{15}{8}\sqrt{\frac{m_B^2a_1^3}{m_A (m_A+m_B)a^3}}\frac{F(e,i)}{(1-e^2)^{3/2}}\right]^2}~,
\label{eq:e_evec}
\end{align}
where $L_1=\mu_1\sqrt{\mathcal{G}m_Aa_1}$ is one of the Delaunay's variables. Note that the amplitude upper bound is just a rough estimate and may not be reached for every $e$.

\subsection{True evection envelope (TEE)}
\label{app:evec_TEE}
The upper bound derived above often overestimate the highest eccentricity, resulting in over-enhanced merger rates. A more accurate way to estimate the eccentricity amplitude due to evection is to find its ``true envelope'', \ie find the minimum of $\Delta G_{1Z'}^{(\rm osc)}$ in Eq.~(\ref{eq:DeltaG1z_prime1}) for $f$ in $[-\pi,\pi]$ at each time step, without using any inequality.

It is important to note that the orbital parameters used in the above derivation only involve a triple system with two orbital planes, so that the parameters such as $g$, $\alpha_1$ are defined with respect to the intersection line of the two orbits and $i$ is actually the mutual inclination. These parameters are not the same as what we use in the main body of this paper, where the parameters are defined with respect to a rest coordinate system (with axes $x,y,z$). To avoid confusion, we will denote the parameters defined in the triple system context with a subscript $_{\scriptscriptstyle \rm A}$, indicating that they are defined for the ``inner orbit A - outer orbit'' system, while other parameters follow our convention in the main body of this paper. We define the direction of the ascending node of the inner orbit with respect to the outer orbit as $\hat{\bm{\Omega}}_{\scriptscriptstyle \rm A}\equiv \widehat{\hat{\bm{G}}\times\hat{\bm{G}_1}}$.

Now we need to express $g_{\scriptscriptstyle \rm A}$ and $\alpha_{1\scriptscriptstyle \rm A}$ (\ie $g$ and $\alpha_1$ in Eq.~\ref{eq:coeff_apbp}) in the coordinate system. Since they are both in $[0,2\pi)$, we need both their cosines and sines. From the definition, we have
\begin{align}
&\cos\alpha_{1\scriptscriptstyle \rm A} = \hat{\bm{X}}'\cdot\hat{\bm{\Omega}}_{\scriptscriptstyle \rm A}~,~~\hat{\bm{G}}_1 = \widehat{\hat{\bm{\Omega}}_{\scriptscriptstyle \rm A}\times\hat{\bm{X}}'}
~,
\nonumber \\ &
\cos g_{\scriptscriptstyle \rm A} = \hat{\bm{X}}\cdot\hat{\bm{\Omega}}_{\scriptscriptstyle \rm A}~,~~{\rm and}~~\hat{\bm{G}} = \widehat{\hat{\bm{\Omega}}_{\scriptscriptstyle \rm A}\times\hat{\bm{X}}}~.
\end{align}
In the $xyz$ coordinate, we have
\begin{align}
\hat{\bm{X}}' &= R_Z(h_1)R_X(i_1)R_Z(\alpha_1)(1,0,0)^{\rm T}\nonumber\\
 & = \left(\begin{array}{c}
\cos h_1\cos\alpha_1 - \sin h_1\cos i_1\sin\alpha_1 \\ \sin h_1\cos\alpha_1 + \cos h_1\cos i_1\sin\alpha_1 \\
\sin i_1\sin\alpha_1 \end{array}\right)~,
\end{align}
\begin{align}
\hat{\bm{X}} &= R_Z(h)R_X(i)R_Z(g)(1,0,0)^{\rm T}\nonumber\\
 & = \left(\begin{array}{c}
\cos h\cos g - \sin h\cos i\sin g \\ \sin h\cos g + \cos h\cos i\sin g \\
\sin i\sin g \end{array}\right)~,\\
\end{align}
After the cosines are evaluated, their sines are determined by
\begin{equation}
\sin\alpha_{1\scriptscriptstyle \rm A} = \pm\sqrt{1-\cos^2 g_{\scriptscriptstyle \rm A}}~~{\rm and}~~\sin g_{\scriptscriptstyle \rm A} = \pm\sqrt{1-\cos^2 g_{1\scriptscriptstyle \rm A}}~,
\end{equation}
where the signs should match with the signs of $(\hat{\bm{\Omega}}_{\scriptscriptstyle \rm A}\times\hat{\bm{X}}')\cdot\bm{G}_1$ and $(\hat{\bm{\Omega}}_{\scriptscriptstyle \rm A}\times\hat{\bm{X}})\cdot\bm{G}$, respectively.

Finally the TEE gives the maximal eccentricity at every step as
\begin{equation}
e_1^{(\rm TEE)} = \sqrt{1-\frac{(G_1+\min\lbrace\Delta G_{1Z'}^{\rm (osc)}\rbrace)^2}{L_1^2}}~.
\label{eq:e_TEE}
\end{equation}

The results of TEE and the ``upper bound'' are shown in Figure \ref{fig:rebound_triple151}.

%%%%%%%%%%%%%%%%%%%%%%%%%%%%%%%%%%%%%%%%%%%%%%%%%%

% Don't change these lines
\bsp	% typesetting comment
\label{lastpage}
\end{document}